\documentclass[aps,prd,twocolumn,nofootinbib,showpacs,superscriptaddress,groupedaddress]{revtex4-1}

\usepackage[T1]{fontenc} 

\usepackage{amsmath}
\usepackage{amssymb}
\usepackage{graphicx}
\usepackage{bm}
\usepackage{epsfig}
\usepackage{amsfonts}
\usepackage{color}
\usepackage{mathtools}
\usepackage[section]{placeins}

\newcommand{\be}{\begin{equation}}
\newcommand{\ee}{\end{equation}}

\newcommand{\bea}{\begin{eqnarray}}
\newcommand{\eea}{\end{eqnarray}}
\newcommand{\bean}{\begin{eqnarray*}}
\newcommand{\eean}{\end{eqnarray*}}

\newcommand{\xv}{{\mathbf x}}

\newcommand{\nn}{\nonumber}

\newcommand{\bB}{{\bf B}}

\newcommand{\bDl}{{\bf\Delta}}
\newcommand{\bE}{{\bf E}}
\newcommand{\bsigma}{{\boldsymbol\sigma}}

\begin{document}

\title {Lattice NRQCD study of S- and P-wave bottomonium states in a
  thermal medium with $N_f=2+1$ light flavors}
\author{Seyong Kim}
\affiliation{Department of Physics, Sejong University, Seoul 143-747, Korea}
\author{Peter Petreczky}
\affiliation{Physics Department, Brookhaven National Laboratory,
  Upton, New York 11973, USA}
\author{Alexander Rothkopf}
\affiliation{Institute for Theoretical Physics,  Heidelberg
  University, Philosophenweg 16, 69120 Heidelberg, Germany}
\date{\today}

\begin{abstract}
We investigate the properties of S-and P-wave bottomonium
  states in the vicinity of the deconfinement transition
  temperature. The light degrees of freedom are represented by
  dynamical lattice quantum chromodynamics (QCD) configurations of the
  HotQCD collaboration with $N_f = 2 + 1$ flavors. Bottomonium correlators are
  obtained from bottom quark propagators, computed in nonrelativistic
  QCD (NRQCD) under the background of these gauge field
  configurations. The spectral functions for the $^3S_1$ ($\Upsilon$)
  and $^3P_1$ ($\chi_{b1}$) channel are extracted from the Euclidean
  time correlators using a novel Bayesian approach in the temperature
  region $140 {\rm MeV} \le T \le 249 {\rm MeV}$ and the results
  are contrasted to those from the standard maximum entropy method. We
  find that the new Bayesian approach is far superior to the maximum
  entropy method.  It enables us to study reliably the presence or absence of
  the lowest state signal in the spectral function of a certain channel,
  even under the limitations present in the
  finite temperature setup. We find that $\chi_{b1}$ survives up to
  $T=249 {\rm MeV}$, the highest temperature considered
  in our study and put stringent constraints on the size of the medium
  modification of $\Upsilon$ and $\chi_{b1}$ states.
\end{abstract}

\maketitle


\flushbottom


\section{Introduction}
 \label{sec:intro}

Heavy quarkonium plays an important role in furthering our
quantitative understanding of quantum chromodynamics (QCD) and has
been the focus of many experimental and theoretical studies. In
particular the large heavy quark mass, relative to the intrinsic
scales of its environment, provides a basis for various effective
field theory (EFT) descriptions and allows us to disentangle the short
distance perturbative aspects from the nonperturbative long distance
effects in QCD \cite{Brambilla:2010cs,Brambilla:2004jw,Bodwin:1994jh}.

The in-medium modification of quarkonium properties, the most dramatic
being its melting, has been suggested as a clean signature of
quark-gluon plasma (QGP) formation in heavy-ion collisions
\cite{Matsui:1986dk, Karsch:1987pv, Satz:2006kba}. In our current
understanding, the medium screens the interactions between the heavy
quark--antiquark pair (Debye screening) and hence weakens the binding
between them. Scattering of medium partons off the gluons, which
mediate the interquark binding (Landau damping)
\cite{Laine:2006ns,Beraudo:2007ky,Brambilla:2008cx} and absorption of
gluons from the medium (singlet-octet transitions)
\cite{Brambilla:2008cx} further disturbs the bound state. Ultimately
the combination of these effects will prevent the existence of
quarkonium states as temperature and/or density increases. Such
mechanisms may experimentally manifest themselves as reduced
production rates of heavy quarkonium. Early experiments indeed
confirmed $J/\psi$ suppression \cite{Baglin:1990iv} but it was soon
discovered that the actual charmonium production in nuclear collision
is far more complicated \cite{Vogt:1991qd} than a simple screening
argument suggests.

There are many competing processes that lead to a modification of the
measured yields: cold nuclear matter effects, shadowing, gluon
saturation and even regeneration of charmonium (instead of
suppression) may occur \cite{Rapp:2008tf}. Thus, careful understanding
of each effect is necessary before charmonium suppression, seen in
experiments, can be attributed to QGP formation. In this regard, the
bottomonium system may turn out to be a more appropriate candidate to
investigate the physics of melting. The bottom quark mass is
significantly heavier than the charm quark mass and hence the effects
of the medium modification are expected to be dominant over
e.g. regeneration.

Interestingly, the CMS experiment at the LHC discovered a clear pattern of
``sequential suppression'' among the $\Upsilon$ states: the dimuon
distribution around the $\Upsilon$ mass in lead-lead collisions,
compared to that in proton-proton collisions, revealed a substantial
reduction of $\Upsilon (2S)/\Upsilon (1S)$ and $\Upsilon (3S)/\Upsilon
(1S)$ production rates \cite{Chatrchyan:2011pe, Chatrchyan:2012lxa}.

One way of exploring the in-medium modification of bottomonium from
first principles is to use a lattice regularized form of
nonrelativistic QCD (NRQCD)
\cite{Thacker:1990bm,Lepage:1992tx,Fingberg:1997qd}, an effective
field theory of QCD. It is formulated using nonrelativistic
Pauli spinors, which propagate in the background of otherwise
relativistic medium gauge fields, defined on a spacetime
lattice. Below the heavy quark mass scale, observables of QCD which
are dominated by infrared physics are well reproduced within this
effective description. The nonperturbative character of lattice NRQCD
is well suited for the regime where $T$ is only slightly larger than
the transition temperature (in which the strong interactions can still
be considered strong \cite{Gyulassy:2004zy, Muller:2007rs}). Being
based on an expansion of the QCD action, the effective description can
be systematically improved to reproduce more and more closely
relativistic QCD itself.

If one further assumes that the inverse size of the heavy quark bound
states is much larger than the binding energy, another effective
theory, namely potential NRQCD (pNRQCD) can be constructed. The
potentials entering in this EFT can be related to Wilson loops and in
the simplest formulation one encounters a quantum mechanical problem
\cite{Laine:2006ns,Beraudo:2007ky,Brambilla:2008cx,Rothkopf:2009pk,Rothkopf:2011db}.
Depending on the relation between inverse size and the temperature of
the system, the potentials can be also temperature dependent. However,
the scale separation required for utilizing pNRQCD is more difficult
to justify and in this sense NRQCD is more robust.

Recent lattice studies of in-medium bottomonium on anisotropic
lattices at nonzero temperature concluded that the ground state
$(1S)$ peak of the $\Upsilon$ spectral function obtained from the
maximum entropy method (MEM) \cite{Jarrell:1996,Asakawa:2000tr}
survives up to $\sim 2 T_c$, and the excited state $(2S, 3S)$ peaks in
this vector channel spectral function disappear gradually as the
temperature increases above $T_c$ \cite{Aarts:2010ek, Aarts:2011sm,
  Aarts:2014cda}. A study of P-wave bottomonium states, such as
$\chi_{b1}$, in lattice NRQCD using the MEM furthermore concluded that
the ground state ``melts'' almost immediately above $T_c$
\cite{Aarts:2010ek,Aarts:2013kaa,Aarts:2014cda}. It has to be kept in
mind however that e.g. the pion mass in these studies remained rather
heavy with $M_\pi\simeq400$ MeV.

In this work, we report on a lattice NRQCD study of bottomonium in 2+1
flavor QCD below and above the chiral transition using HotQCD
configurations, generated with highly improved staggered Quark (HISQ)
action on isotropic $48^3 \times 12$ lattices \cite{Bazavov:2011nk}
(see a preliminary report in \cite{Kim:2013seh}). The gauge
configurations have been generated for a physical value of the
strangle quark mass $m_s$ and light quark masses $m_l=m_s/20$ that in
the continuum limit correspond to the pion mass of $M_\pi=161$ MeV, only
slightly above the physical pion mass of $140$ MeV.

Previous lattice NRQCD studies at nonzero temperature used a fixed
lattice scale approach, where the inverse temperature is available
only at integer steps. Here we achieve a finer temperature scan by
changing the lattice spacing instead. Note that at each coupling an
accompanying zero temperature lattice simulation is required to set
the absolute energy scale.

An accurate and precise extraction of spectral functions using the standard MEM is
difficult, since the number of lattice points in imaginary time
direction is typically small. More importantly it is the extent
of the imaginary time in physical units that decreases with increasing
temperature. The underlying technical reason is that the small number of lattice sites in temporal direction
$(N_\tau = 12)$ limits the number of basis functions available in the
MEM search space, which relies on a singular value decomposition.

Instead, we deploy a new Bayesian method
\cite{Burnier:2013nla,Burnier:2013esa} and compare its results with
those obtained using conventional Bryan's MEM. Based on the same data,
the new method produces in general sharper, i.e. more highly resolved
spectral features for both $\Upsilon$ and $\chi_{b1}$ and allows
better control over most of the systematic errors associated with the
spectral function reconstruction. While systematic uncertainties
resulting from a small $N_\tau$ still prevents us from providing
quantitative estimates of the in-medium mass shifts and widths, the
higher precision allows us to put stringent upper limits on these
effects.

Despite the difference, $\Upsilon$ spectral functions
from the improved Bayesian method and those from MEM both show a
similar qualitative temperature dependence: the ground state peak of
the $\Upsilon$ channel survives up to the highest temperature studied $(T =
249 {\rm MeV})$. For the P-wave channel, the
difference is more substantial. The first peak in the $\chi_{b1} $
spectral function obtained from the improved Bayesian method survives
to $249{\rm MeV}$ but in the result based on the MEM it disappears for
$T \gtrsim 211 {\rm MeV}$.

In the first part of Sec. \ref{sec:nrqcd} we briefly review the
lattice formulation of NRQCD and specify pertinent technical details
underlying the measurements of the bottomonium correlators. The second
part contains the basics of the novel Bayesian method, which we deploy
in the extraction of the bottomonium spectra. Section \ref{sec:zeroT}
describes the calibration of the NRQCD mass scale carried out on low
temperature lattices, while in Sec. \ref{sec:nonzeroT} we present
the central results of our study, the spectral properties of in-medium
bottomonium at temperatures around the deconfinement transition
temperature. We end the main part of the manuscript in Sec.
\ref{sec:conclusion} with concluding remarks. The systematic
uncertainties of the spectral reconstructions, as well as dependencies
on the NRQCD discretization are investigated and discussed in 
Appendixes \ref{sec:systematics} and \ref{sec:systematics2}
respectively.

\section{Numerical Methods}
\label{sec:nrqcd}

A pronounced separation of scales invites the use of effective field
theoretical methods for the description of in-medium
bottomonium. Indeed, the heavy quark mass $M_b\simeq4.6{\rm GeV}$
\cite{Beringer:1900zz} and the intrinsic scale of QCD $\Lambda_{\rm
  QCD}\sim 200{\rm MeV}$, as well as the typical momentum exchange
within a possible bound state lie widely apart. The presence of a
thermal medium introduces an additional scale $T$, which however also
lies well below the bottom quark rest mass in current heavy-ion
collision experiments.

In an EFT, the physics above the energy of interest is
integrated out \cite{Brambilla:2004jw}, which requires one to
determine the relevant degrees of freedom within the hierarchy of
scales present in the system. To avoid the intricacies of relative
scale ordering in the presence of finite temperature, we will deploy
here a lattice regularized version of NRQCD, where only the hard scale
$M_b$, i.e. the bottomonium rest mass is integrated out. The heavy
quarks appear as nonrelativistic Pauli spinors, which propagate under
the background of the light medium degrees of freedom (in our case
gluons, up, down and strange quarks), which themselves are in thermal
equilibrium.

The nonperturbative character of the EFT description is of the essence,
since close to the transition temperature
the QGP still remains strongly correlated and temperature effects on the
binding dynamics ($ \sim M_bv$), which are our main interest, will not
be perturbative.

\subsection*{Lattice NRQCD}

In order to investigate the properties of bottomonium in a thermal
medium, we compute the correlators of heavy quarkonium using a lattice
discretization of the ${\cal O} (v^4)$ NRQCD Lagrangian
\cite{Lepage:1992tx, Davies:1994mp, Gray:2005ur} for bottom quarks,
\begin{equation}
\label{LNRQCD}
{\cal L} = {\cal L}_0 + \delta {\cal L},
\end{equation}
with
\begin{equation}
\label{LNRQCD_1}
{\cal L}_0 = \psi^\dagger \left(D_\tau - \frac{{\bold D^2}}{2M_b} \right) \psi +
\chi^\dagger \left(D_\tau + \frac{{\bold D^2}}{2M_b} \right) \chi,
\end{equation}
and
\begin{widetext}
\begin{eqnarray*}
\label{LNRQCD_2}
\delta {\cal L} = &&\hm - \frac{c_1}{8M_b^3} \left[\psi^\dagger ({\bold D^2})^2
  \psi - \chi^\dagger ({\bold D^2})^2 \chi \right] + c_2 \frac{ig}{8M_b^2}\left[\psi^\dagger \left({\bold
    D}\cdot{\bold E} - {\bold E}\cdot{\bold D}\right) \psi +
  \chi^\dagger \left({\bold D}\cdot {\bold E} - {\bold E}\cdot{\bold
    D} \right) \chi \right] \nonumber \\
&&\hm - c_3 \frac{g}{8M_b^2}\left[\psi^\dagger
  \bsigma\cdot\left({\bold D}\times{\bold E}-{\bold E}\times{\bold
    D}\right)\psi + \chi^\dagger {\bold
    \sigma}\cdot\left({\bold D}\times{\bold E}-{\bold E}\times{\bold
    D}\right)\chi \right] \nonumber - c_4 \frac{g}{2M_b} \left[\psi^\dagger {\bold \sigma}\cdot{\bold B} \psi -
    \chi^\dagger {\bold \sigma} \cdot {\bold B} \chi \right].
\end{eqnarray*}
\end{widetext}

Here $D_\tau$ and ${\bold D}$ are gauge covariant temporal and
spatial derivatives, $\psi$ denotes the heavy quark and $\chi$ denotes
the heavy antiquark. From the discretized version of Eq. \eqref{LNRQCD},
the lattice NRQCD propagator for the bottom quark is computed as
an initial value problem.
\begin{eqnarray}
G (\xv, \tau_0) = &&\hm S(\xv), \nn\\
G (\xv, \tau_1) = &&\hm  \left(1 - \frac{H_0}{2n}\right)^n
U_4^\dagger(\xv, 0) \left(1 - \frac{H_0}{2n}\right)^n G(\xv,0), \nn \\
G (\xv, \tau_i ) = &&\hm  \left(1 - \frac{H_0}{2n}\right)^n
U_4^\dagger(\xv, \tau) \left(1 - \frac{H_0}{2n}\right)^n \nonumber\\
&&\times \left(1 -\delta H \right)  G (\xv, \tau_{i-1}).\label{NRQCDEvolEq}
\end{eqnarray}
$S(\xv)$ denotes an appropriate complex valued random point source,
diagonal in spin and color, which is used to improve the
signal-to-noise ratio. In the continuum formulation the initial
condition for $G (\xv, \tau)$ corresponds to a delta function, which
we approximate on the lattice through averaging multiple correlators,
started from random sources on different slices $\tau_{\rm start}$
along Euclidean time e.g.
\begin{align}
 S^{\Upsilon}(\xv,\tau_{\rm start})=\eta(\xv,\tau_{\rm start}), \quad \langle \eta^\dagger(\xv) \eta(\xv') \rangle_{\tau_{\rm start}}=\delta_{\xv\xv'}
\end{align}

The lowest-order Hamiltonian reads
\be
H_0 = - \frac{\Delta^{(2)}}{2M_b},
\ee while
\begin{eqnarray}
\delta H = &&\hm - \frac{(\Delta^{(2)})^2}{8 M_b^3} + \frac{ig}{8 M_b^2}
(\bDl^{\pm}\cdot \bE - \bE\cdot \bDl^{\pm})\nonumber \\
&&- \frac{g}{8 M_b^2} \bsigma \cdot
  (\bDl^{\pm} \times \bE - \bE\times \bDl^{\pm})   \nonumber \\
&&\hm
- \frac{g}{2 M_b} \bsigma\cdot\bB
 + \frac{a^2\Delta^{(4)}}{24 M_b} - \frac{a (\Delta^{(2)})^2}{16 n M_b^2}.
\label{eq:deltaH}
\end{eqnarray}
Here, $n$ denotes a parameter, which controls the effective temporal
step size in Euclidean time, when propagating the Green's function on
the lattice. Choosing an appropriate value is essential to the
stability of the high momentum behavior of the propagator $G$. We use
$n=2$, in anticipation of the characteristic values of $M_b a_s$,
which will arise on the lattices used in this study. We have checked
by varying the parameter $n$ up to values of four that the choice
$n=2$ already allows us to capture the bottomonium bound state physics
in a robust manner. A detailed description of these tests can be found
in Appendix \ref{sec:systematics2}.

The lattice covariant derivative $\Delta$ is defined as
\begin{eqnarray}
a \Delta_i^{+} \psi (\xv, \tau) &=& U_i (\xv, \tau) \psi (\xv + \hat{i} a ,
\tau) - \psi (\xv, \tau) \nonumber \\
a \Delta_i^{-} \psi (\xv, \tau) &=& \psi (\xv, \tau) - U_i^\dagger (\xv
- \hat{i} a, \tau) \psi (\xv - \hat{i} a , \tau) \nonumber \\
\Delta^{(2)} &=& \sum_{i=1}^{3} \Delta_i^{+} \Delta_i^{-}, \;
\Delta^{(4)} = \sum_{i=1}^{3} (\Delta_i^{+} \Delta_i^{-})^2,
\end{eqnarray}
and the chromoelectric $({\bold E})$ and the magnetic field $({\bold
  B})$ are defined from clover-leaf plaquettes. The last two terms of
Eq. \eqref{eq:deltaH} correct for finite lattice spacing
errors. Tadpole improvement of the gauge link variable using the
fourth root of a single link plaquette \cite{Lepage:1992xa} (listed in Table
\ref{tab:parameters}) is adopted and $c_i$ in Eq. \eqref{LNRQCD_2} is
set to the tree-level value ($ = 1$). The bottom quark mass is not
tuned using the NRQCD dispersion relation. Instead, $M_b a$ for each
lattice spacing is set using $M_b = 4.65$ GeV in the computation
(listed in Table \ref{tab:parameters}).

In this work, a partial set of gauge configurations in 2+1 flavor QCD
is used, which is based on the highly improved staggered Quark (HISQ)
action with physical strange quark mass, $m_s$ and light quark masses
$m_l=m_s/20$ corresponding to a pion mass of $161$ MeV in the
continuum limit. For the calculations at nonzero temperature the
lattice size amounts to $48^3 \times 12$. These configurations were
generated for the study of the finite temperature QCD phase transition
described in \cite{Bazavov:2011nk}. The chiral transition temperature
in the continuum limit was determined to be $154(9)$ MeV
\cite{Bazavov:2011nk}. On $N_\tau=12$ lattices used in this study, the
central value of the transition temperature is slightly larger,
$T_c=159(3)$ MeV, but still is compatible with the above number within
errors. In the following discussion we will use the value $T_c=154
{\rm MeV}$ and often quote the temperature in units of $T_c$. The
lattice parameters are given in Table
\ref{tab:parameterT0}-\ref{tab:parameters}. At $T>0$ we estimate the
correlators on 400 gauge configurations, while at $T\simeq0$ we use
100 configurations at our disposal.

After gauge fixing into Coulomb gauge, we calculate the bottom quark
Green function and subsequently determine the bottomonium correlators
\begin{align}
\nonumber D({\mathbf x},\tau)&=\sum_{{\mathbf x}_0}\langle O({\mathbf x},\tau)
G({\mathbf x},\tau) O^\dagger({\mathbf x}_0,\tau_0) G^\dagger({\mathbf
  x},\tau) \rangle_{\rm med}
\end{align}
for each channel
\begin{align}
O(^3S_1;{\mathbf x},\tau)=\sigma_i,\quad O(^3P_1;{\mathbf
  x},\tau)=\overset{\leftrightarrow_s}{\Delta}_i\sigma_j-\overset{\leftrightarrow_s}{\Delta}_j\sigma_i,
\end{align}
with $\chi^\dagger\overset{\leftrightarrow_s}{\Delta}_i\psi=-\Big[\frac{1}{4}\big(\Delta^+_i+\Delta^-_i\big)\chi\Big]^\dagger\psi +\chi^\dagger\Big[\frac{1}{4}\big(\Delta^+_i+\Delta^-_i\big)\psi\Big]$
as defined (unfortunately with a typo) in \cite{Thacker:1990bm}. For
example, the $\Upsilon$ channel correlator is computed explicitly as

\begin{widetext}
\begin{eqnarray}
\langle ({\chi^\dagger}_a  (\sigma_x)_{ab} \psi_b (x^\prime))^\dagger
\chi^\dagger_c (\sigma_x)_{cd} \psi_d (x) \rangle &=& 2 \langle G_{+ +}
(x^\prime;x)^\dagger G_{- -} (x^\prime;x) + G_{+ -} (x^\prime;x)^\dagger G_{- +}
(x^\prime;x) \rangle \nonumber \\
\langle (\chi^\dagger_a  (\sigma_y)_{ab} \psi_b (x^\prime))^\dagger
\chi^\dagger_c (\sigma_y)_{cd} \psi_d (x) \rangle &=& 2 \langle G_{+ +}
(x^\prime;x)^\dagger G_{- -} (x^\prime;x) - G_{+ -} (x^\prime;x)^\dagger G_{- +}
(x^\prime;x) \rangle \nonumber \\
\langle (\chi^\dagger_a  (\sigma_z)_{ab} \psi_b (x^\prime))^\dagger
\chi^\dagger_c (\sigma_z)_{cd} \psi_d (x) \rangle &=& \langle G_{+ +}
(x^\prime;x)^\dagger G_{+ +} (x^\prime;x) + G_{-
  -}(x^\prime;x)^\dagger G_{- -} (x^\prime;x) \nonumber \\
& & - G_{+ -}(x^\prime;x)^\dagger G_{+ -} (x^\prime;x) - G_{-
  +}(x^\prime;x)^\dagger G_{- +} (x^\prime;x) \rangle, \nonumber \\
\end{eqnarray}
\end{widetext}

where $+(-)$ denotes the spin-up (spin-down) component and $\langle
\cdots \rangle$ refers to the average over the thermal
ensemble. Similarly by using  $\chi^\dagger \stackrel{\leftrightarrow}{\Delta}_i \psi \equiv
\chi^\dagger \Delta_i \psi - \Delta_i \chi^\dagger
\psi$ we have for the $\chi_{b1}$ channel

\begin{widetext}
\begin{eqnarray}
\langle \left[\chi^\dagger_a
  \left(\stackrel{\leftrightarrow}{\Delta}_i (\sigma_j)_{ab} -
  \stackrel{\leftrightarrow}{\Delta}_j (\sigma_i)_{ab} \right)\psi_b
(x^\prime)\right]^\dagger \chi^\dagger_c \left(\stackrel{\leftrightarrow}{\Delta}_i
(\sigma_j)_{cd} - \stackrel{\leftrightarrow}{\Delta}_j (\sigma_i)_{cd}
\right) \psi_d
(x) \rangle \nonumber \\
= \langle {\rm tr} \left[G^\dagger (x^\prime;x) \sigma_i
  \Delta_j G_V (x^\prime;x,j) \sigma_i \right] \rangle
- \langle {\rm tr} \left[{G_V}^\dagger (x^\prime;x,j) \sigma_i
  \Delta_j G (x^\prime;x) \sigma_i \right] \rangle
\nonumber \\
+  \langle {\rm tr} \left[G^\dagger (x^\prime;x) \sigma_j
  \Delta_i G_V (x^\prime;x,i) \sigma_j \right] \rangle
- \langle {\rm tr} \left[{G_V}^\dagger (x^\prime;x,i) \sigma_j
  \Delta_i G (x^\prime;x) \sigma_j \right] \rangle
\nonumber \\
- \langle {\rm tr} \left[G^\dagger (x^\prime;x) \sigma_i
  \Delta_j G_V (x^\prime;x,i) \sigma_j \right] \rangle
+ \langle {\rm tr} \left[{G_V}^\dagger (x^\prime;x,j) \sigma_i
  \Delta_i G (x^\prime;x) \sigma_j \right] \rangle
\nonumber \\
- \langle {\rm tr} \left[G^\dagger (x^\prime;x) \sigma_j
  \Delta_j G_V (x^\prime;x,i) \sigma_i \right] \rangle
+ \langle {\rm tr} \left[{G_V}^\dagger (x^\prime;x,j) \sigma_j
  \Delta_i G (x^\prime;x) \sigma_i \right] \rangle
\nonumber
\end{eqnarray}
\end{widetext}
 Computational cost is reduced by using point split sources
(along the $i$th direction), from which the propagator $G_V(x^\prime ;
x, i)$ is evolved. ``${\rm tr}$'' refers to a color and spin trace.

\subsection*{Bayesian spectral reconstruction}

While our goal is to determine the spectral properties of in-medium
bottom quark-antiquark pairs, lattice QCD only provides us with
correlation functions in Euclidean time. In the particular case of
NRQCD bottomonium correlators $D(\tau)$, the sought after spectral
functions $\rho(\omega)$ can be extracted via the inversion of the
following integral relation
\begin{align}
 D(\tau)=D({\bf p}=0,\tau)=\sum_{\bf x}D({\bf x},\tau)=\int_{-2M_q}^\infty
 d\omega \; e^{-\omega\tau}\;\rho(\omega).\label{Eq:SpecConv}
\end{align}
The exponentially damped integral Kernel $K(\tau,\omega)={\rm
  exp}[-\omega\tau]$ is temperature independent, which alleviates the
``constant contribution'' problem
\cite{Umeda:2007hy,Aarts:2002cc,Petreczky:2008px}. Since NRQCD is
an effective nonrelativistic description of bottomonium physics, it
implicitly contains a shift in energies to the two-quark threshold
$2M_q$ (up to renormalization).

Obviously the introduction of a spacetime regularization as the basis
for lattice QCD simulations only allows us to obtain the correlation
function at $N_\tau$ discrete steps of Euclidean time. Furthermore the
stochastic character of the Monte Carlo algorithm involved in
generating the lattice configurations entails that only estimates of
finite precision can be calculated for each observable. Hence the
inversion of Eq. \eqref{Eq:SpecConv} becomes inherently ill defined,
as we attempt to extract the spectral features of thermal bottomonium
along $N_\omega\gg N_\tau$ frequencies from a finite number of noisy
data points.

One possibility to give meaning to the inversion is by making use
of prior knowledge in addition to the measured data. Bayes theorem
\begin{align}
 P[\rho|D,I]\propto P[D|\rho,I]P[\rho|I]
\end{align}
provides the mathematical framework to do so. Here the likelihood
probability $P[D|\rho,I]={\rm exp}[-L]$ is given by
\begin{align}
 L[\rho]=\frac{1}{2}\sum_{ij}(D_i-D^\rho_i)C_{ij}(D_j-D^\rho_j)
\end{align}
where $C_{ij}$ denotes the covariance matrix of the measured data and
\begin{align}
 D^\rho_i=\sum_{l=1}^{N_\omega}\;{\rm exp}[-\omega_l\tau_i]\;\rho_l\;\Delta\omega_l
\end{align}
denotes the Euclidean correlator, which would result from our current
choice $\rho$. We also enforce the additional constraint $L=N_\tau$,
since the correct spectral functions lead to an $L$ value of
comparable magnitude. If we were to attempt a naive $\chi^2$ fit,
i.e. to maximize the likelihood alone, one encounters an infinite
number of degenerate solutions that all reproduce the data points
$D_i$ within their errors. Taking into account prior knowledge allows
us to regularize the $\chi^2$ fit. The particular functional form of
the prior probability $P[\rho|I]={\rm exp}[S]$ selects from the
degenerate maximum likelihood solutions a single one that represents
the most likely spectrum given the measured data and our prior
information.

In this study we use a prior functional $S$, which has recently been
proposed, based on the following three conditions: (1) it enforces
that the spectrum is positive definite, (2) it guarantees that the
choice of units for $\rho(\omega)$ remains irrelevant for the end
result, and (3) that the reconstructed spectrum is a smooth function
except where data introduce peaked structures
\begin{align}
 S[\rho]=\alpha\sum_l\;\Big( 1-\frac{\rho_l}{m_l}+{\rm
   log}\Big[\frac{\rho_l}{m_l}\Big]\Big)\Delta \omega_l.
\end{align}
Note that this expression differs from the Shannon-Jaynes entropy used
in the maximum entropy method.

The function $m(\omega_l)=m_l$ residing within $S$ is called the
default model and by definition corresponds to the correct spectrum in
the absence of data. Since we do not assume any additional knowledge
beyond positive definiteness and smoothness, we will use a constant
$m(\omega)={\rm const.}$ in the following. As we also measure the data
point at $\tau=0$, which encodes the normalization of the spectrum, we
adjust the overall magnitude of the default model accordingly. The
dependence of the results on different choices of $m(\omega)$ is
investigated in Appendix \ref{sec:systematics}.

The hyperparameter $\alpha$, which controls the weighting of the data
versus prior information is taken into account in a Bayesian fashion.
In the new Bayesian method deployed here, it is integrated out using
a flat overall hyperprior.
\begin{align}
 P[\rho|D,I]=P[D|I]\int_0^\infty d\alpha P[\rho|m,\alpha]
\end{align}
Hence we do not rely on the usually poor Gaussian approximation
entering the $\alpha$ probability estimation of the MEM (for details
see Ref.\cite{Burnier:2013nla}).

The Bayesian reconstruction (BR), given the measured data and our
prior knowledge, is thus obtained from finding the most probable
solution for the combined likelihood and prior probability
\begin{align}
 \left.\frac{\delta}{\delta \rho}P[\rho|D,I]\right|_{\rho=\rho^{\rm
     BR}}=0.
\label{BayesStationary}
\end{align}
In contrast to the standard MEM, we do not restrict the search space a
priori but allow the full $N_\omega$ degrees of freedom to vary. This
not only allows us to resolve spectral peaks with a much smaller width
but also removes the strong dependence of the reconstructed peaks on
the choice of the starting point of the discretized frequency interval
$\omega_{\rm min}$ as shown in Appendix \ref{sec:systematics}.

A crucial qualitative difference to the MEM exists in that our prior
functional does not possess the flat directions inherent in the
Shannon-Jaynes entropy deployed there. This leads to the optimizer
algorithm actually converging to a unique extremum up to any desired
tolerance. Thus we do not arbitrarily stop the minimizer at a
predefined step size but require convergence to machine precision.

To estimate the statistical uncertainty in our results, we perform
jackknife analyses for each spectrum, by repeating the reconstruction
procedure and excluding successive blocks of $10$ ($T\simeq0$) or $40$
($T>0$) configurations from the averaging of the underlying
correlators. The variation between each of these results is then used
to estimate a jackknife error bar which we attach to all quantities
plotted in the following section. As is known from prior work with the
MEM, the reconstruction, especially of the width of peak structures
depends on the amount of available data points as well as the
signal-to-noise ratio. The uncertainty related to these two factors is 
explicitly assessed in Appendix \ref{sec:systematics}.

Once a spectral reconstruction is performed, we need to interpret the
outcome in terms of bottomonium physics. On the one hand we need to
ascertain whether peaked structures arising in the Bayesian result are
actually encoded in the supplied data points and whether such peaks
are indeed an indication of a bound state available to the heavy quark
antiquark pair. The former question is related to the limited amount
of data points available along the temporal axis in this study. Indeed
even if we discretize the frequency domain between $-2M_q<\omega_{\rm
  min}<\omega_{\rm max}<\infty$ with a large number of $N_\omega\gg
N_\tau$ steps of length $\Delta \omega_l$, we will encounter finite
resolution artifacts, as is known from the closely related
inverse-fourier transform. If the spectrum contains a region in which
it essentially vanishes before abruptly changing into either peaked
structures or at high temperatures into a continuum, any
reconstruction based on a finite number of data points will contain
some kind of numerical ringing usually referred to as the Gibbs
phenomenon. The intensity of these artifacts is furthermore related to
the choice of default model, which if incompatible to the encoded
spectral structures will exacerbate ringing. In the following we
intentionally use the most neutral, i.e. a flat default model and
hence need to prepare for the encounter with wiggly structures in the
numerical extraction of spectral functions that do not have a physical
counterpart encoded in the data.

In addition, we need to understand whether a peak encoded in the data
can actually be attributed to the strong interactions binding together
bottom quarks into a bound state. Both aspects are addressed in this
study through a comparison of the fully interacting spectra to those
based on noninteracting NRQCD correlators (similar in spirit to
\cite{Karsch:2003wy, Aarts:2005hg} however without taking the
thermodynamic limit). These free spectra are obviously devoid of bound
state features and hence any peaked structures arising in their
Bayesian reconstruction need to be attributed to numerical
ringing. Our criterion for accepting a peak as a sign of the presence
of a bound state is that it needs to be quantitatively larger than the
corresponding wiggly feature in the reconstructed free spectral
function\footnote{Note that the standard MEM often does not show these
  ringing structures, since in Bryan's implementation it does not
  possess enough degrees of freedom to do so. In essence the singular
  value decomposition deployed there acts as an additional low-pass
  filter that is applied to the Bayesian solution. This in general
  however does not mean that the MEM answer resembles the correct
  spectrum more closely.}.

\section{Low Temperature Calibration}
\label{sec:zeroT}

\begin{table}[ht]
\begin{center}
\vspace*{0.2cm}
\begin{tabular}{|c|c|c|r|r|r|r|r|}
\hline
$\beta$	& Volume      &$T$[MeV]  & $a$[fm]& $u_0$   &$M_b a$ \\
\hline
6.664 &$32^3\times 32$&52.7& 0.117 &0.87025 & 2.76 \\
6.800 &$32^3\times 32$&59.9& 0.103 &0.874849 & 2.42 \\
6.950 &$32^3\times 32$&69.0& 0.0893&0.879442 & 2.11 \\
7.280 &$48^3\times 64$&46.6& 0.0660&0.88817 & 1.56 \\
\hline
\end{tabular}
\vspace*{0.2cm}
\caption{List of parameters for the $T\simeq0$ lattice configurations,
  used to calibrate the NRQCD energy shift.}
\label{tab:parameterT0}
\end{center}
\end{table}

In NRQCD, the heavy quark mass scale is integrated out so that the
origin of the energy axis is shifted. To retrieve the physical masses
of the bottomonium states from these binding energies computed in the
EFT, it is necessary to reintroduce the energy shift. In particular
we have
\begin{equation}
M_{\Upsilon (1S)}^{\rm exp} = E^{\rm sim}_{\Upsilon (1S)} + 2 (Z_{M_b} M_b - E_0) ,
\end{equation}
where $E^{\rm sim}_{\Upsilon (1S)}$ is the $(1S)$ energy computed in
the $\Upsilon$ channel, $Z_{M_b}$ is the mass renormalization, $M_b$
denotes the bare lattice heavy quark mass, while $E_0$ is an
additional energy shift \cite{Davies:1994mp}. $M_{\Upsilon}^{\rm exp}$
refers to the experimental value for the mass of the $\Upsilon$
state. The energy shift and the renormalization factor depend on the
lattice cutoff scale. For this reason we introduce a
$\beta$-dependent energy shift parameter
\begin{equation}
C_{\rm shift}(\beta)=2 (Z_{M_b} M_b - E_0),
\end{equation}
that will be determined numerically by matching the calculated
$\Upsilon$ mass to its experimental value. Since in this work the
simulation temperature is changed by changing the lattice spacing, we
measure $100$ low temperature correlators using the configurations
with parameters listed in Table \ref{tab:parameterT0}, in order to
determine the ground state peak position of the $^3S_1$ channel
$E_{\Upsilon (1S)}$.

\begin{figure*}
\centering
 \includegraphics[scale=0.3, angle=-90]{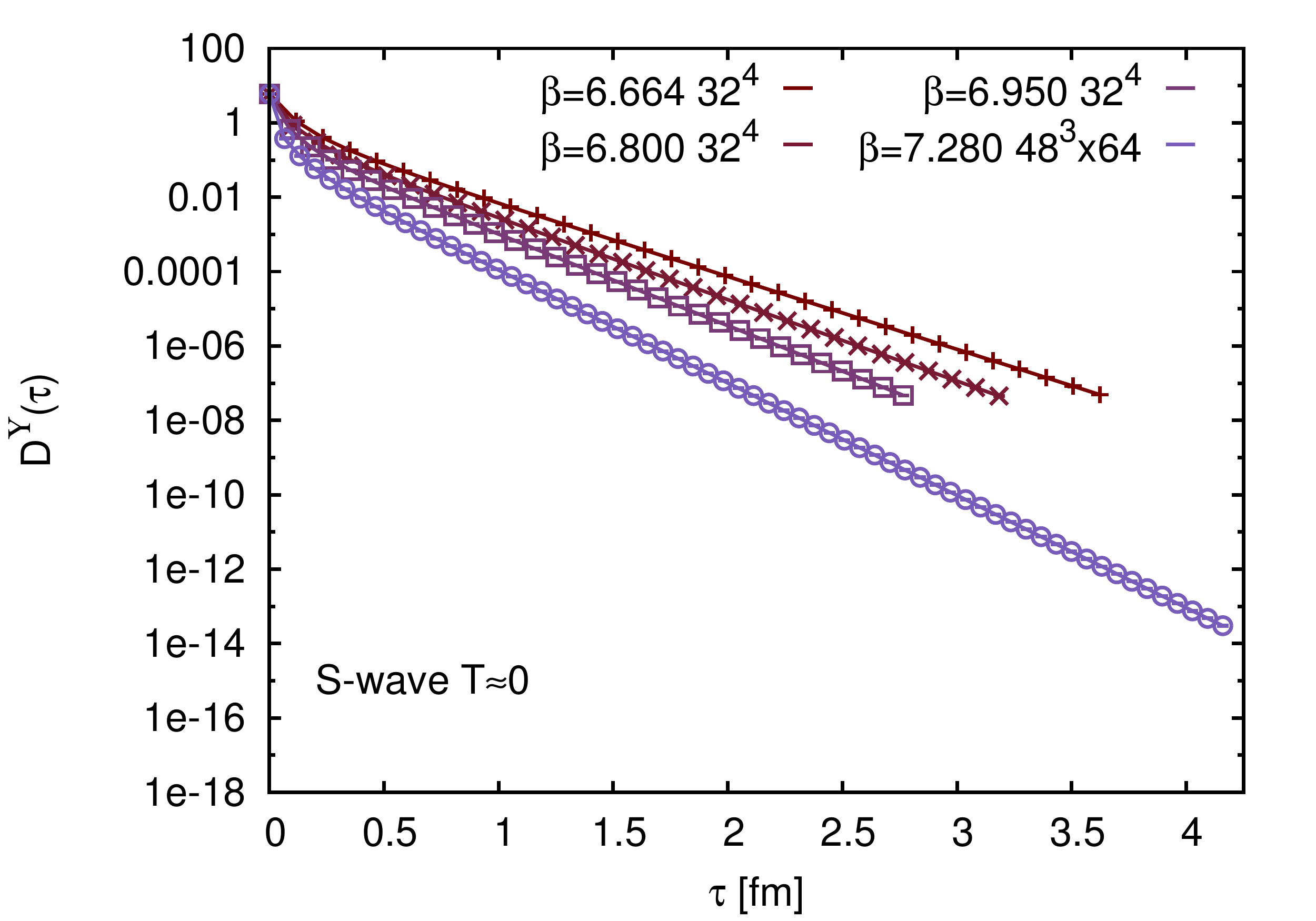}
 \includegraphics[scale=0.3, angle=-90]{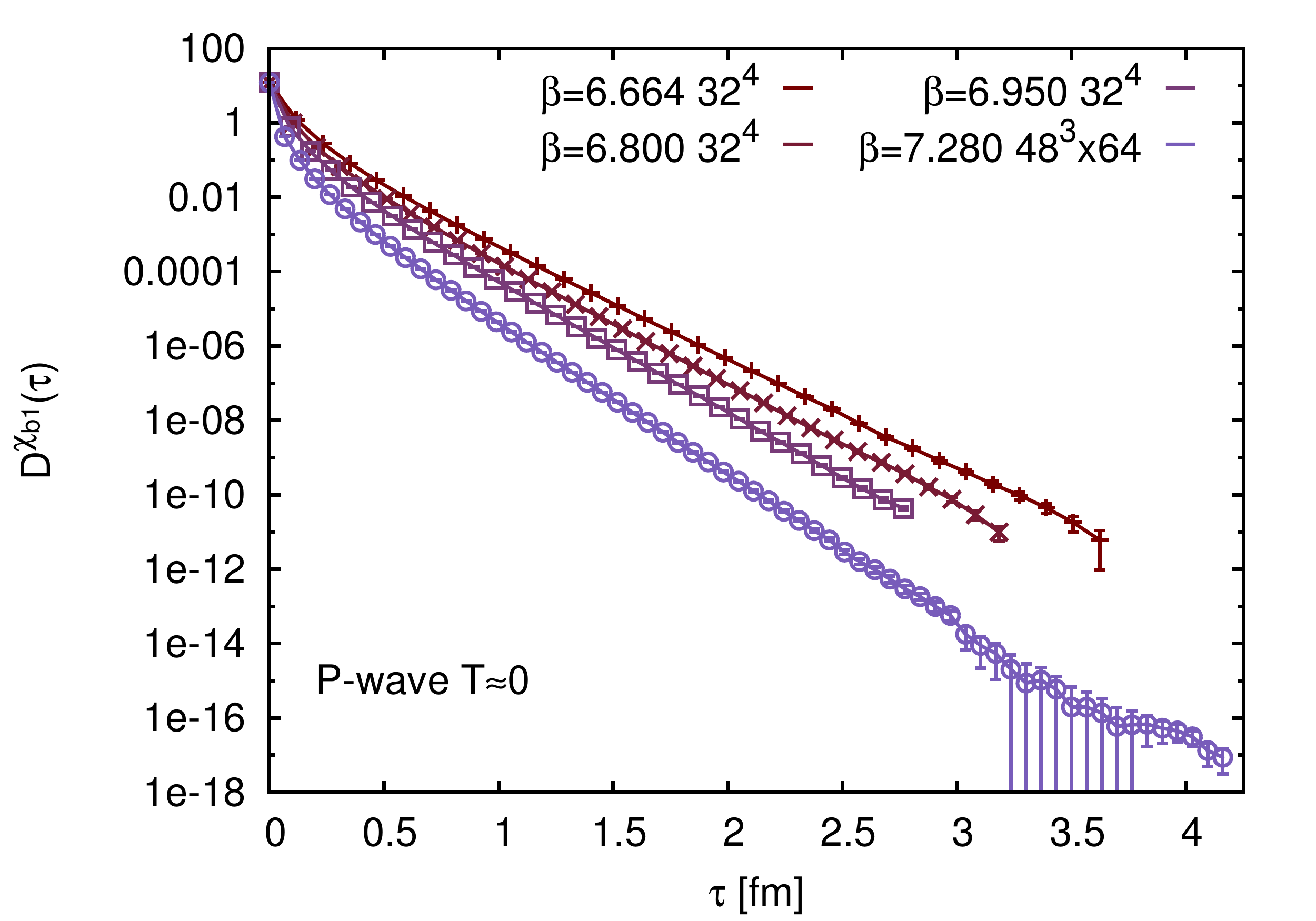}
 \includegraphics[scale=0.3, angle=-90]{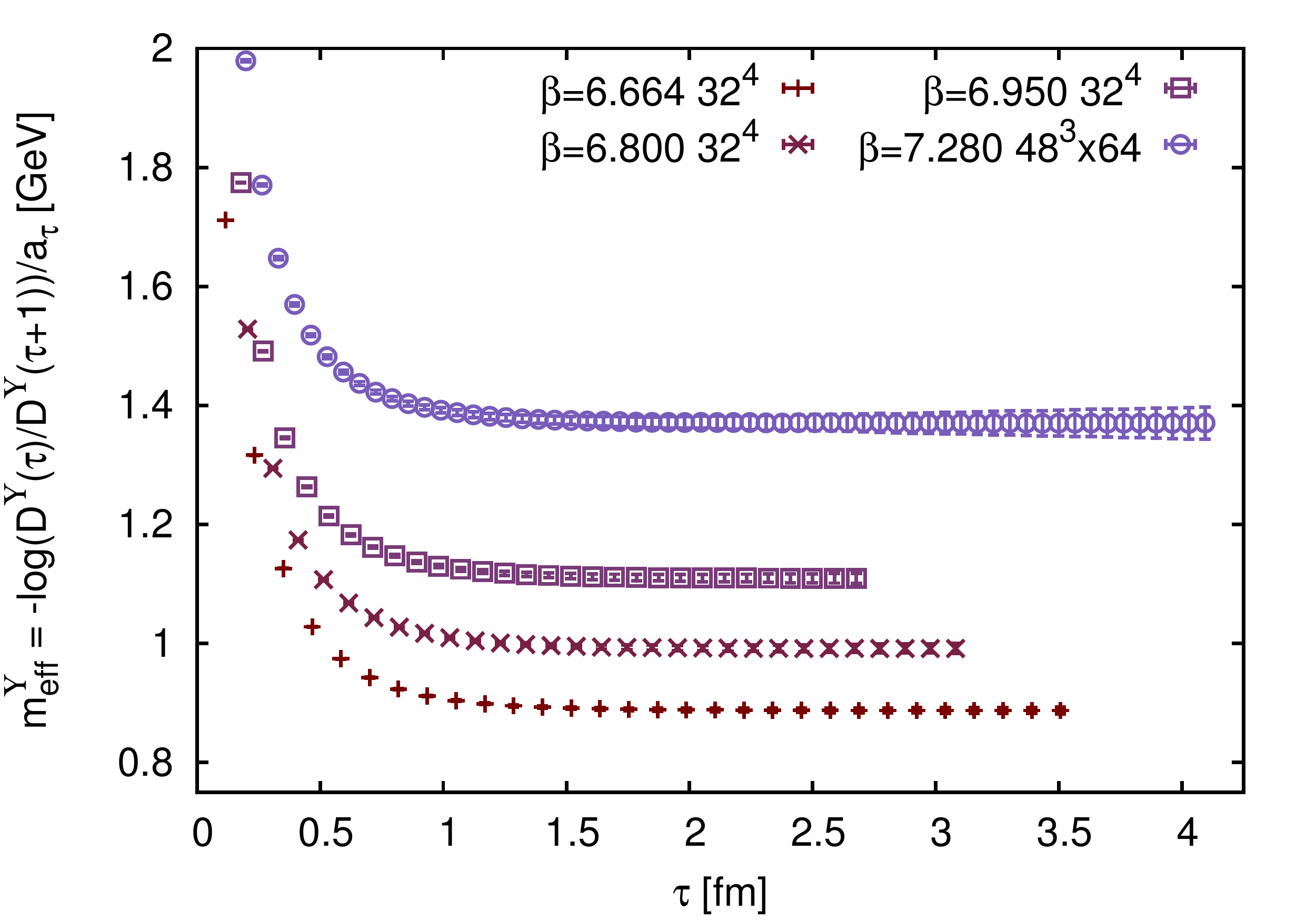}
 \includegraphics[scale=0.3, angle=-90]{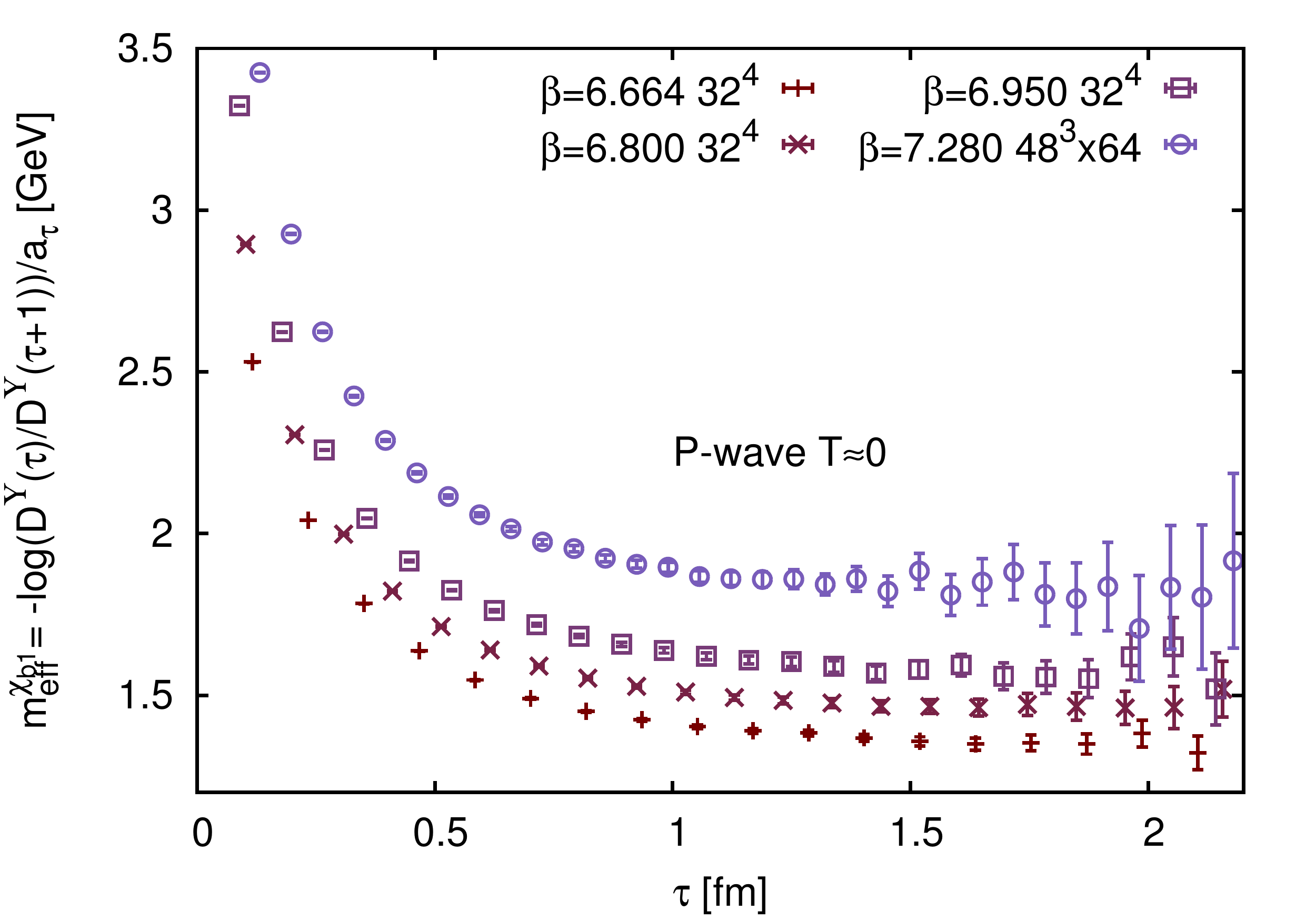}
\caption{The lattice NRQCD correlators (top) and corresponding
  effective masses (bottom) $m_{\rm eff}(\tau)=-{\rm
    log}[D(\tau)/D(\tau+a_\tau)]/a_\tau$ in the S-wave channel (left)
  and P-wave channel (right) at $T\simeq0$ for $\beta = 6.664, 6.800,
  6.950,$ and $7.280$}
 \label{zeroTcorr}
\end{figure*}

Figure \ref{zeroTcorr} (top) shows the typical behavior of S-wave
($\Upsilon$) and P-wave ($\chi_{b1}$) channel correlators. They
exhibit, as expected, an exponential falloff at late Euclidean time,
which is manifest in a flattening off of the corresponding effective
mass parameter (bottom). This tells us that at these low temperatures
the ground state only carries a negligible width and is clearly
separated from any other excited state peaks in the spectrum. Hence we
can carry out a standard exponential fit to determine the NRQCD ground
state masses in each channel.
\begin{figure*}
\centering
 \includegraphics[scale=0.3, angle=-90]{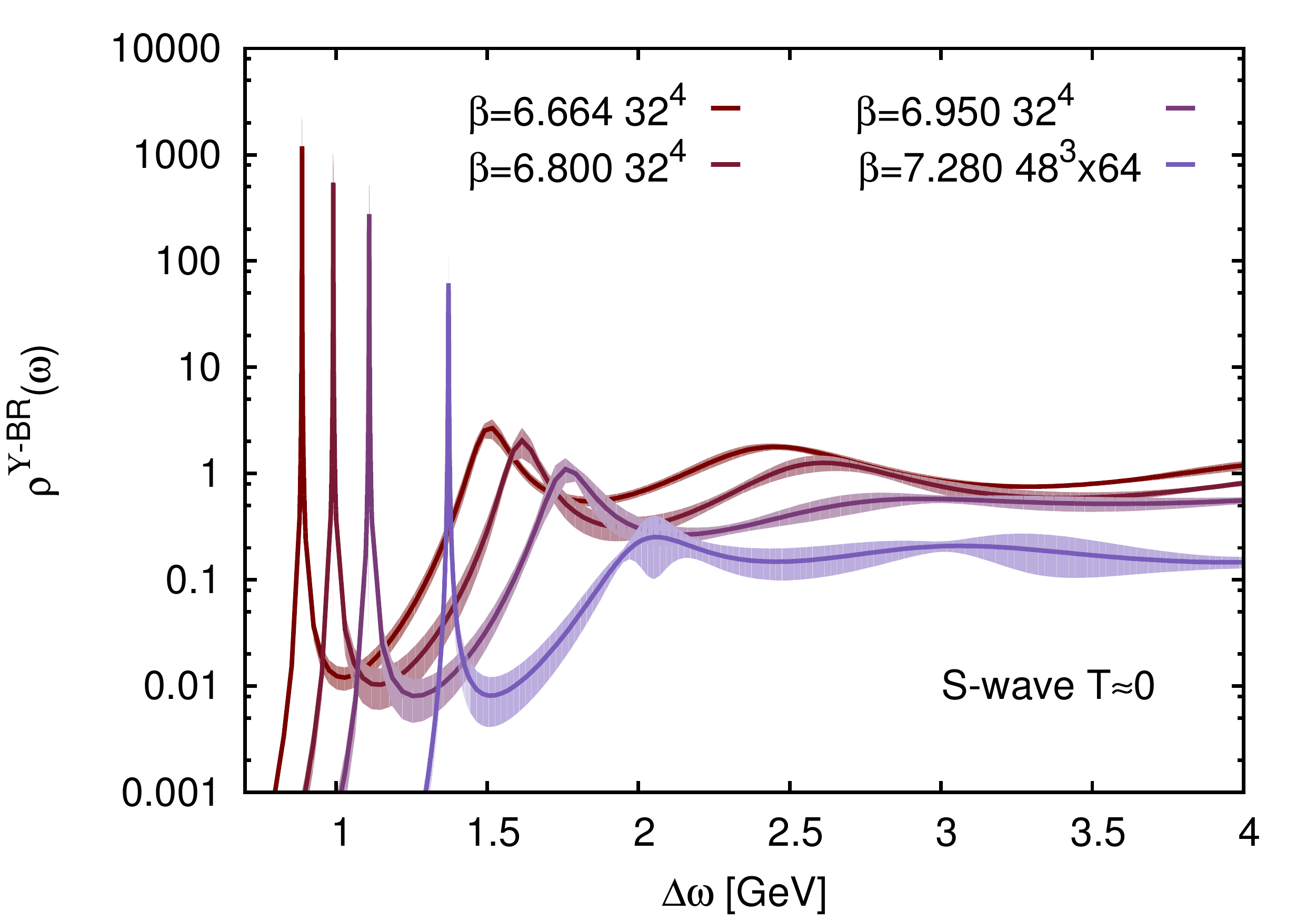}
 \includegraphics[scale=0.3, angle=-90]{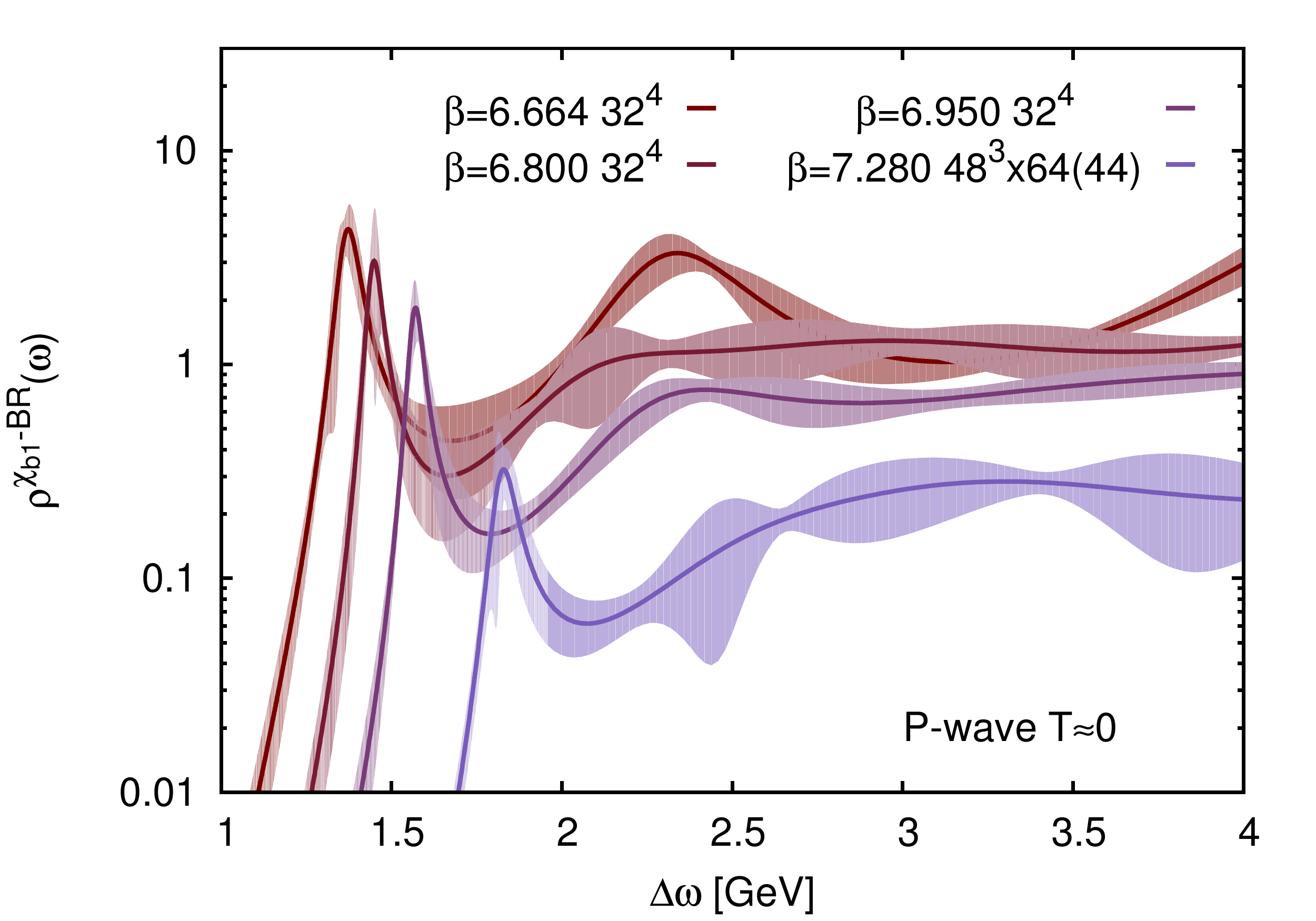}
\caption{Bayesian reconstruction of the nonshifted spectra in the
  S-wave channel (left) and P-wave channel (right) at $T\simeq0$
  $\beta = 6.664$ (dark red solid), $6.800$ (dashed), $6.950$ (solid) and $7.280$ (light blue dashed). Note
  that at $\beta=7.280$ we only use the first $44$ data points, i.e. where
  the signal is not yet lost to roundoff errors. }
 \label{zeroTspect}
\end{figure*}

The Bayesian spectral reconstruction of the $T\simeq0$ data is
performed using at each of the $\beta$ values the same numerical
frequency interval $I_\omega=[-0.5,30]$ discretized in $N_\omega=1800$
steps. A high precision interval of $N_{\rm hp}=550$ points is chosen
around the lowest lying peak to resolve its narrow width. The values
$\tau_{\rm max}$ are rescaled, so that in each case the algorithm uses
$\tau_{\rm max}^{\rm num}=20$. This reduces systematic uncertainties,
since each reconstruction proceeds based on the same relation between
$\tau$ and $\omega$. For $\beta=7.280$ we discard the last 19 data
points, as their values are dominated by noise probably arising from
finite rounding errors. Taking a constant default model $m(\omega)=\rm
const.$, which is normalized according to $D(\tau=0)$ and enforcing
the condition $|L-N_\tau|<10^{-5}$, we find the optimal solution
according to Eq. \eqref{BayesStationary} using the Limited-memory Broyden-Fletcher-Goldfarb-Shanno algorithm. We
repeat the reconstruction ten times using in each instance a different
set of $90$ of the $100$ measured correlators and determine from the
variation between individual results the jackknife error bars shown in
the figures below. The arithmetic used in the evaluation of the
likelihood and prior probability is taken to be $768$ bits. The
results of the Bayesian reconstruction of the $\Upsilon$ and $\chi_{b1}$
spectral functions are shown in Fig. \ref{zeroTspect}. 
$\Upsilon(1S)$ is very well determined from the reconstruction. The
second bump corresponds to excited states, mostly $\Upsilon(2S)$. The
difference between the position of the first and the second peak is
$608,~617,~640$ and $664$ MeV for $\beta=6.664,~6.800,~6.95$ and
$7.28$, respectively. These values are reasonably close to the
experimental value for the $2S-1S$ $\Upsilon$ mass splitting of $563$
MeV, and are smaller than the $3S-1S$ $\Upsilon$ mass splitting, which
is equal to $895$ MeV \cite{Beringer:1900zz}. The integral under the
peak is proportional to the wave function at the origin squared,
$|R(0)|^2$. We find that the ratio of the integral of the first peak
to the integral of the second peak is about one, whereas the
corresponding ratio of $|R(0)|^2$ is expected to be about half
\cite{Eichten:1995ch}. The area under the peak is more affected by
contamination from higher states compared to the peak position. This
is similar to the situation in the charmonium sector
\cite{Jakovac:2006sf}.

In the case of P-wave spectral function the first peak corresponding
to the $\chi_{b1}(1P)$ state is broader and the statistical errors on
the spectral functions are larger. This is due to several
effects. First, the mass of the $\chi_{b1}(1P)$ state is larger than
the mass of $\Upsilon(1S)$ so the signal-to-noise ratio is smaller.
Second, the amplitude of the ground state is proportional to
$|R'(0)|^2/M_b^2$ and thus is smaller compared to the S-wave
amplitude. In addition the continuum part of the spectral function
scales like $\omega^{1/2}$ for the S-wave, while it scales like
$\omega^{3/2}$ for the P-wave. As the result the relative
contribution of the lowest peak versus the continuum part of the
spectral functions is much smaller for $P$-waves. This makes the
reconstruction of the P-wave spectral function more difficult,
especially at high temperatures.

\begin{figure}
\centering
 \includegraphics[scale=0.3, angle=-90]{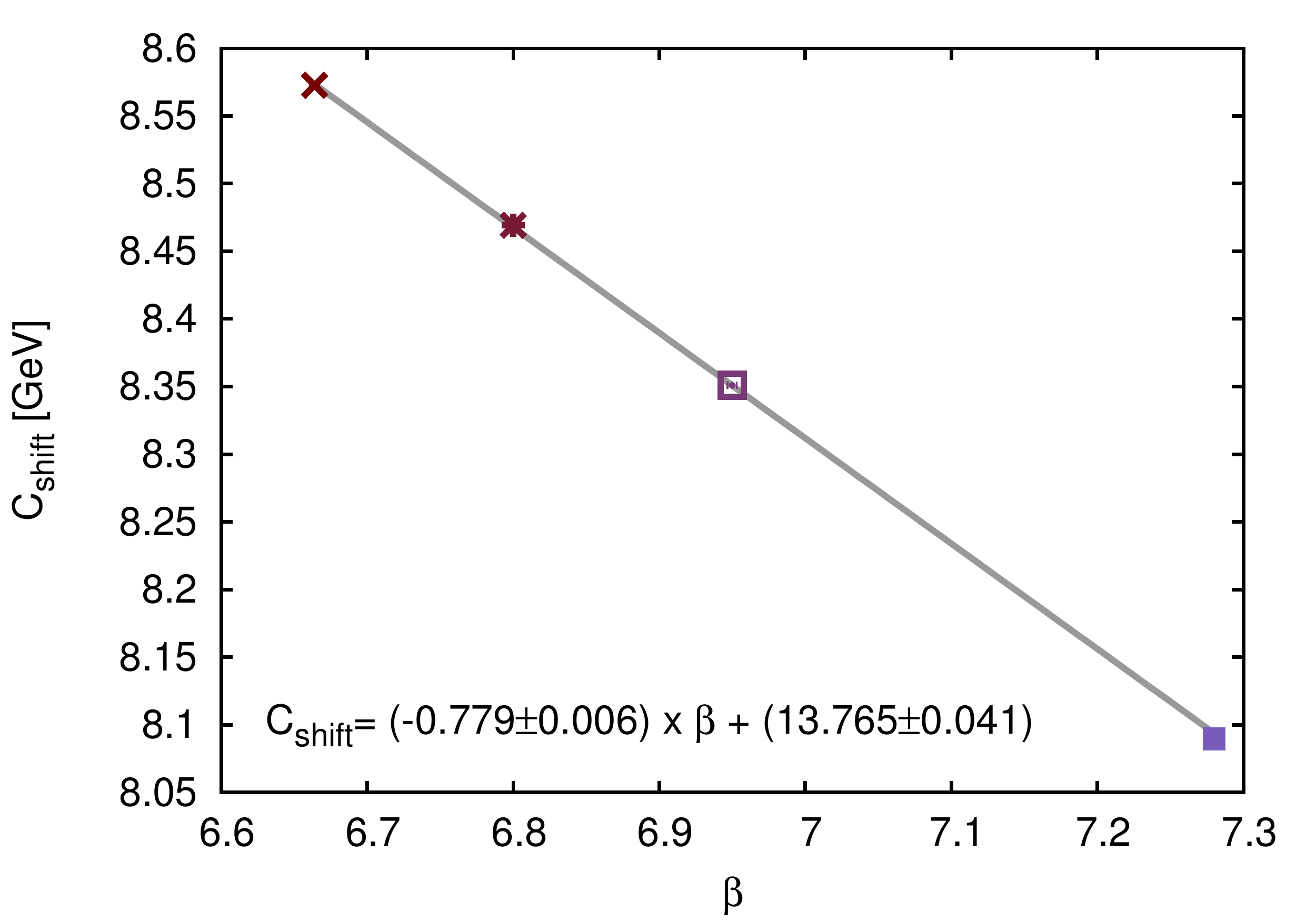}
\caption{``Energy shift constant $(C_{\rm shift})$'' determined from
  the position of the lowest lying peak in the S-wave channel spectral
  function for $\beta = 6.664, 6.800, 6.950,$ and $7.280$.}
 \label{Eoffset}
\end{figure}

Both exponential fitting and the determination of the ground state
peak position in the S-wave spectrum give a consistent value for
$E_{\Upsilon (1S)}$. Hence we are able to determine the constant,
$C_{\rm shift}$,
\begin{equation}
M_{\Upsilon (1S)} = E_{\Upsilon (1S)} + C_{\rm shift}(\beta)\label{ShiftEq}
\end{equation}
where the experimental value, $M_\Upsilon (1S) = 9.46030(26)$
\cite{Beringer:1900zz} is used as input, at each lattice
spacing. Figgure \ref{Eoffset} shows the obtained values of $C_{\rm
  shift}$, plotted against the lattice coupling $\beta$, where $\beta
= 6.664$ corresponds to the coupling underlying the lowest temperature
lattices ($T = 140$ MeV) and $\beta = 7.280$ that for the highest
temperature ($T = 249$ MeV).

Figure \ref{Eoffset} shows that the data points are well described by a
linear fit. Hence we use the linearly interpolated values for $C_{\rm
  shift}$ to calibrate the energy scale in the nonzero temperature
runs of Table \ref{tab:parameters}.

With this calibration in place we can check the consistency of our
approach by extracting the known vacuum ground state mass of the
P-wave channel. To this end we fit the lowest lying spectral peaks
shown in the right panel of Fig. \ref{zeroTspect} with a Lorentzian
and shift the obtained value according to Eq. \eqref{ShiftEq}. The
jackknifed estimates for $M_{\chi_{b1}}$ obtained at each individual
$\beta$ are in turn fitted with a constant, from which we obtain a
$\chi_{b1}$ mass of $M_{\chi_{b1}}=9.917(3){\rm GeV}$. The assigned
error is understood to represent a combination of the statistical
errors due to jackknife variation and systematic errors from variation
between different beta values. This value lies slightly above the
Particle Data Group value of $M_{\chi_{b1}} (1P) = 9.89278(26)(31)$
GeV \cite{Beringer:1900zz} but is consistent with the one obtained in
other recent lattice NRQCD studies with $N_f=2+1$ flavors
\cite{Aarts:2014cda}, i.e. $M_{\chi_{b1}}=9.921(15){\rm GeV}$.

\section{Spectral Functions at Finite Temperature}
\label{sec:nonzeroT}

\begin{table}[ht]
\begin{center}
\vspace*{0.2cm}
\begin{tabular}{|c|c|c|r|r|r|r|}
\hline
$\beta$	& T    & $T/T_c$& $a$(fm)& $u_0$   &$M_b a$ \\
\hline
6.664	&140& 0.911  & 0.117& 0.87025  & 2.76  \\
6.700	&145& 0.944  & 0.113& 0.87151  & 2.67  \\
6.740	&151& 0.980  & 0.109& 0.87288  & 2.57  \\
6.770	&155& 1.01  & 0.106& 0.87388  & 2.50  \\
6.800	&160& 1.04  & 0.103& 0.87485  & 2.42  \\
6.840	&166& 1.08  & 0.0989& 0.87612 & 2.34  \\
6.880	&172& 1.12  & 0.0953& 0.87736 & 2.25  \\
6.910	&177& 1.15  & 0.0926& 0.87827 & 2.19  \\
6.950	&184& 1.19  & 0.0893& 0.87945 & 2.11  \\
6.990	&191& 1.24  & 0.086 & 0.88060 & 2.03  \\
7.030	&198& 1.29  & 0.0829& 0.88173 & 1.96  \\
7.100	&211& 1.37  & 0.0777& 0.88363 & 1.84  \\
7.150	&221& 1.44  & 0.0743& 0.88493 & 1.75  \\
7.280	&249& 1.61  & 0.0660& 0.88817 & 1.56  \\
\hline
\end{tabular}
\vspace*{0.2cm}
\caption{List of parameters for the $T>0$ lattice configurations used
  in extracting the in-medium bottomonium spectral functions.}
\label{tab:parameters}
\end{center}
\end{table}
\begin{figure}
\begin{center}
\vspace{-0.4cm}
 \includegraphics[scale=0.3,angle=-90]{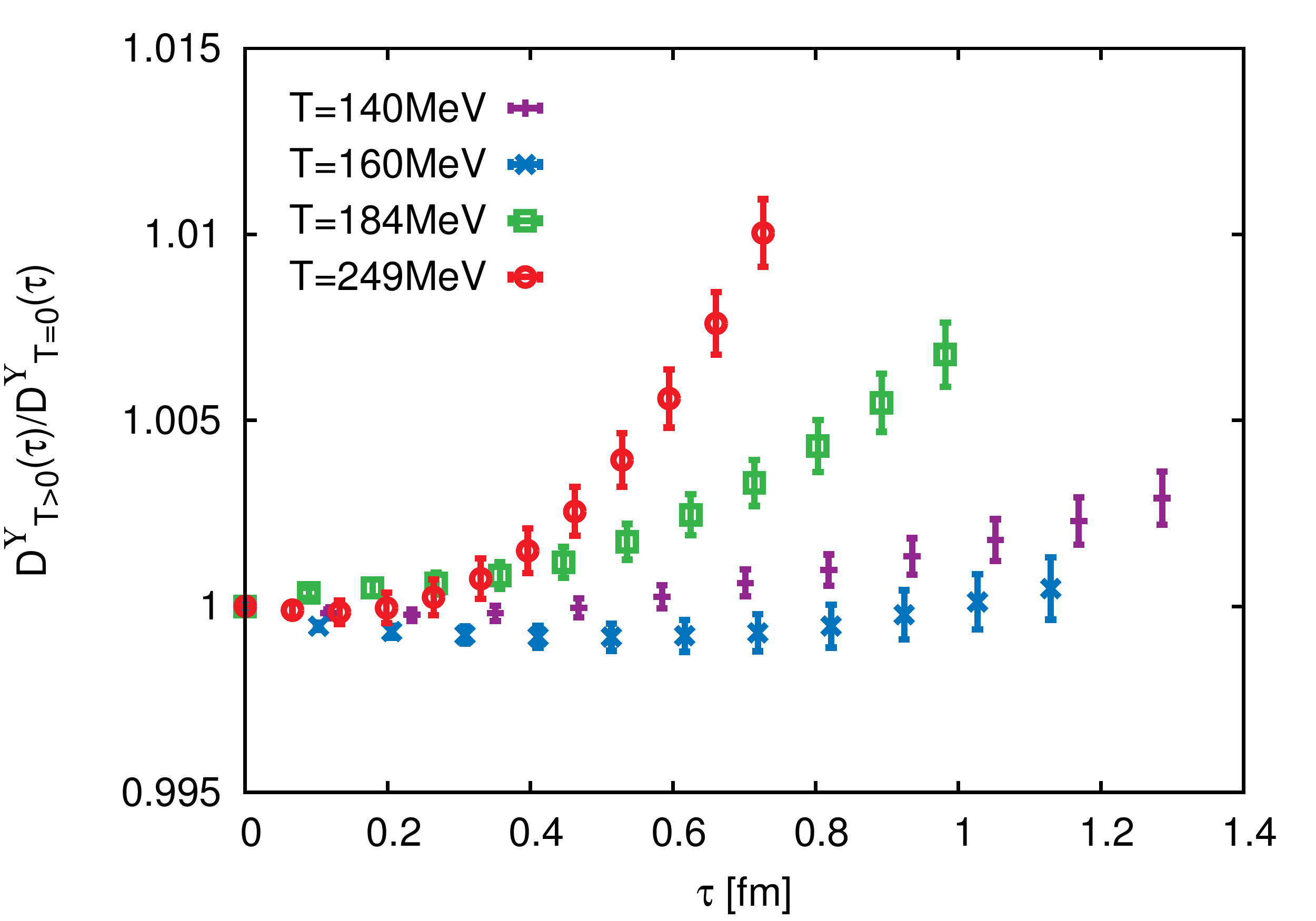}
 \includegraphics[scale=0.3,angle=-90]{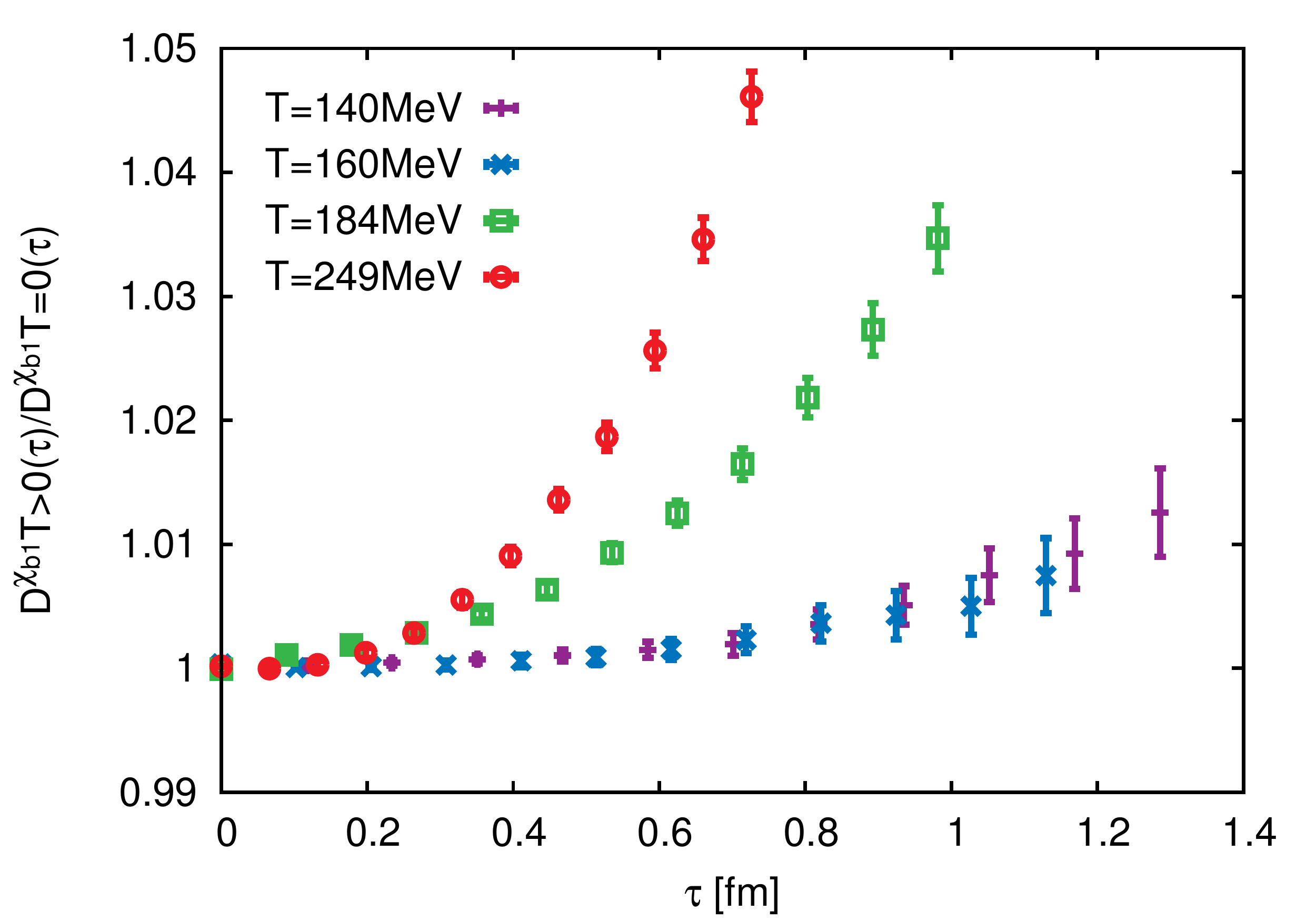}
\end{center}\vspace{-0.2cm}
\caption{Ratios of the Euclidean correlator at finite temperature and
  close to $T=0$ for the $\Upsilon$ (top) and $\chi_{b1}$ (bottom) channel at
  the lattice spacings, where $T\simeq0$ measurements were carried
  out.}
 \label{Tcorr}
\end{figure}

\begin{figure}
\begin{center}
\vspace{-0.4cm}
\includegraphics[scale=0.3,angle=-90]{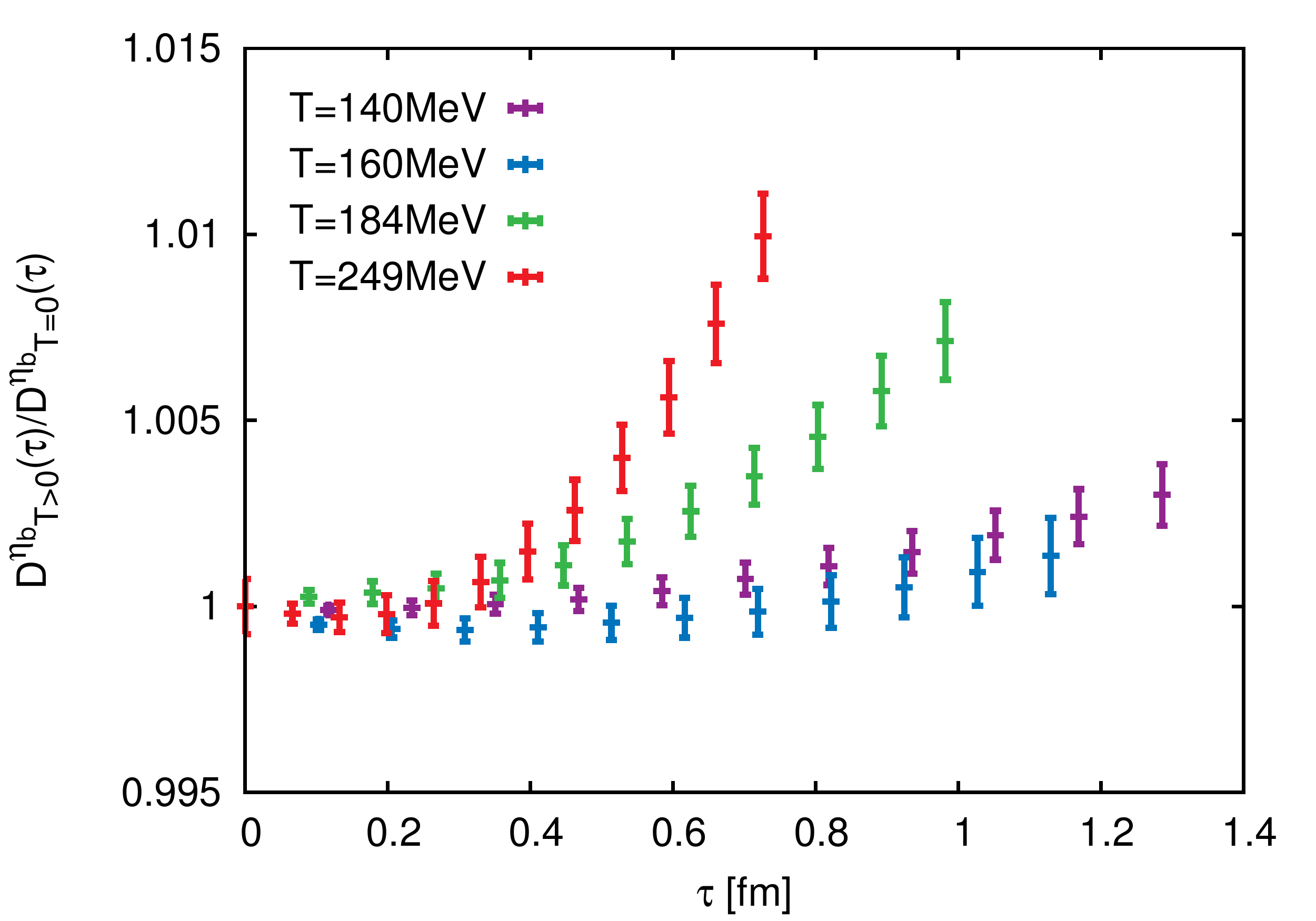}
\includegraphics[scale=0.3,angle=-90]{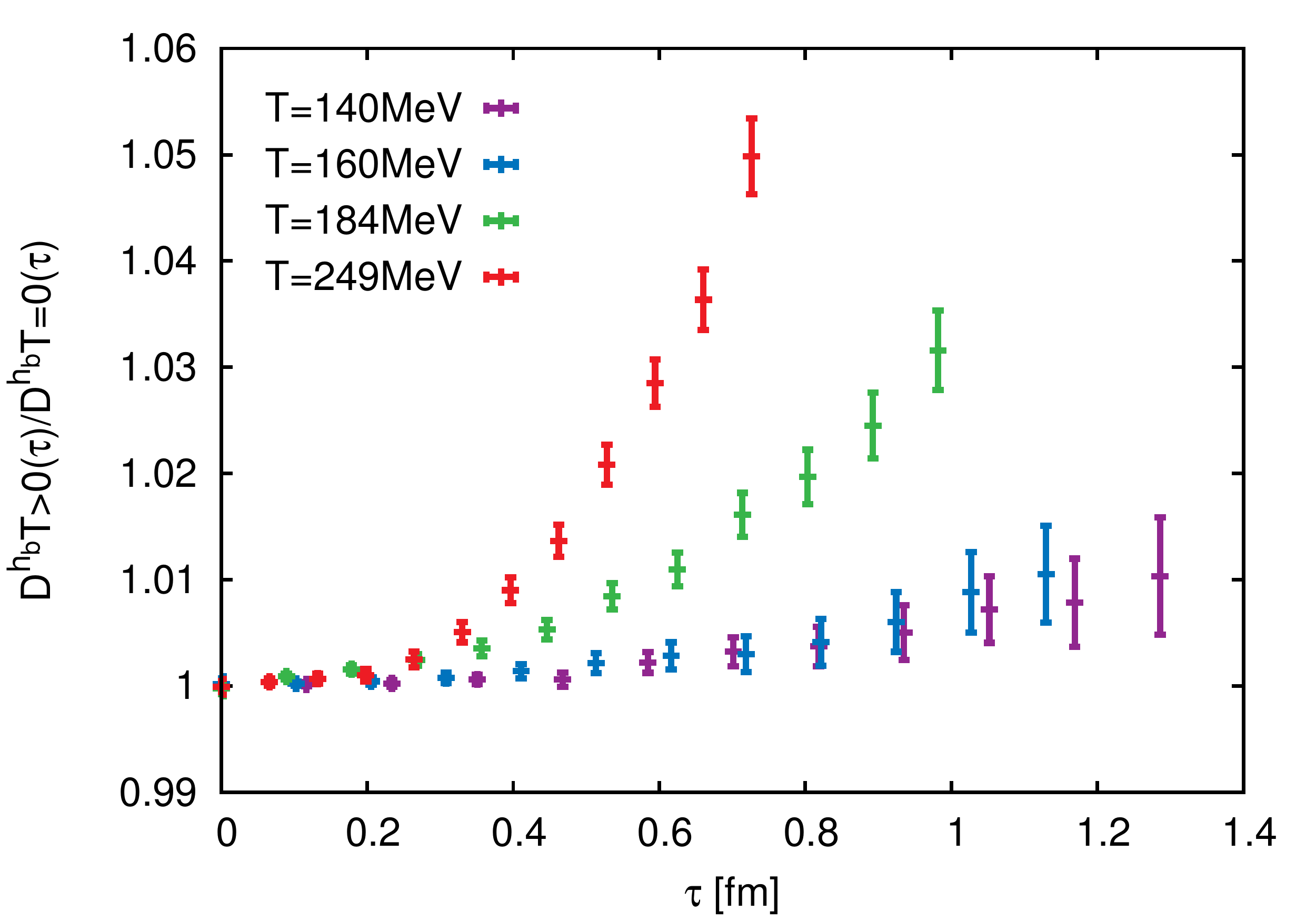}
\end{center}\vspace{-0.2cm}
\caption{Ratios of the Euclidean correlator at finite temperature and
  close to $T=0$ for the $\eta_b$ (top) and $h_{b}$ (bottom) channel at the
  lattice spacings, where $T\simeq0$ measurements were carried
  out. Note the qualitative as well as quantitative similarity to the
  $\Upsilon$ and $\chi_{b1}$ ratios shown in Fig. \ref{Tcorr}}
 \label{TcorrAdditional}
\end{figure}

At each of the temperatures listed in Table \ref{tab:parameters}, we
solve Eq. \eqref{NRQCDEvolEq} to obtain $400$ estimates of the S-wave
and P-wave lattice NRQCD correlators at $N_\tau=12$ data points. In
Fig. \ref{Tcorr} the ratio between the finite temperature averaged
correlators and their low temperature counterparts is plotted for the
four lattice spacing at which a $T\simeq0$ measurement is available.
We observe statistically significant in-medium modification of both
$\Upsilon$ and $\chi_{b1}$ correlators starting at temperature
$T_1=160$ MeV. While the medium modification for the S-wave leads to
changes of at most $1\%$ the P-wave correlators exhibit a stronger
effect, i.e. up to $5\%$. This is expected because the larger size of
the $\chi_{b1}$ state makes is more susceptible to medium effects. We
also calculated the above ratio for $n=3$ and $4$ and found that it is
independent of $n$ within the statistical accuracy. Note that we see a
smaller modification of the ratio of the correlators compared to
previous NRQCD studies \cite{Aarts:2011sm,Aarts:2014cda}. The ratio
of the finite and zero temperature correlators for $\eta_b$ and $h_b$
was also determined and their temperature dependence is shown in
Fig. \ref{TcorrAdditional}. It is similar to that of the $\Upsilon$
and $\chi_{b1}$ correlators, respectively. This is expected since the
sign and the magnitude of the binding energy of $\eta_b$ and $h_b$ are
almost the same as those of $\Upsilon$ and $\chi_{b1}$.

From the measured correlators, we extract the corresponding spectral
functions using both the new Bayesian method and the conventional
MEM. The NRQCD energy shift obtained in Sec. \ref{sec:zeroT} is
used to set the absolute physical energy scale.

The Bayesian spectral reconstruction of the $T>0$ data is performed
using at each of the $\beta$ values a common numerical frequency
interval $I_\omega=[-1,25]$ discretized in $N_\omega=1200$ steps. A
high precision interval of $N_{\rm hp}=550$ points is chosen around
the lowest lying peak to resolve its narrow width. The value
$\tau_{\rm max}=12$ is rescaled, so that in each case the algorithm
uses $\tau_{\rm max}^{\rm num}=20$. Taking a constant default model
$m(\omega)=\rm const.$, which is normalized according to $D(\tau=0)$
and enforcing the condition $|L-N_\tau|<10^{-5}$, we find the optimal
solution according to Eq. \eqref{BayesStationary} using the LBFGS
algorithm. We repeat the reconstruction ten times using in each
instance a different set of $360$ of the $400$ measured correlators
and determine from the variation between individual results the
jackknife error bars shown in the figures below. The arithmetic used
in the evaluation of the likelihood and prior probability is taken to
be $512$ bits.

The MEM reconstructions are performed with almost the same settings,
only a different frequency interval $I_\omega=[-0.15,25]$ is
chosen. The reason is related to the restricted nature of the search
space, due to which the reconstruction success depends strongly on
choosing $\omega_{\rm min}$ close enough to the relevant spectral
features (see also the discussion of the MEM systematics in Appendix
\ref{sec:comparison}). We select an $\alpha$ range that covers the
peak in the probability distribution $P[\alpha]$ and due to the
absence of true convergence stop the deployed Levenberg Marquardt
minimizer if it reaches a step size of $5\times 10^{-9}$.

\begin{figure*}
 \includegraphics[scale=0.3,angle=-90]{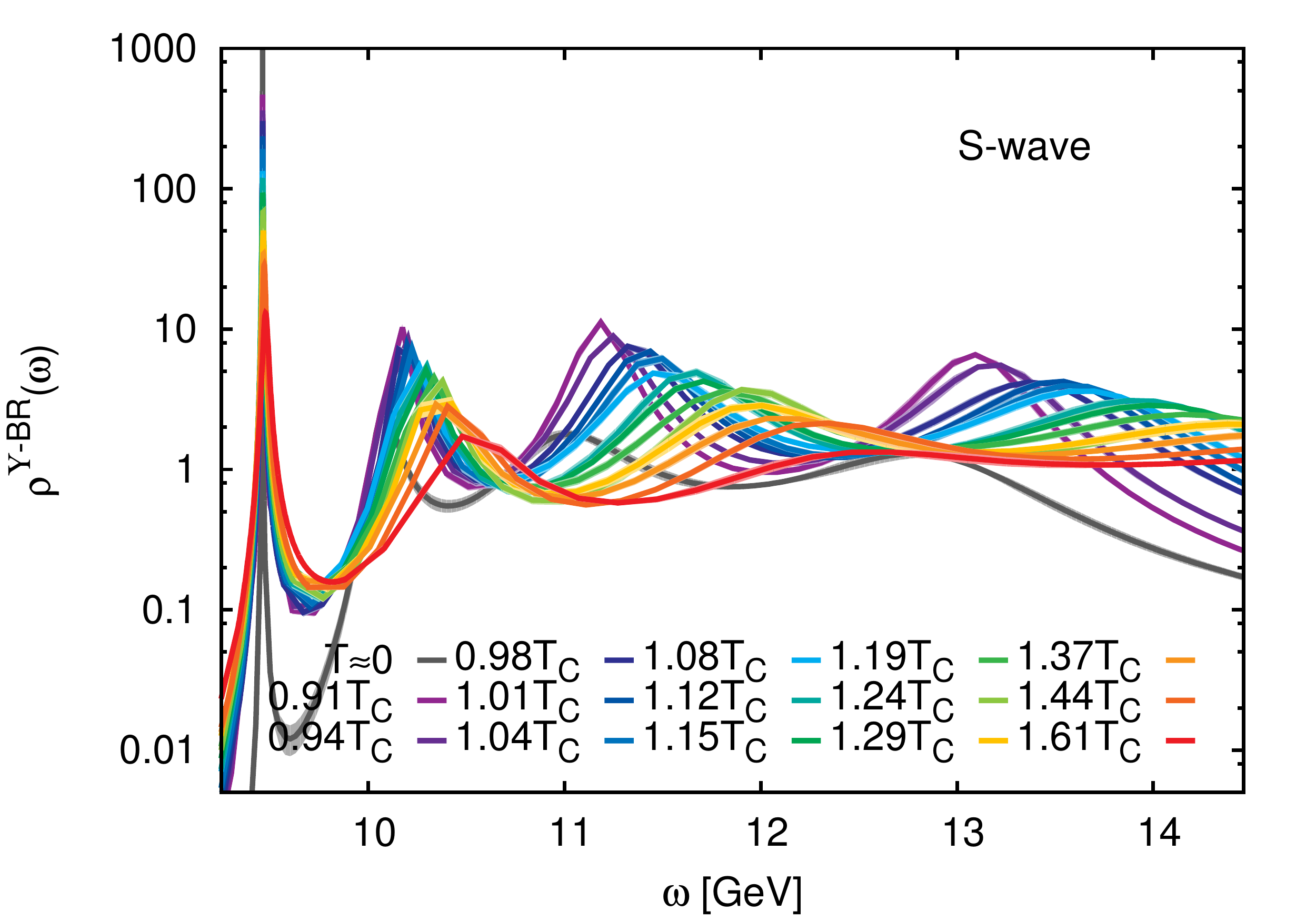}
 \includegraphics[scale=0.3,angle=-90]{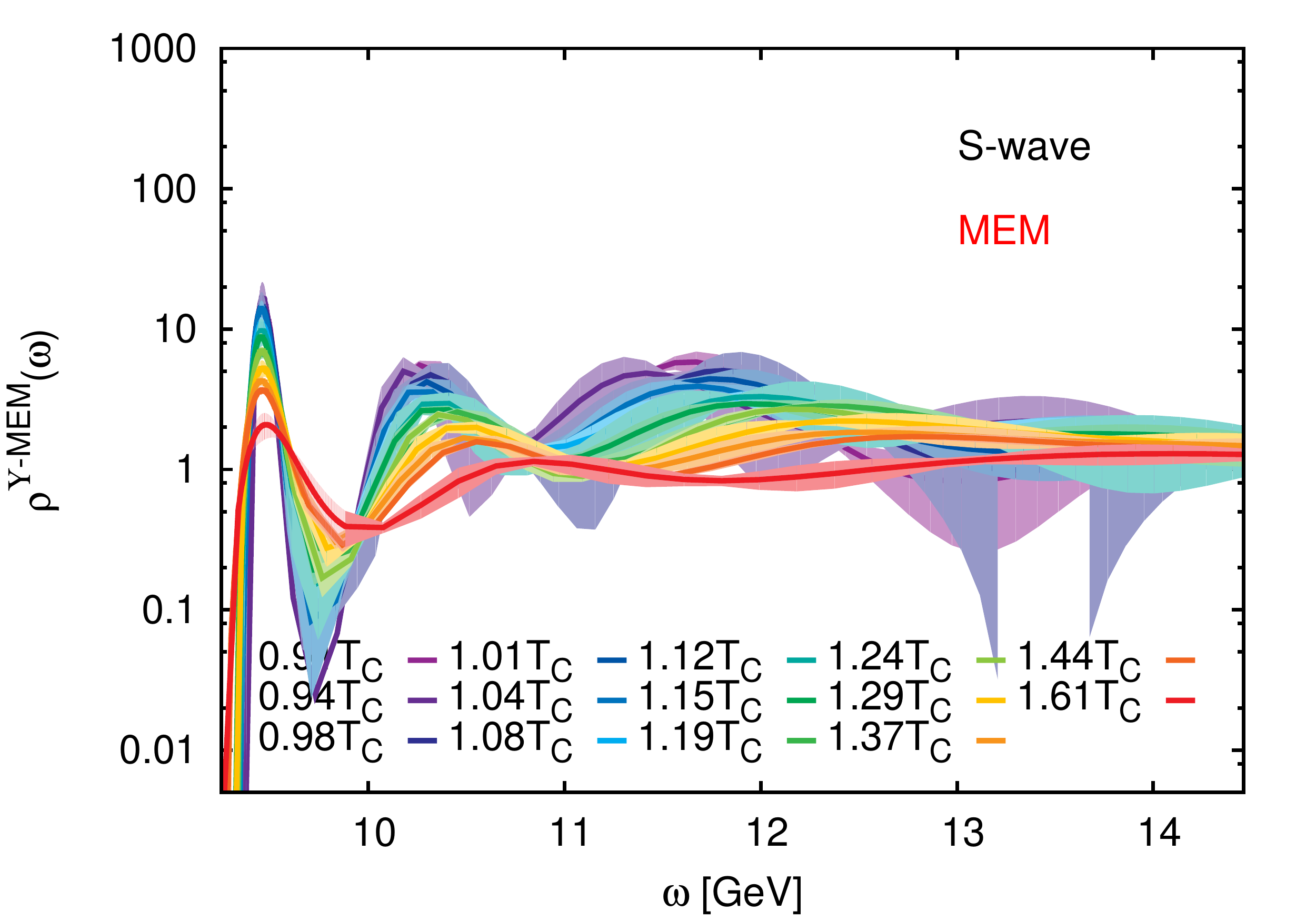}
 \includegraphics[scale=0.3,angle=-90]{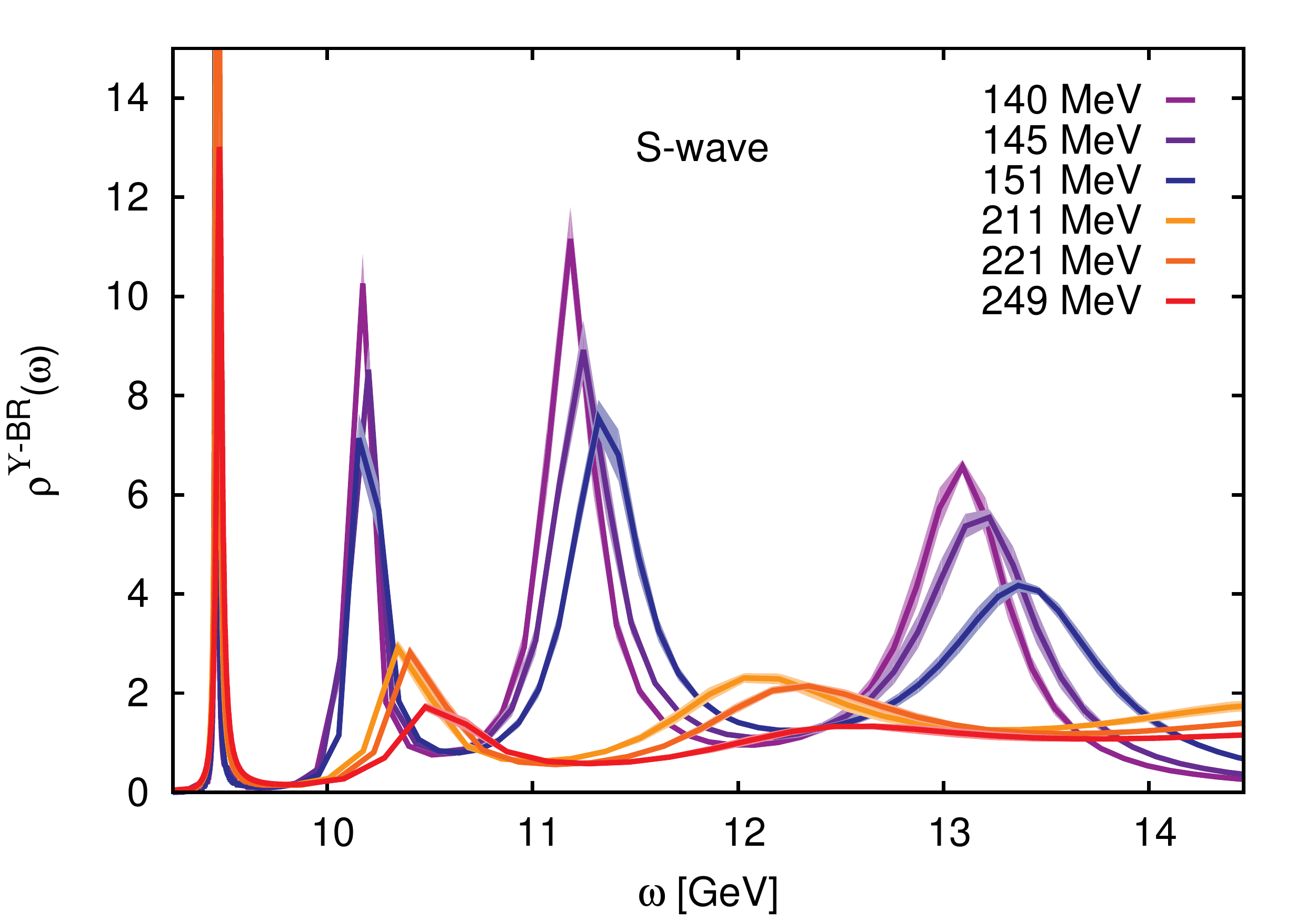}
 \includegraphics[scale=0.3,angle=-90]{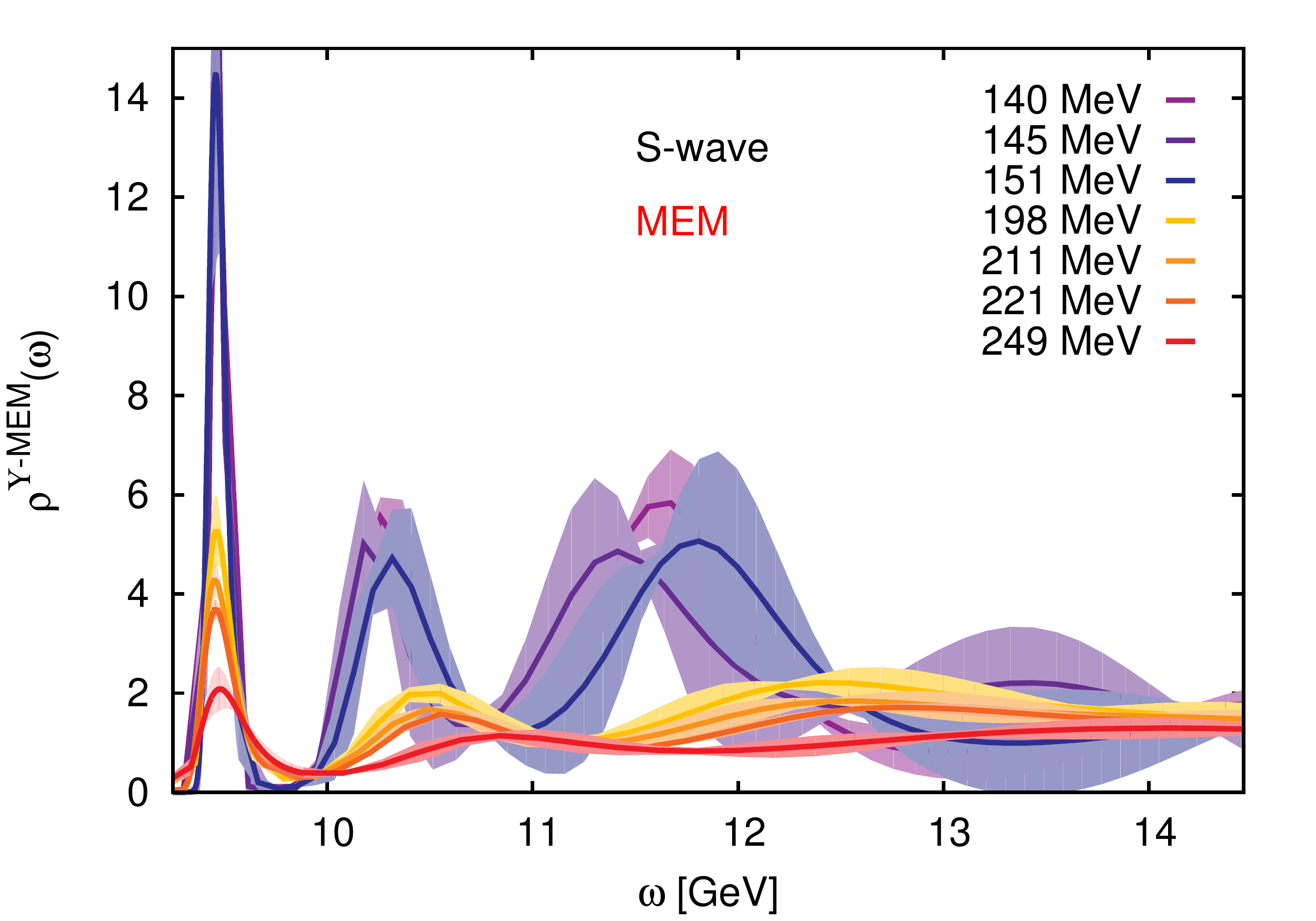}
 \includegraphics[scale=0.3,angle=-90]{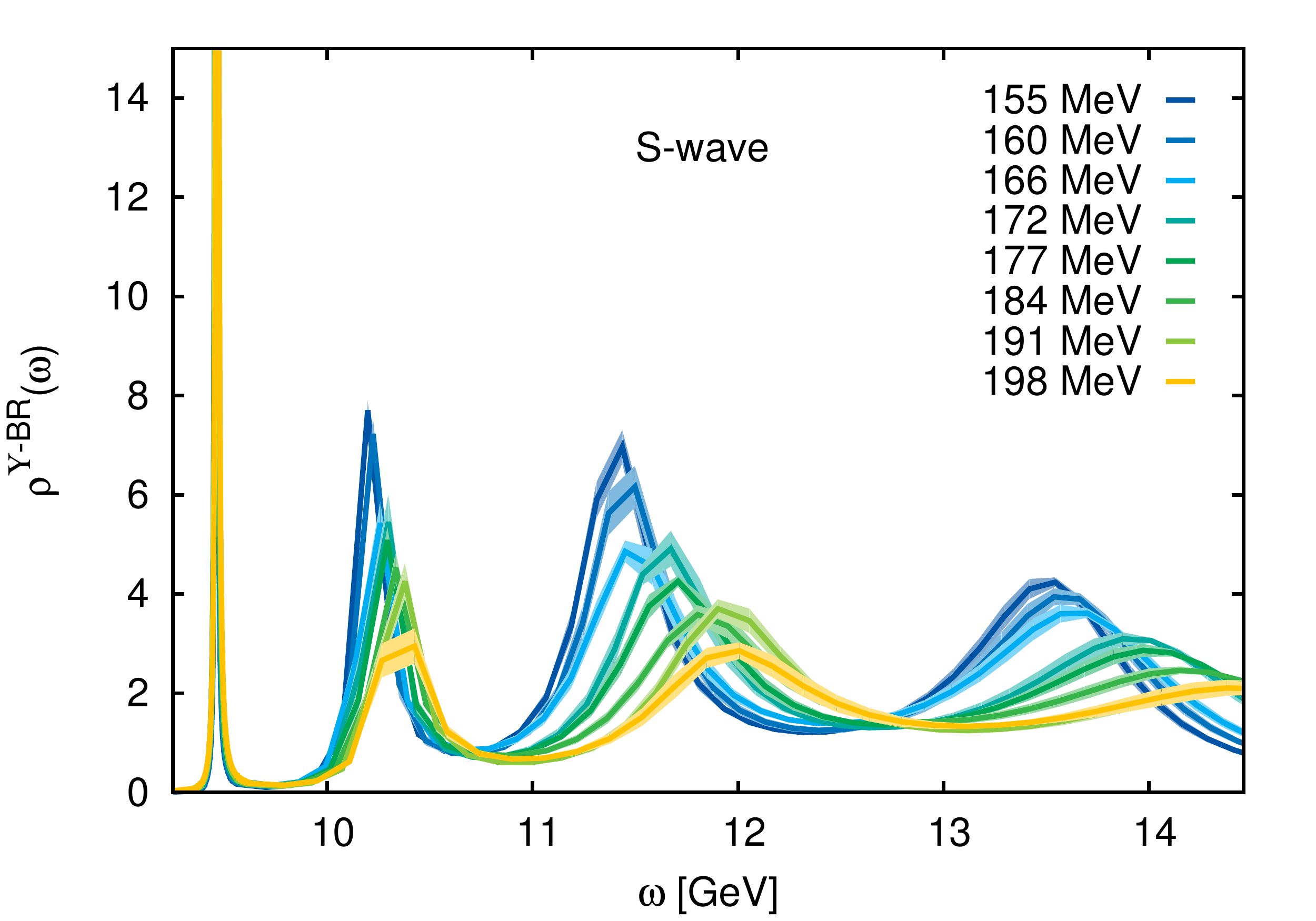}
 \includegraphics[scale=0.3,angle=-90]{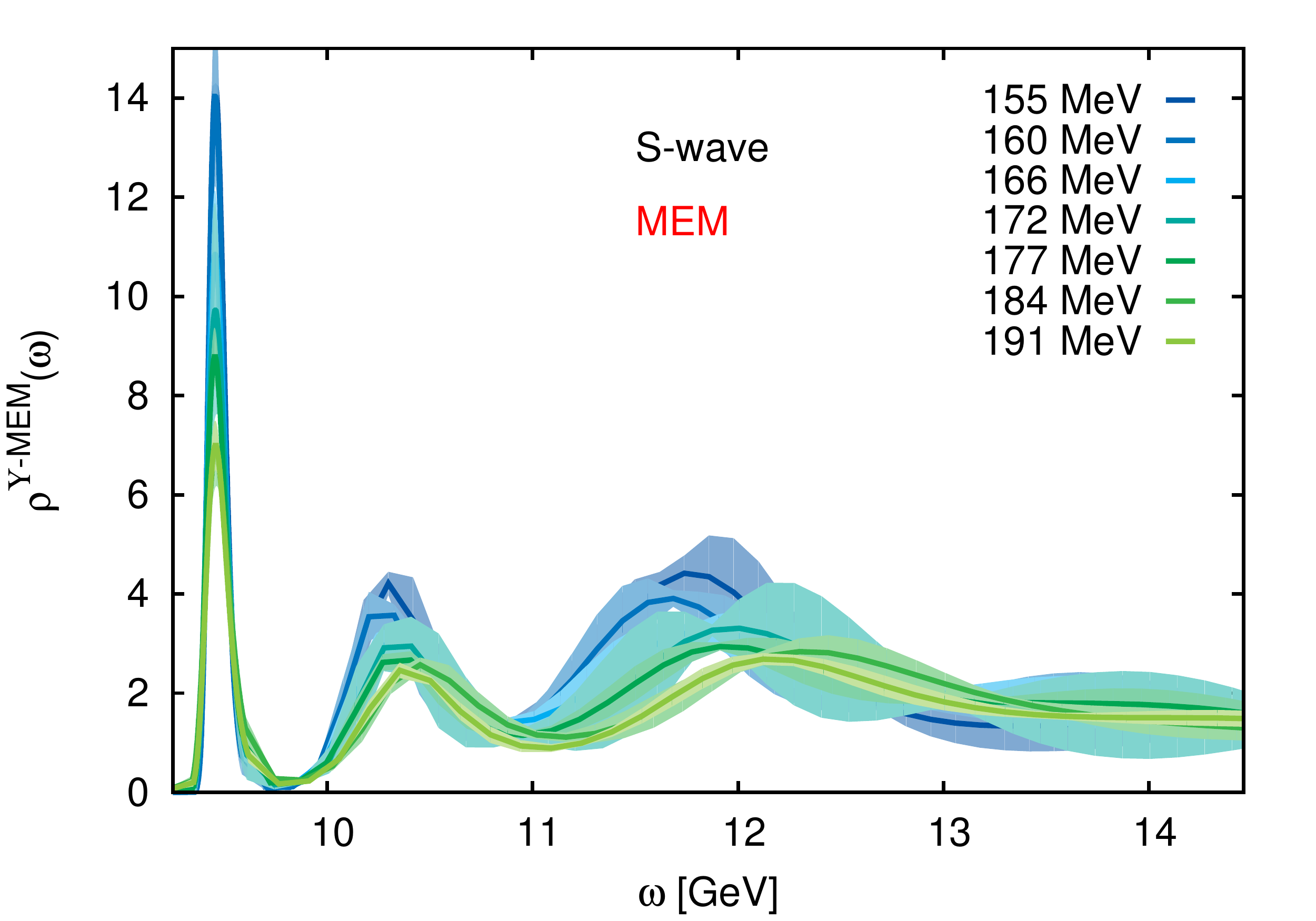}
 \caption{The mass-shift calibrated S-wave spectra from the new
   Bayesian approach (left column) and those from the MEM (right column)
   for fourteen different temperatures between $140$ (dark
   violet) and $249$MeV (red). The middle row shows spectra for the
   three lowest temperatures (below $T_c$) and the three highest
   temperatures. The bottom figures contain spectra just above the
   deconfinement transition $T = 155, \cdots, 198 {\rm
     MeV}$. Note that even though on a linear scale the ground state peak
   from the novel method does not appear to change (bottom left) its height
   decreases as can be seen in the top left panel.} \label{Fig:ThermalSpectraOverviewSwave}
\end{figure*}

\subsection{The upsilon channel -- S-wave}

In Fig. \ref{Fig:ThermalSpectraOverviewSwave} we show several
different visualizations of the reconstructed S-wave channel spectra
for a qualitative inspection. The left column contains the results of
the Bayesian reconstruction, while in the right column the MEM results
are presented.To obtain an overview of the different orders of
magnitude between the ground state peak and higher lying features, the
top row figures are given in logarithmic scale spanning the relevant
frequencies above the $2 M_b$ threshold. All fourteen spectra between
$140$ (dark violet) and $249$MeV (red) are included. The middle row
compares in linear scale three spectra at the lowest and the
highest temperatures investigated. The bottom row contains the spectra
just above the deconfinement transition.

We find that the new Bayesian approach allows us to extract the
features of the spectral functions with a much higher resolution than
the MEM. Based on exactly the same data set, we manage to obtain a
width of the lowest lying peak, which is consistently at least an
order of magnitude smaller than that of the MEM. Furthermore we
observe that the functional form of the ground state peaks in the MEM
resembles a Gaussian. This behavior is qualitatively different from
the Lorentzian observed with the new approach, which from general
arguments, is expected to be the correct shape for a particle of
finite lifetime.

\begin{figure*}
\centering
 \hspace{-0.3cm}\includegraphics[scale=0.3,
   angle=-90]{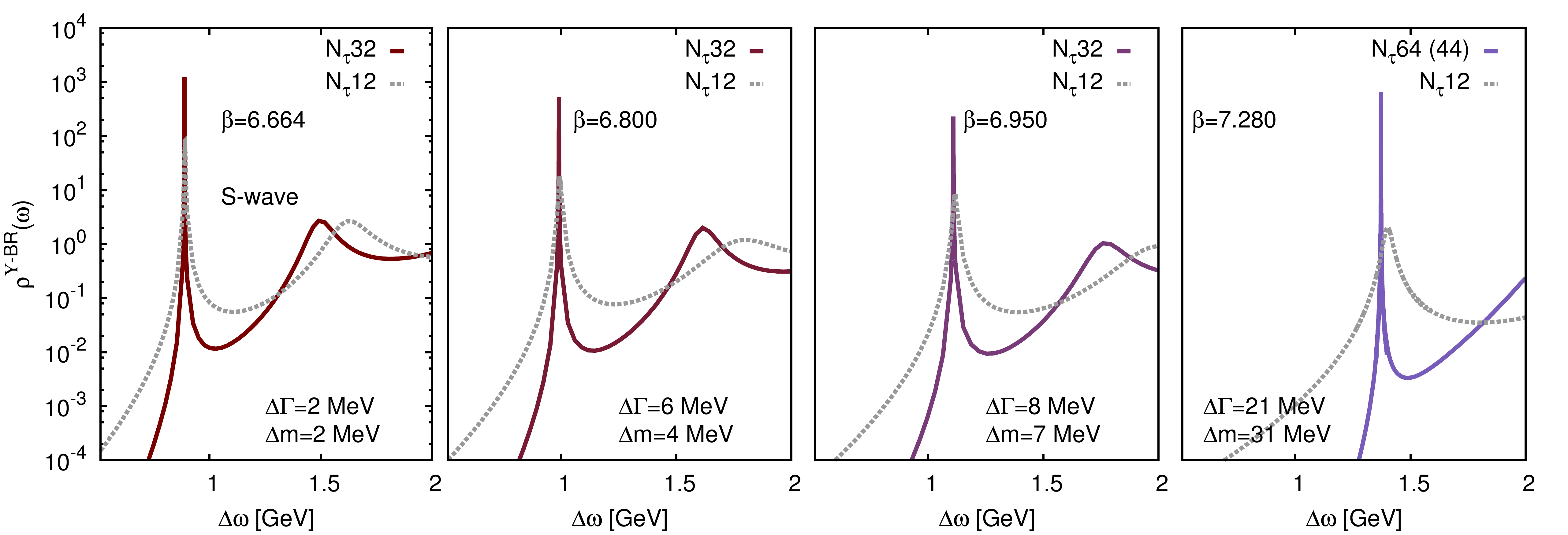}\hspace{0.2cm}
 \caption{Change in S-wave peak position and width when reconstructing
   the $T=0$ spectra at $\beta=6.664, \cdots, 7.280$ (from the left to
   the right) from the full $\tau_{\rm max}=32/64$ data set, as well
   as from the subset with $\tau_{\rm max}=12$. Since for a fixed
   $\beta$ the high frequency regime remains unchanged when going from
   $T\simeq 0$ to $T>0$ the observed differences can serve as a
   measure of the limits to the reconstruction
   accuracy.}\label{Fig:T0DatPRemovalEffectS}
\end{figure*}
\begin{figure*}
 \includegraphics[scale=0.3,
   angle=-90]{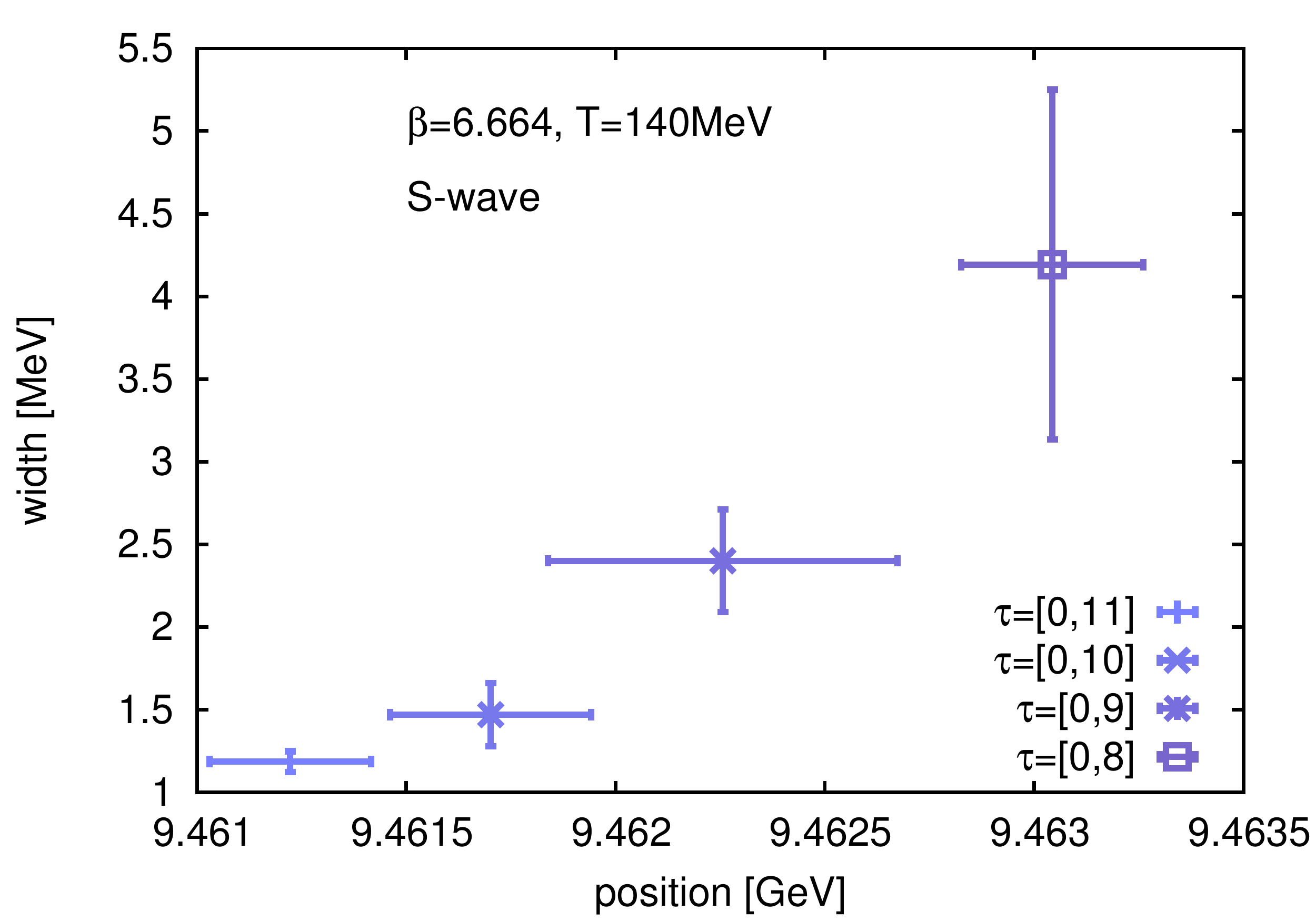}
 \includegraphics[scale=0.3,
   angle=-90]{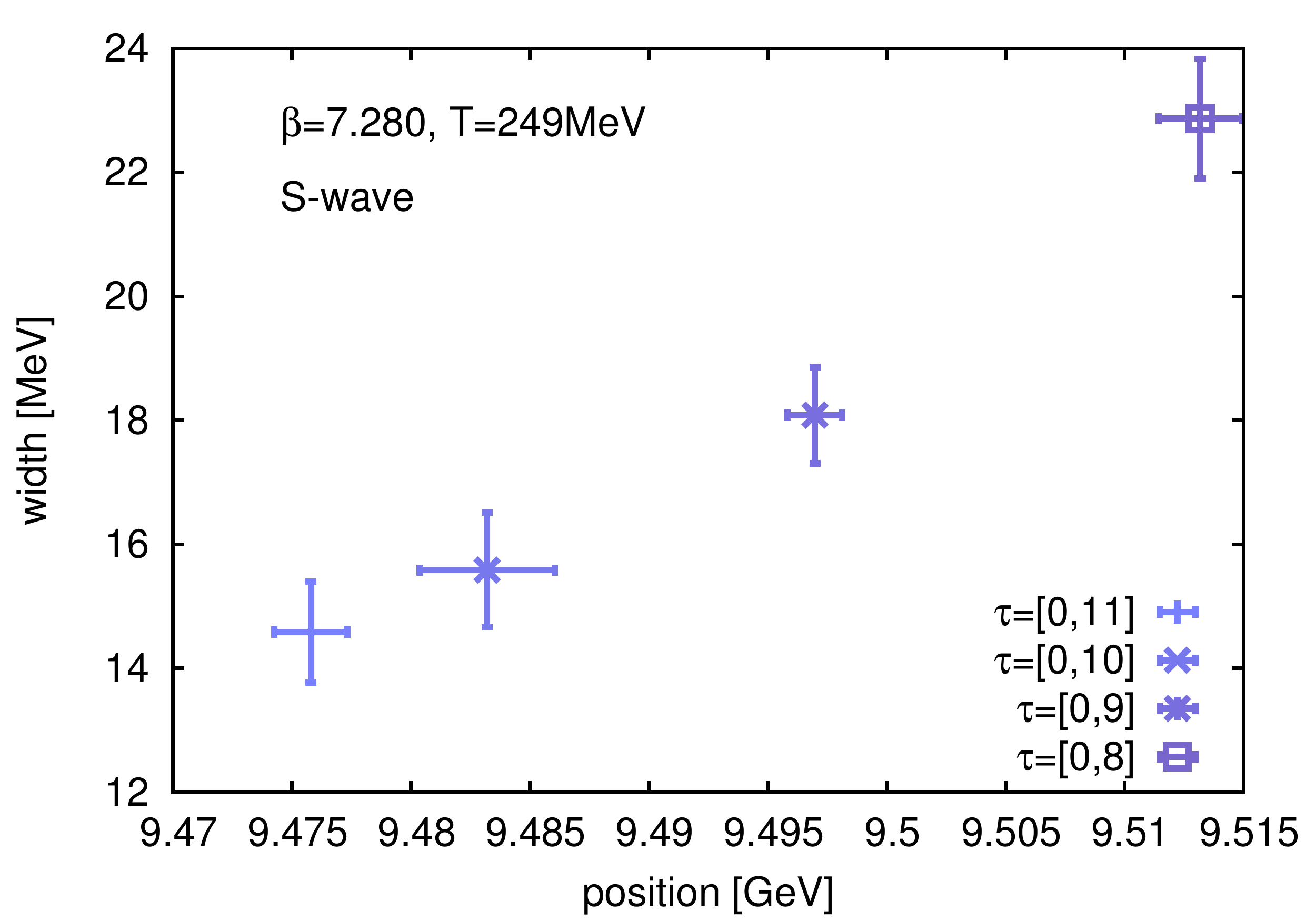}
 \caption{S-wave at $140$ (left) and $249$MeV (right): Dependence
   of the reconstructed lowest peak position and peak width on
   choosing different subsets of data points along the $\tau$
   axis. Note that the systematic effect of removing data points leads
   to a shift of the peak width and a broadening of the peak beyond
   the jackknife error bars}\label{Fig:6664DatapointDependeceS}
\end{figure*}

If we take a look at the ground state peak reconstructed from the new
approach, we see that with increasing temperature there appears to set
in a monotonous, albeit relatively small shift of the peak position to
higher frequencies as well as a broadening of the width. While it is
tempting to attribute these changes to the in-medium modification of
bottomonium itself we have to first ascertain how far the systematic
uncertainties due to the limited number of data points underlying the
reconstruction do influence the outcome.

Two different comparisons help us to understand the limitations of the
accuracy of our results. The first is shown in
Fig. \ref{Fig:T0DatPRemovalEffectS}, where we compare the effect of
removing all but twelve data points from the $T\simeq0$ correlator
data sets used in the calibration performed in
Sec. \ref{sec:zeroT}. Since the kernel in our case is temperature
independent this comparison tells us how the same spectrum encoded at
different temperature is resolved by our method\footnote{Since in the
  new Bayesian approach the dimensionality of the solution space is
  not restricted a priori to the number of data points as in the
  standard MEM, changing the number of underlying data points only
  affects the amount of information available and does not change the
  reconstruction prescription.}. As we can see the accuracy differs
for different lattice spacing, the change in mass and width being
$2$MeV for $\beta=6.664$ and maximal $31$MeV for the width at
$\beta=7.280$.

The second comparison is shown in
Fig. \ref{Fig:6664DatapointDependeceS}, where we remove from the $T>0$
correlator data set itself up to four of the points closest to
$\tau_{\rm max}=1/T$. We find that the result is a monotonous shift of
the reconstructed peak position to higher frequencies and a
significant increase in the broadening of the width. Both the
influence on the position and width go beyond the error bars shown,
which represent the statistical uncertainty and are obtained from the
jackknife variance. The strength of the effects is comparable to that
observed in Fig. \ref{Fig:T0DatPRemovalEffectS}.

From these two checks we conclude that the size of our data set does
not allow us to make a quantitative statement about the changes in
peak position and width with temperature as the systematic errors
dominate our results \footnote{Note that the default model dependence
  and other systematic error are smaller than the dependence on
  $N_\tau$ as can be seen in Appendix \ref{sec:systematics}}. We
however are in a position, due to the superior resolution of the new
Bayesian approach, to give stringent upper limits on the size of the
in-medium modification of $\Upsilon$ ranging between
\begin{align}
\nonumber  \Delta m_{\rm T}(140{\rm MeV})&<2{\rm MeV}, \quad \Delta \Gamma_{\rm T}(140{\rm MeV})<5{\rm MeV}\\
 \Delta m_{\rm T}(249{\rm MeV})&<40{\rm MeV}, \quad \Delta \Gamma_{\rm T}(249{\rm MeV})<21{\rm MeV}.
\end{align}

Another question we can address, is to determine an upper bound for
the temperature up to which the $\Upsilon$ channel ground state
survives. As we have seen in the previous discussion, the effect of a
reduced number of data points is a monotonous broadening of the ground
state peak. If a well-defined peak is visible in the reconstruction
the true spectrum should also possess such a structure of at least the
same strength. In that sense
Fig. \ref{Fig:ThermalSpectraOverviewSwave} allows us to confirm the
qualitative findings of previous lattice NRQCD studies
\cite{Aarts:2011sm,Aarts:2014cda} in that the ground state peak
retains a narrow width up to the highest temperature studied ($T =
249$ MeV). While the survival of the ground state peak in the S-wave
channel is a robust feature of our analysis, no definite statements
can be made about the fate of $\Upsilon(2S)$ state due to the
systematic errors related to the number of data points and small
extent of the time direction. This will become apparent in the
discussions below.

The determination of the actual survival or dissolution of the ground
state from an inspection by eye of the lowest lying peak shape of
the reconstructed spectra however is highly nontrivial. Indeed it
needs to be understood, whether a peak visible in the Bayesian spectra
can be attributed to a bound state held together by the strong
interactions. Uncertainty here arises from the fact that we
reconstruct spectra from a finite number of data points, which will
inevitably lead to the presence of numerical ringing. This in turn
needs to be distinguished from actual physical peak structures.

Here we propose to give the decision about survival or melting a solid
footing through a comparison of the fully interacting spectra to those
obtained from noninteracting correlators. To this end we set all
links on our lattices to unity when calculating the Euclidean NRQCD
Green's function. The same random sources as in the interacting case
enter the initial conditions for Eq.\eqref{NRQCDEvolEq}. Since the
free correlators are not uniquely defined in lattice NRQCD, it is our
choice of $S(\mathbf{x})$ that determines their values at
$\tau=0$. Their time evolution depends on the NRQCD mass parameter
$M_b a$, for which we choose the same $\beta$-dependent values
obtained on the interacting lattices we wish to compare to. To enable
a meaningful comparison between our reconstructed spectra, we
normalize in the following the free correlator at the first time step
to the value of the corresponding interacting spectral function.
Since the free correlators also possess different, i.e. much smaller
errors than the interacting ones, we add Gaussian noise of a similar
strength as their errors to the free data points before the
reconstruction. Note that spectral functions from free NRQCD
correlators do not contain a physical scale. We introduce the scale in
the free theory calculation by using the lattice spacing that
corresponds to the value of $\beta$ for which the comparison with free
spectral functions is performed.

\begin{figure}
\centering
 \includegraphics[scale=0.3,angle=-90]{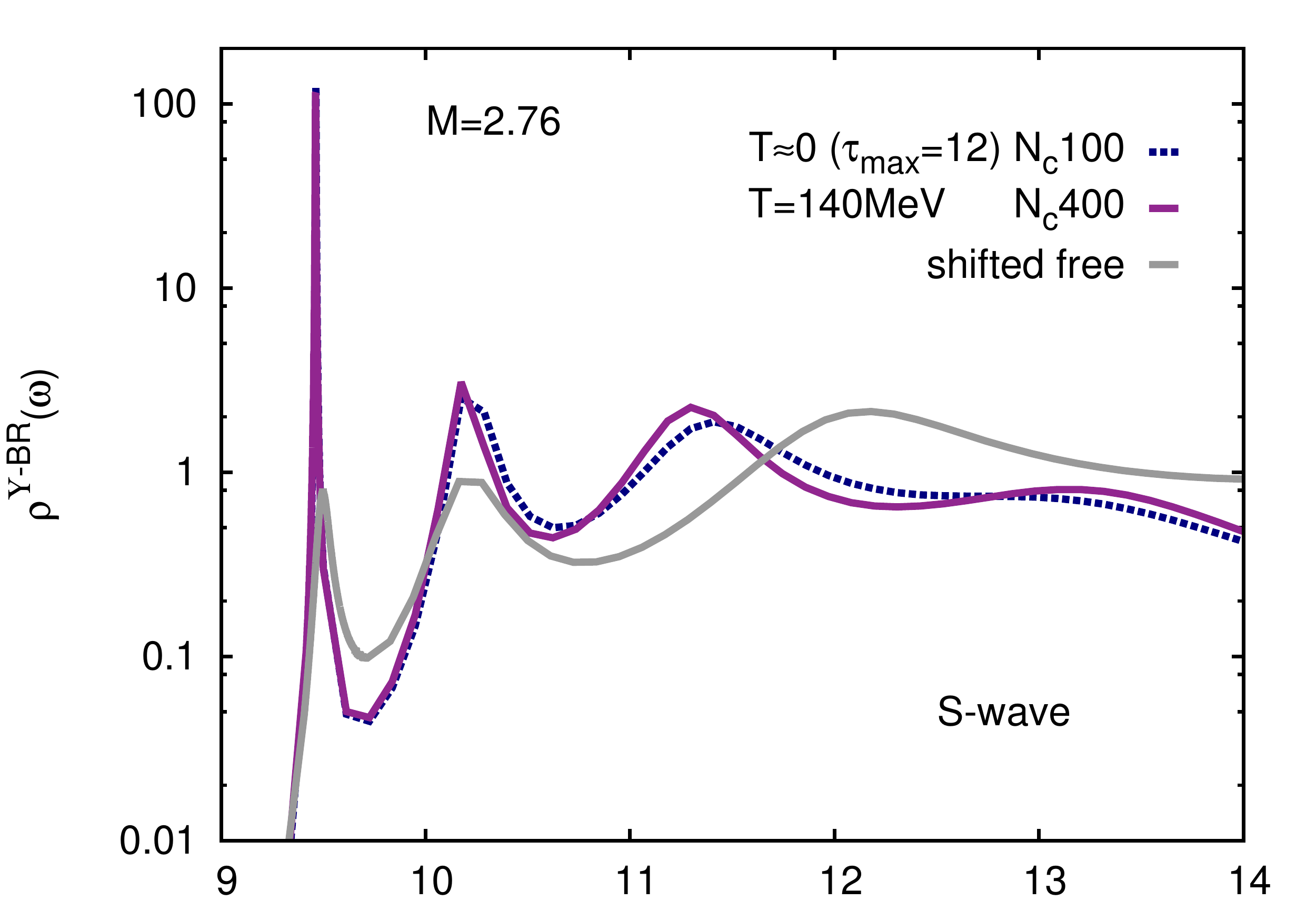}
 \includegraphics[scale=0.3,angle=-90]{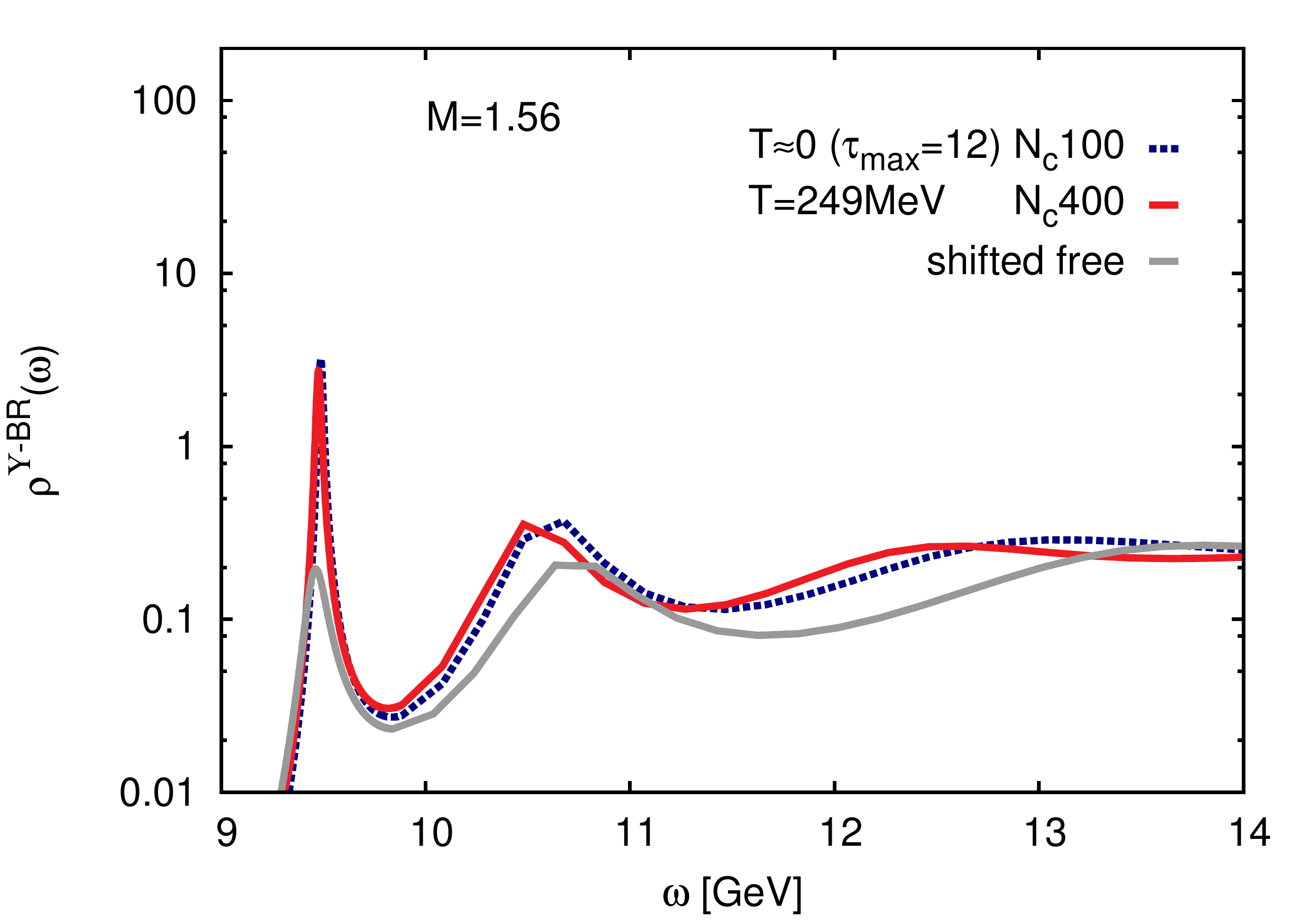}
 \caption{S-wave spectral reconstructions at $T = 140$MeV (top) and
   $T = 249$MeV (bottom). To disentangle systematic and medium effects
   we compare the Bayesian $T>0$ spectra from all $N_\tau=12$ data
   points (colored solid) to the spectra obtained from the first
   $\tau_{\rm max}=12$ points of the $T\simeq0$ correlator (dashed
   gray). To verify the presence of a bound state the spectra from
   noninteracting correlators are plotted (solid gray). $N_c$ denotes
   the number of correlators used in the analysis.
 } \label{Fig:CompFreeIntS}
\end{figure}

In Fig. \ref{Fig:CompFreeIntS} we present this comparison in the
S-wave channel at the largest (left) lattice spacing $\beta=6.664$
($T=140$MeV) and the smallest (right) $\beta=7.280$ ($T=249$MeV). The
colored solid curves show the result of the finite $T$ reconstruction
from $N_\tau=12$ data points, while the dark blue, dashed curve
represents the reconstruction from the first $\tau_{\rm max}=12$ data
points of the $T\simeq0$ correlator. As we discussed before with the
limited number of data points available to us, the reconstructed
spectra at $T\simeq0$ and $T>0$ are very similar especially around the
lowest lying peak. On the other hand the free spectra obtained from
the same number of data points do differ. The Bayesian
reconstructions of the free spectra in Fig. \ref{Fig:CompFreeIntS}
show peak structures even though there are no such features present in
the analytic form of free NRQCD spectral function. This is reminiscent
of the Gibbs phenomenon mentioned in Sec. \ref{sec:nrqcd}. Here in
the case of $\Upsilon$, the ground state peak at $T>0 $ is easily
distinguished from this numerical ringing, as it is at least one order
of magnitude larger. From these results we conclude that the ground
state of the $\Upsilon$ channel survives at least up to $T=249$MeV.

\begin{figure} \centering
  \includegraphics[scale=0.3,angle=-90]{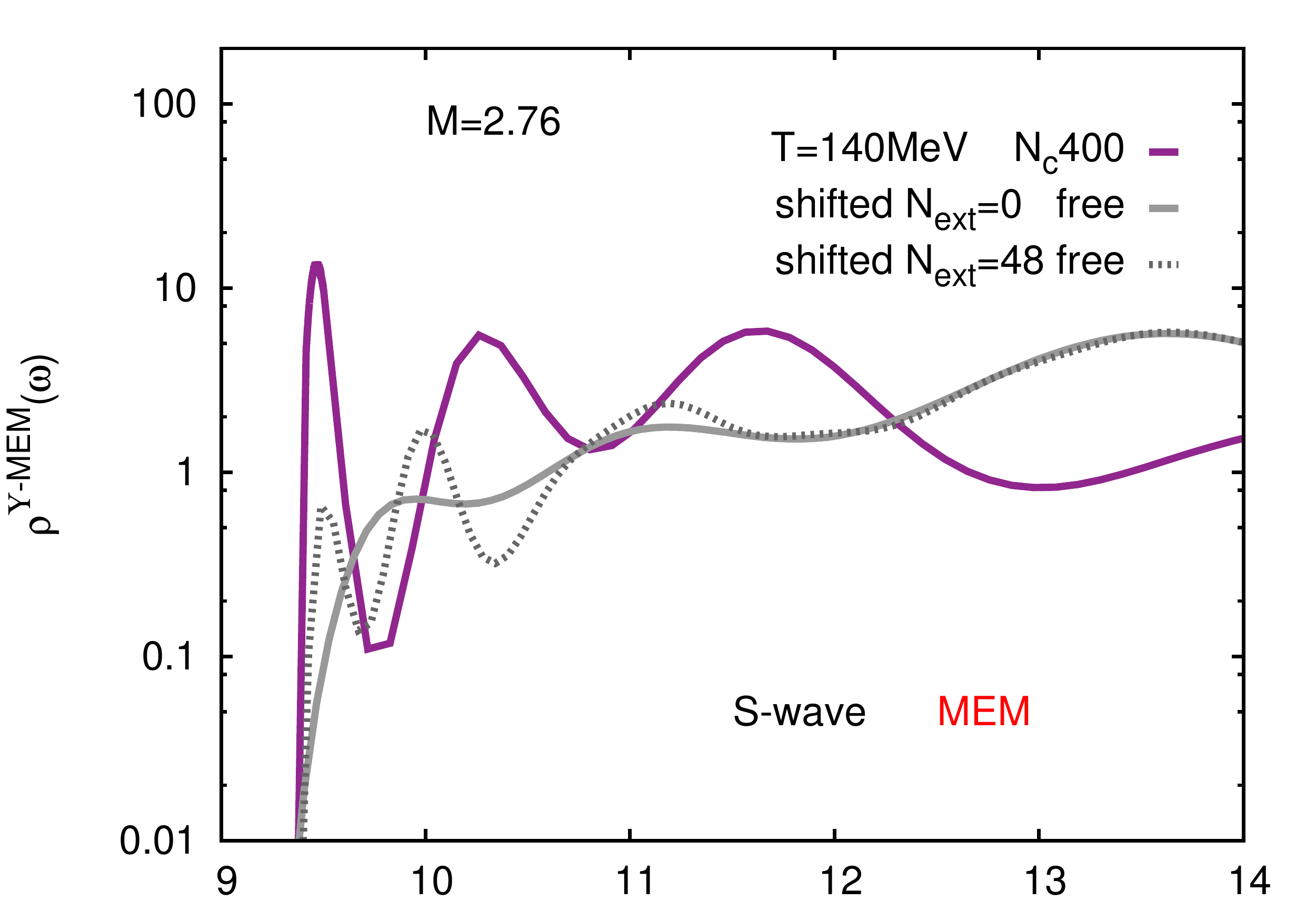}
  \includegraphics[scale=0.3,angle=-90]{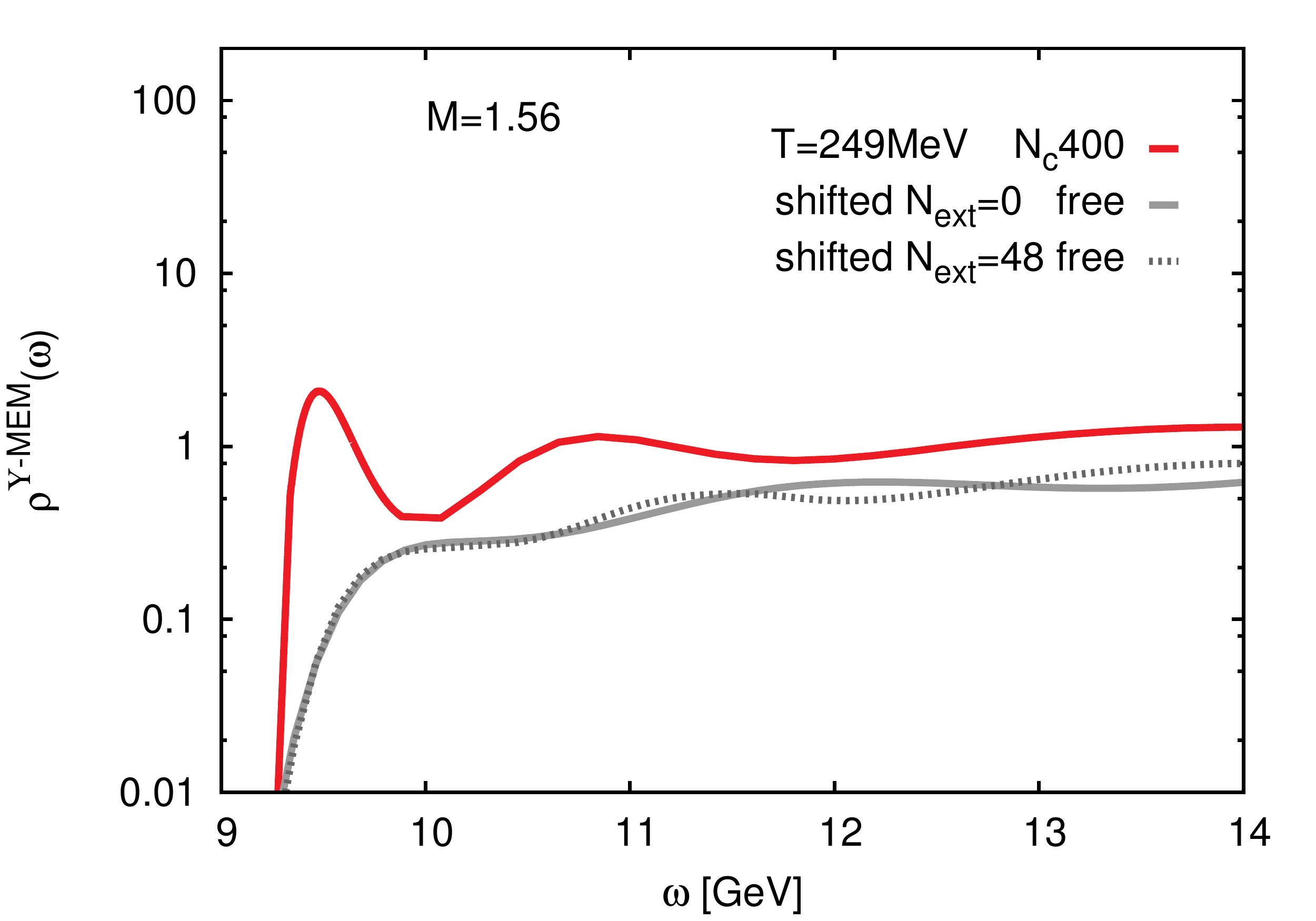}
  \caption{S-wave MEM spectral reconstructions at $T = 140$ (top)
    and $T = 249$MeV (bottom). Finite T spectrum (colored) as well as
    free spectrum with standard (solid gray) and extended $(N_{\rm
      ext}=48)$ search space (dashed gray).
  } \label{Fig:CompFreeIntSMEM}
\end{figure}

For comparison purposes we show in Fig. \ref{Fig:CompFreeIntSMEM} the
interacting (colored gray) and free spectra (gray) obtained from a MEM
reconstruction. As we already saw in
Fig. \ref{Fig:ThermalSpectraOverviewSwave}, the lowest lying peaked
features are much more shallow here than in the novel Bayesian
approach and their functional form is not Lorentzian, contrary to what
is expected for a particle of finite lifetime. Consistent with the
previous discussion, we find that even at the highest temperature a
ground state peak appears to survive. Note that the smoothness of the
MEM free spectra does not necessarily mean that an accurate
reconstruction has been achieved. Indeed as we increase the number of
basis functions to $N_{\rm ext}=48$, the number of wiggles, at least
for $\beta=6.664$, significantly increases and the result approaches a
similar form as the Bayesian result obtained by the novel method.

\FloatBarrier
\subsection{The $\chi_{b1}$ channel -- P-wave}
\begin{figure*}
  \includegraphics[scale=0.3,angle=-90]{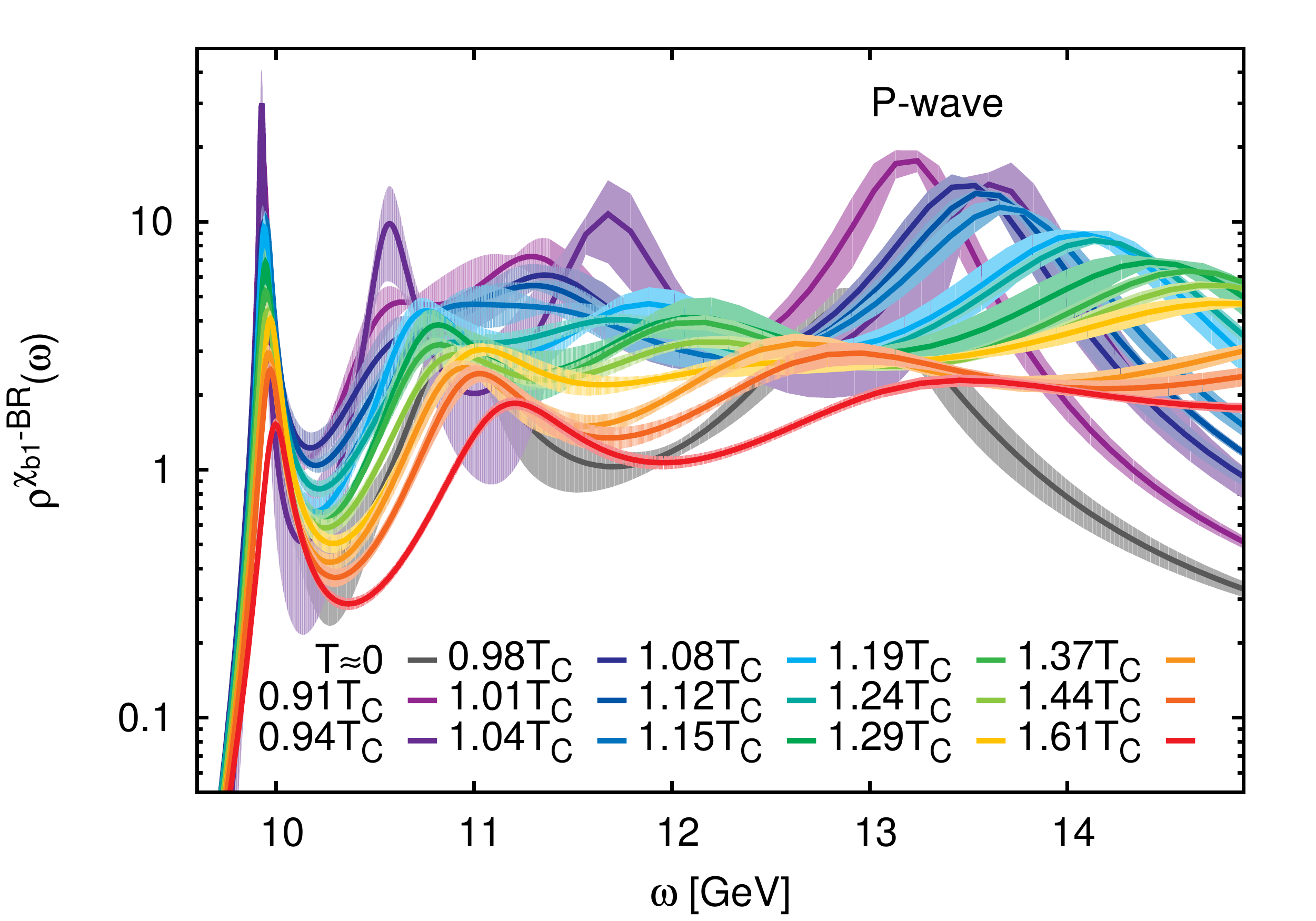}
  \includegraphics[scale=0.3,angle=-90]{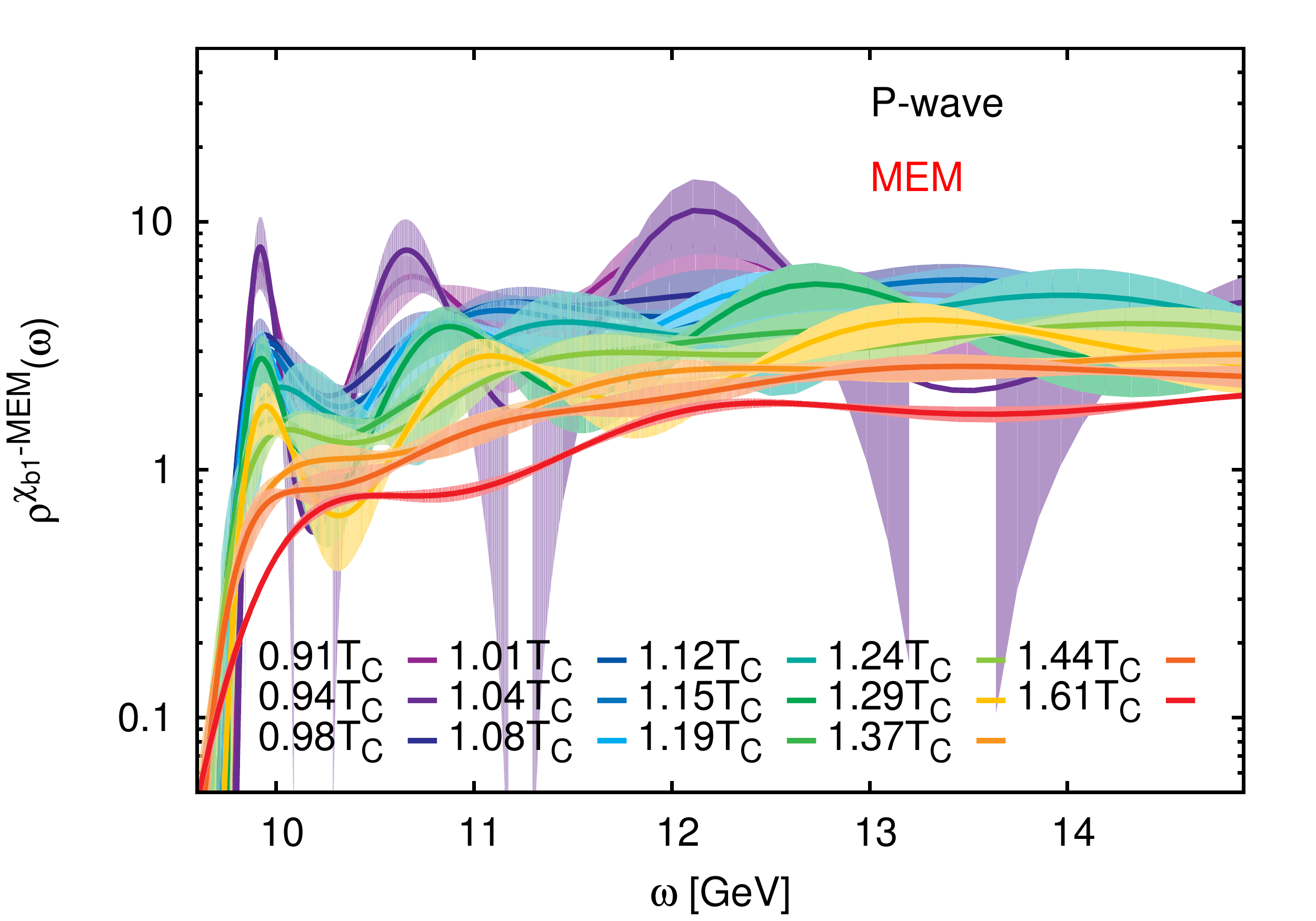}
  \includegraphics[scale=0.3,angle=-90]{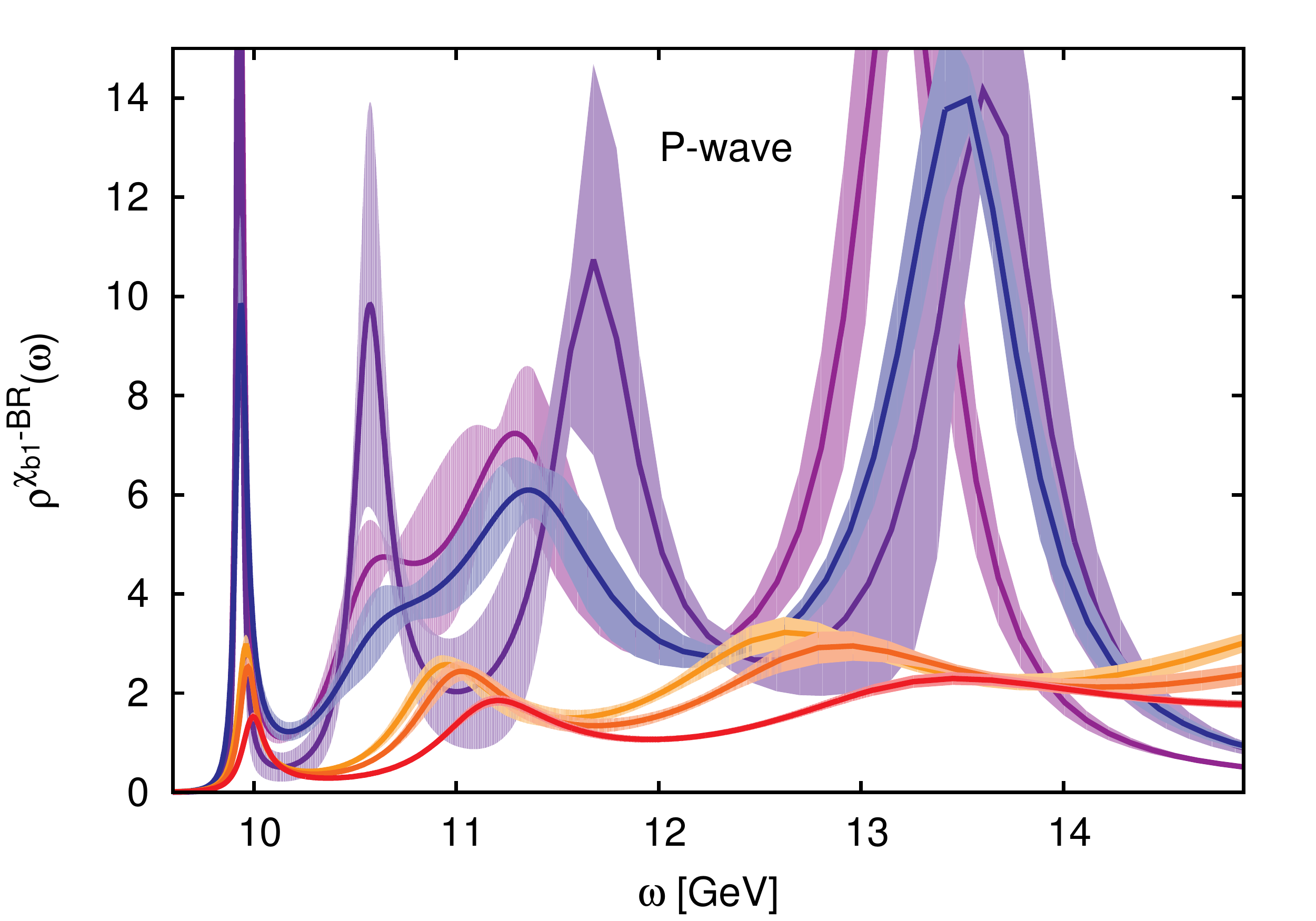}
  \includegraphics[scale=0.3,angle=-90]{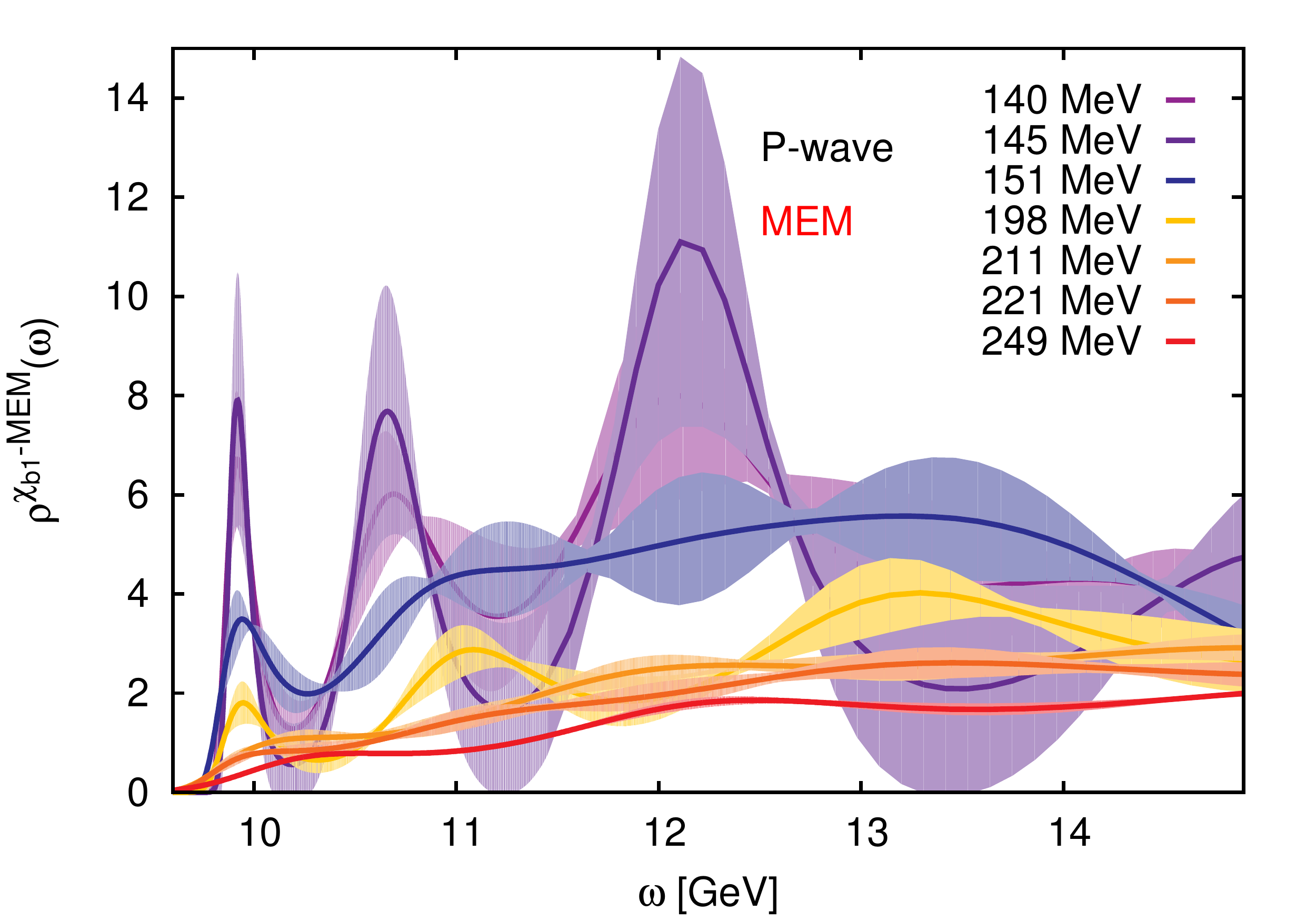}
  \includegraphics[scale=0.3,angle=-90]{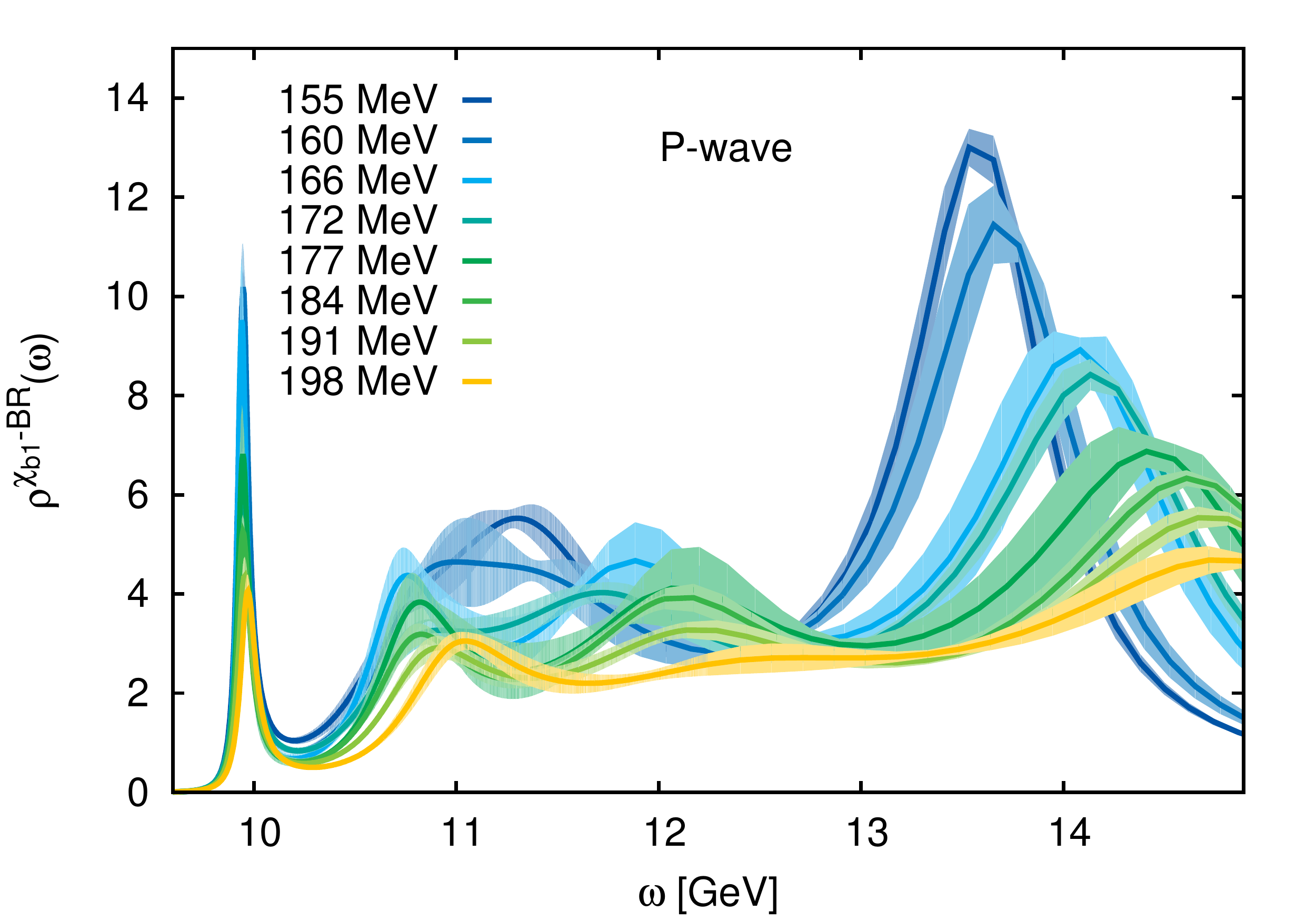}
  \includegraphics[scale=0.3,angle=-90]{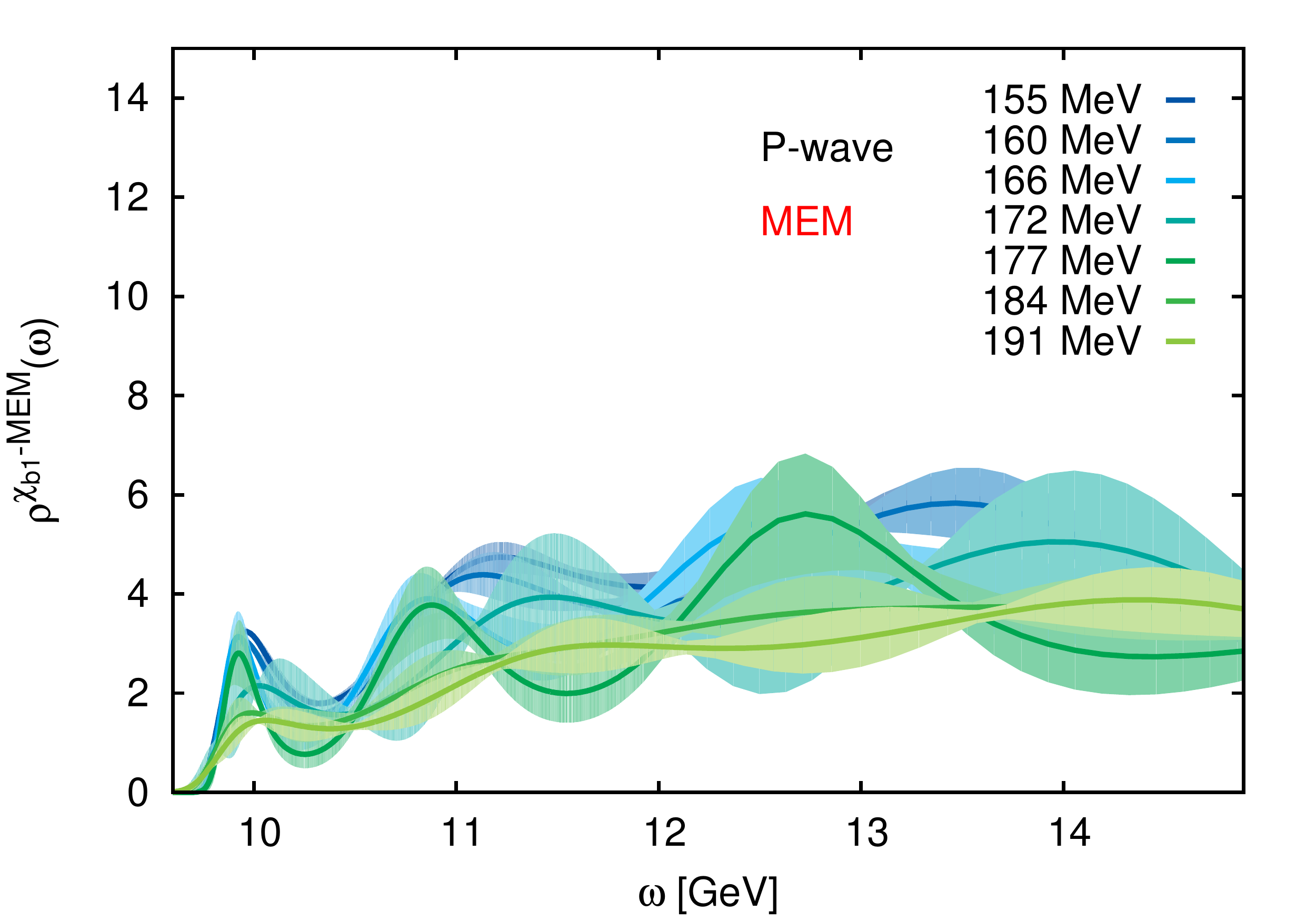}
  \caption{The
    mass-shift calibrated P-wave spectra from the new Bayesian
    approach (left column) and those from the MEM (right column) for
    fourteen different temperatures between $140$ (dark violet) and
    $249$MeV (red). The middle row shows spectra for the three lowest
    temperatures (below $T_c$) and the three highest temperatures. The
    bottom figures contain spectra just above the deconfinement
    transition $T = 155 {\rm MeV}, \cdots, 198 {\rm MeV}$.
  } \label{Fig:ThermalSpectraOverviewPwave}
\end{figure*}

We continue with the P-wave ($\chi_{b1}$ channel) spectral functions
shown in Fig. \ref{Fig:ThermalSpectraOverviewPwave}, where again the
results of the new Bayesian approach on the left are contrasted to the
maximum entropy method on the right. The P-wave spectra, due to the
larger ground state mass of $\chi_{b1}$, start at higher frequencies
and thus the underlying correlators suffer more strongly from the
finite number of measurements and lead to less reliable
reconstructions, which is reflected in larger error
bars.
The top row contains the spectral functions at all temperatures, while
the middle row shows those from the lowest three (all below $T_c$) and
for the highest three ($T = 211, 221, 249$ MeV) respectively. The
bottom row features the spectra just above the phase transition. As
for the S-wave, the new Bayesian approach allows us to obtain much
sharper resolved peaks than the MEM, using the same data set. This
difference turns out to lead to a pronounced qualitative difference in
the P-wave case, as can be seen in the middle row of
Fig. \ref{Fig:ThermalSpectraOverviewPwave}. The MEM spectra at the
highest temperatures appear almost featureless, while the new Bayesian
approach manages to resolve a well-defined ground-state peak. Hence,
while a naive inspection by eye in the case of the MEM suggests P-wave
melting at $T \gtrsim 211$MeV $(=1.37 T_c)$, i.e. slightly above the
deconfinement temperature ($T_c = 154$MeV), no such conclusion can be
drawn from the result of the new Bayesian approach.

The outcome of the MEM is consistent with the findings of other MEM
based studies that suggested P-wave melting already at ($T = 201$MeV
$=1.09T_c$). However the spectral functions reconstructed using the
new Bayesian approach at the same temperatures, show well-resolved
narrow peak structures and do not hint at melting of the P-wave
bottomonium ground state. Perhaps this is not surprising since the
limited search space of the standard MEM artificially restricts the
resolution of the reconstructed spectra and thus produces only washed
out features while the new Bayesian method operates directly in the
full search space and hence can produce well-defined peaks.

\begin{figure*}
  \centering \hspace{-0.3cm}\includegraphics[scale=0.3,
    angle=-90]{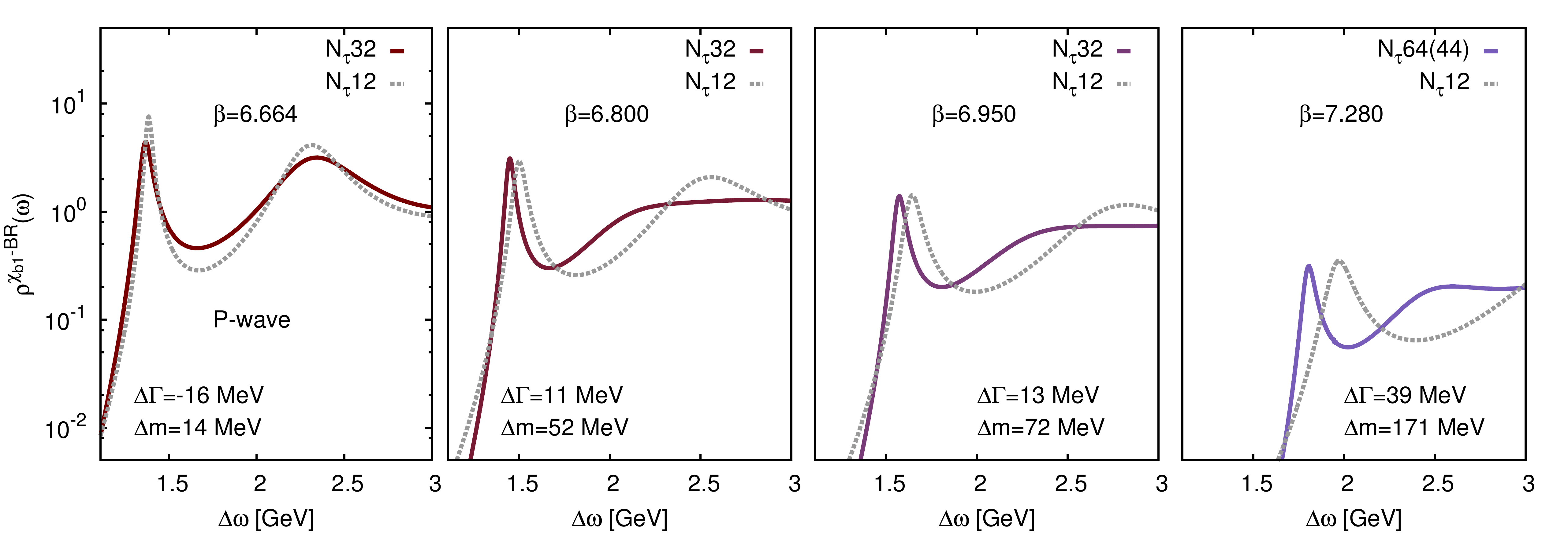}\hspace{0.2cm}
  \caption{Change in P-wave peak position and width when
    reconstructing the $T=0$ spectra from the full $\tau_{\rm
      max}=32/64$ data set, as well as from the subset with $\tau_{\rm
      max}=12$. Since for a fixed $\beta$ the high frequency regime
    remains unchanged when going from $T\simeq0$ to $T>0$ the observed
    differences can serve as a measure of the limits to the
    reconstruction
    accuracy.}\label{Fig:T0DatPRemovalEffectP}
\end{figure*}

\begin{figure*}
  \includegraphics[scale=0.3,
    angle=-90]{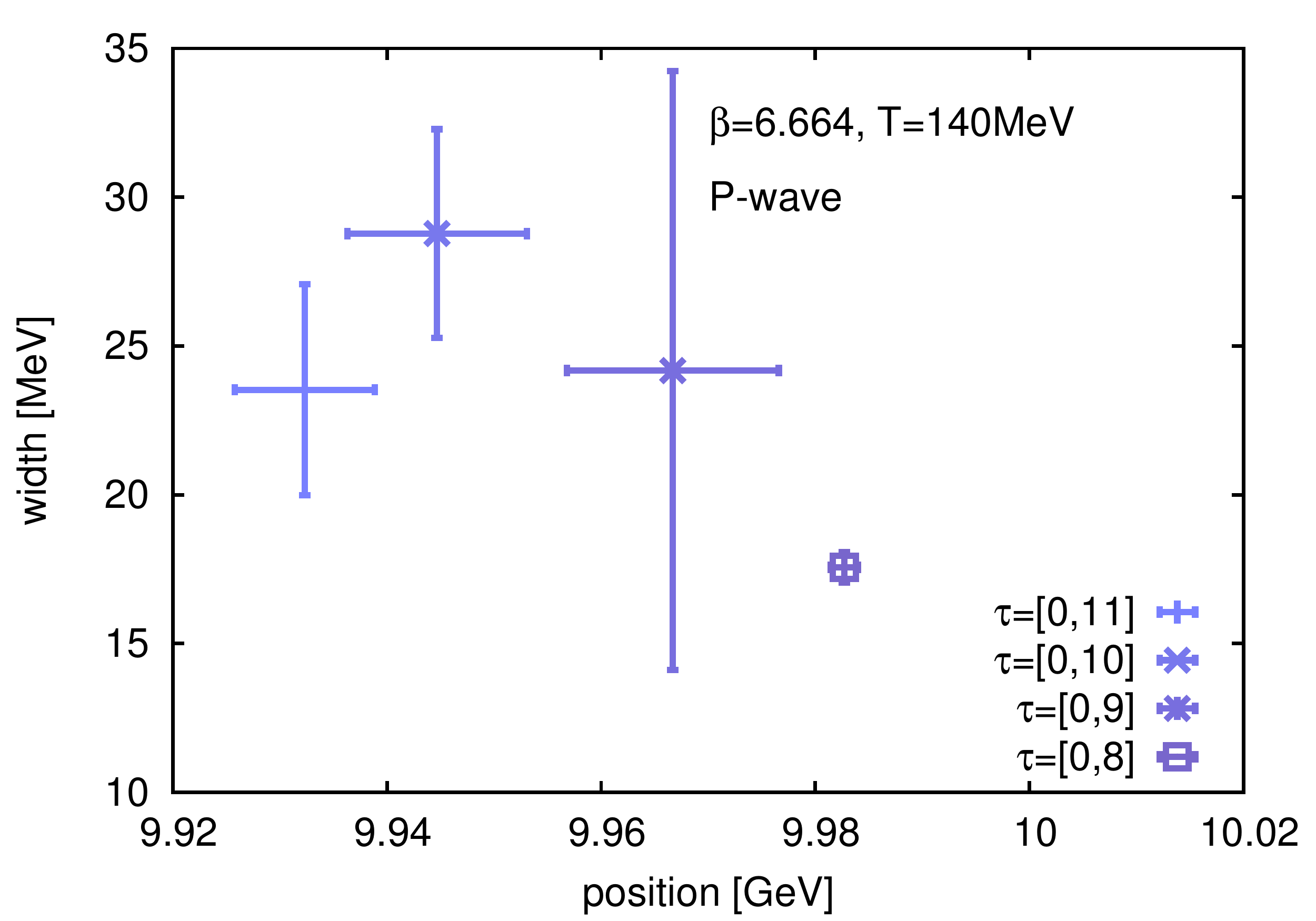}
  \includegraphics[scale=0.3,
    angle=-90]{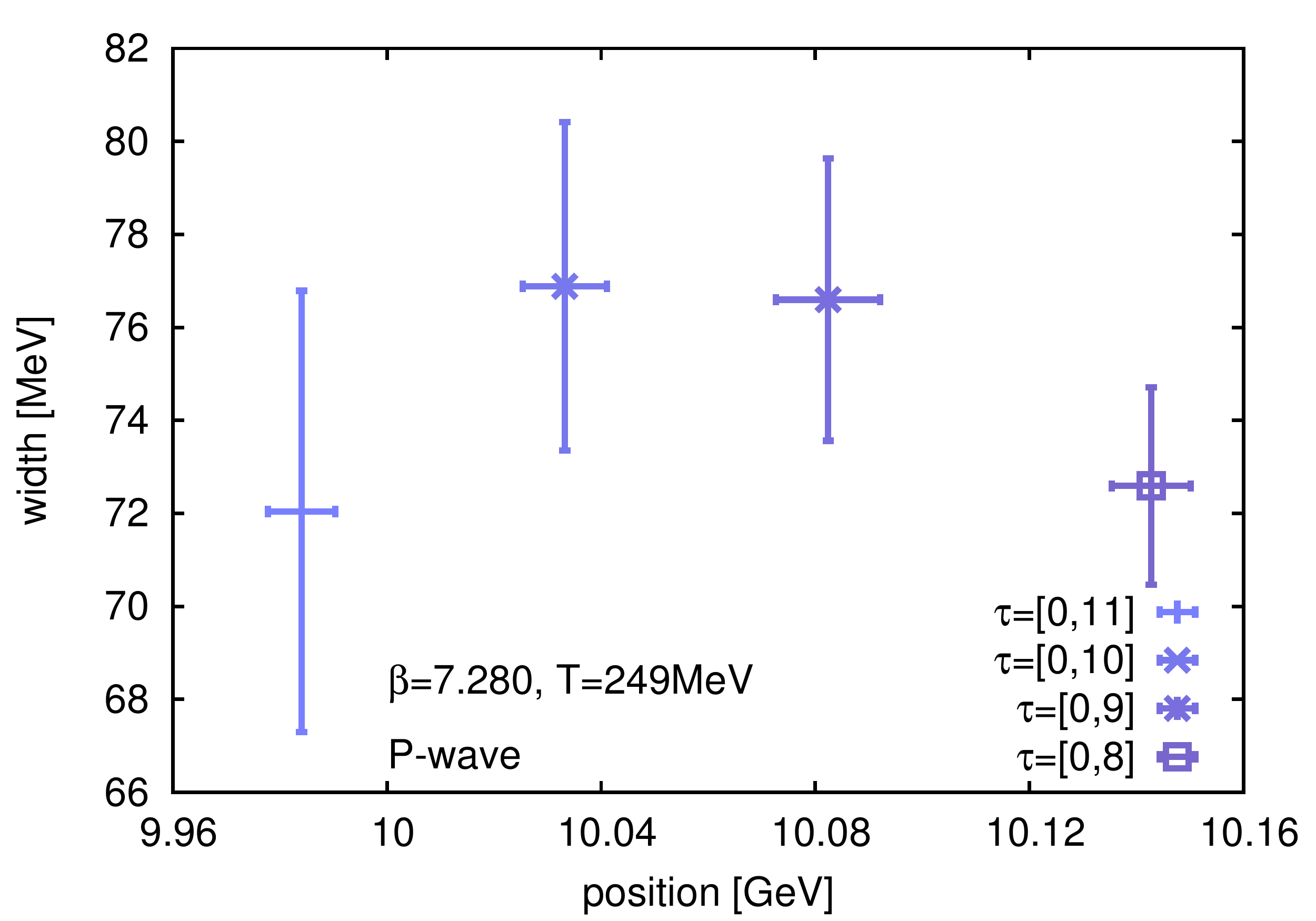}
  \caption{$140$ (left) and $249$MeV (right): Dependence of the
    reconstructed lowest peak position and peak width on choosing
    different subsets of data points along the $\tau$ axis. For the
    P-wave with worse signal-to-noise ratio, removing data points
    leads to an even stronger systematic shift of the peak width and a
    broadening of the peak than in the S-wave
    case.}\label{Fig:6664DatapointDependeceP}
\end{figure*}

Before attempting to clarify the fate of the $\chi_{b1}$ state, a look
at the systematics for the P-wave is in order. We begin with
Fig. \ref{Fig:T0DatPRemovalEffectP}, where we compare the effect of
removing all but twelve data points from the $T\simeq0$ correlator
data sets presented in Sec.\ref{sec:zeroT}. The fact that the
$\chi_{b1}$ state is located at higher frequencies leads to a stronger
exponential falloff in the correlators and thus leads to a smaller
signal-to-noise ratio than that for S-wave with fixed statistics. Thus
we expect that discarding data points will affect the outcome of our
reconstruction even more strongly in the P-wave case, which is also what we
find. The change in mass and width is $\simeq15$MeV at
$\beta=6.664$. At $\beta=7.280$ the change in mass is $171$MeV and the
change in width is $\simeq 40$MeV.
In Fig. \ref{Fig:6664DatapointDependeceP} the second pertinent
comparison is shown, where we remove from the $T>0$ correlator data
set up to four of the points closest to $\tau_{\rm max}=1/T$. Again we
find that the result is a monotonous shift of the reconstructed peak
position to higher frequencies while the width does not seem to be
affected beyond the relatively large jackknife error bars. We find the
strength of these effects to be stronger than those observed in
Fig. \ref{Fig:T0DatPRemovalEffectP} for the peak position but
comparable for its width.
From these previous two tests, we conclude that also for the P-wave,
the size of our data set does not allow us to make a quantitative
statement about the changes in peak position and width as temperature
is changed. Even though the modification of the peak position and
width is more pronounced than in the S-wave, also our systematic
uncertainty is larger. The upper limits for the in-medium modification
of $\chi_{b1}$ we can provide are
\begin{align} \nonumber |\Delta
  m_{\rm T}|(140{\rm MeV})&<60{\rm MeV}, \quad \Delta \Gamma_{\rm
    T}(140{\rm MeV})<20{\rm MeV}\\ \Delta m_{\rm T}(249{\rm
    MeV})&<200{\rm MeV}, \quad \Delta \Gamma_{\rm T}(249{\rm
    MeV})<40{\rm MeV}.
\end{align}
Returning to the question of the fate of $\chi_{b1}$ at high
temperatures, the need for a robust criterion to distinguish a melted
state from a bound state is even more evident for the P-wave. Simply
looking at the lowest lying peak in
Fig. \ref{Fig:ThermalSpectraOverviewPwave} one might conclude melting
of $\chi_{b1}$ at $T=249$MeV, since the second peak structure is as
large as the ground state feature. To reach a judgment on $\chi_{b1}$
on a systematic basis, we deploy the same strategy as laid out for the
S-wave, i.e. comparing the interacting reconstructed spectra to those
from free NRQCD correlators with the same bottom quark mass
parameter.
\begin{figure*}
  \centering
  \includegraphics[scale=0.3,angle=-90]{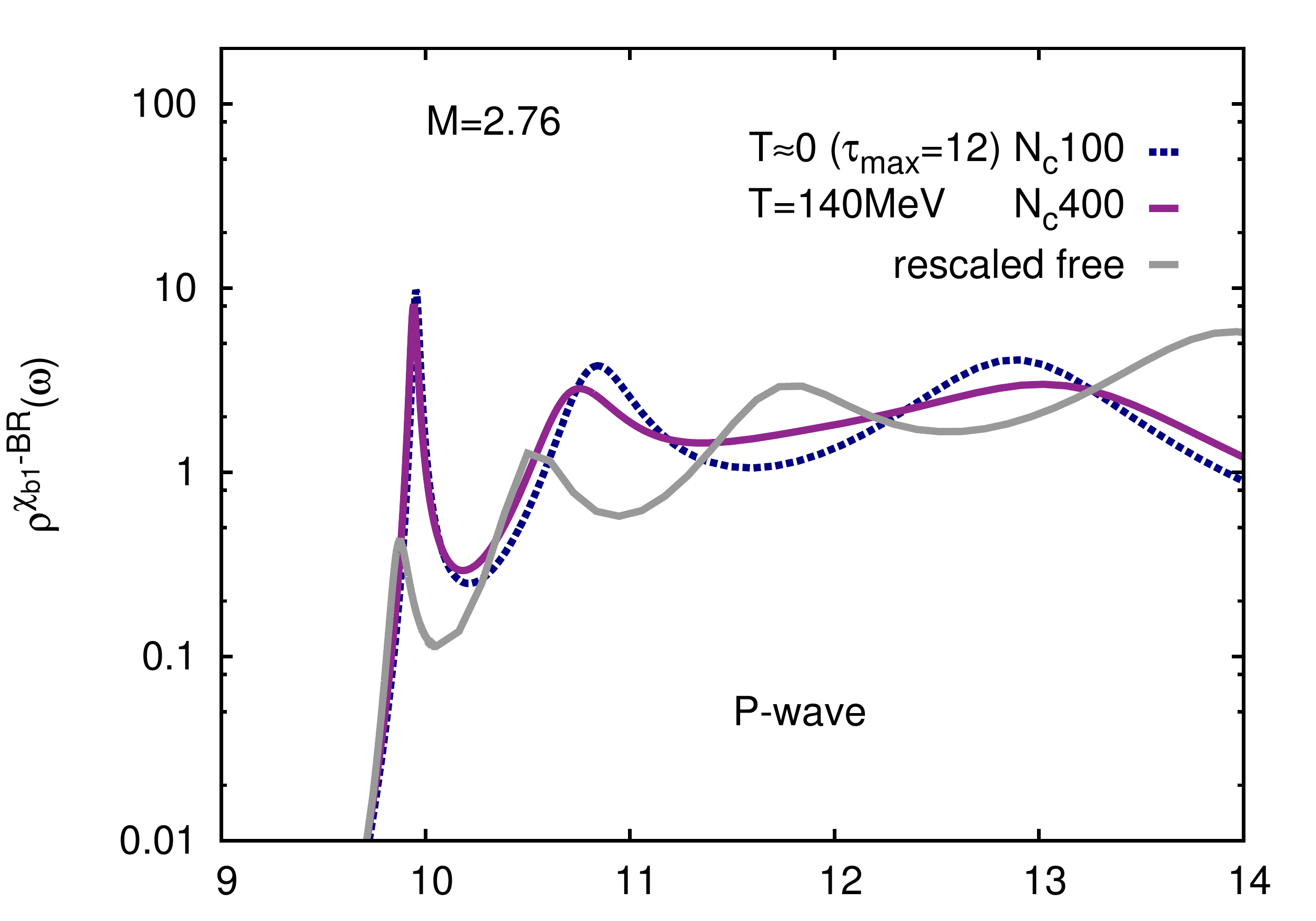}
  \includegraphics[scale=0.3,angle=-90]{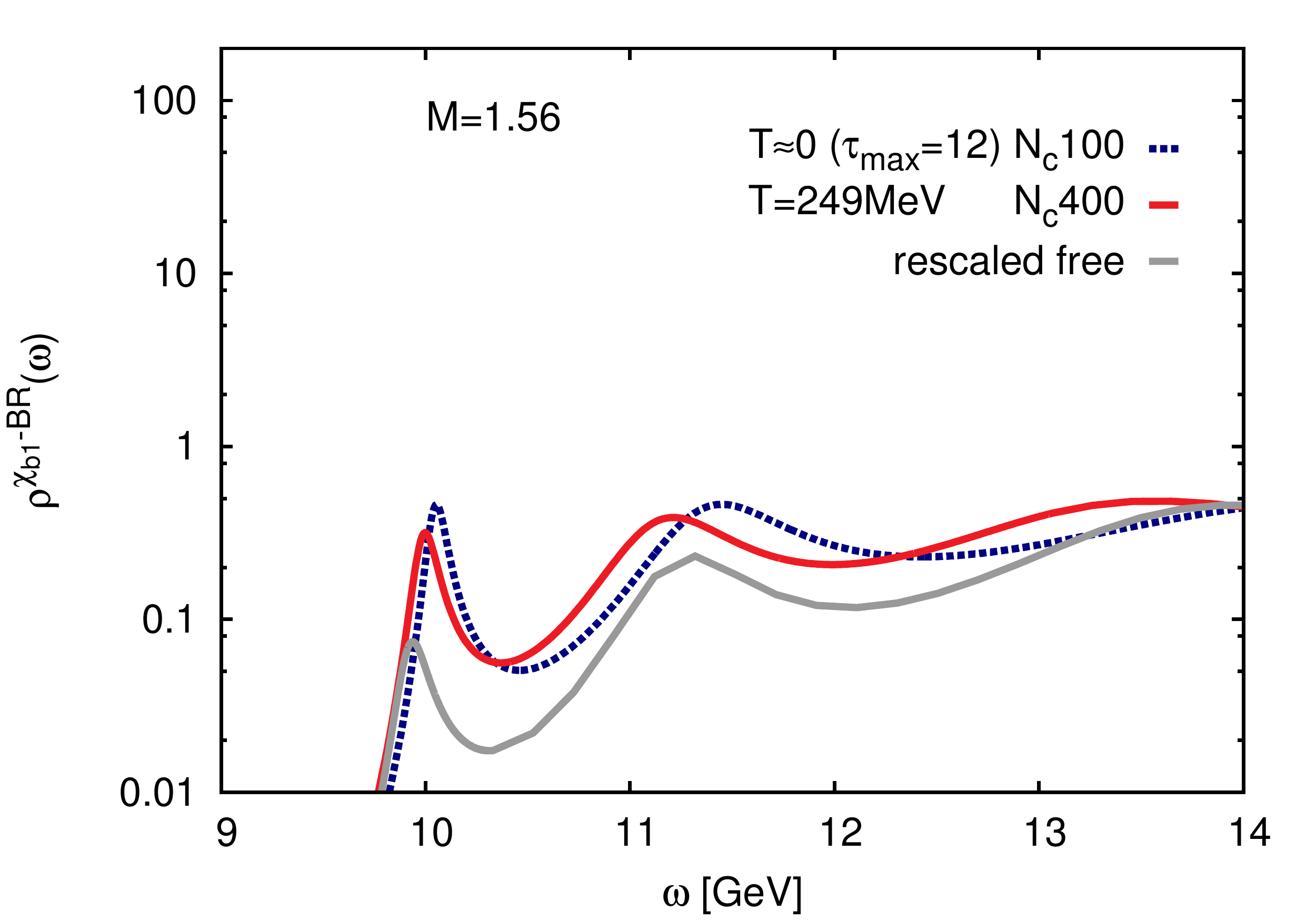}
  \caption{P-wave
    spectral reconstructions at $T = 140$MeV (left) and $T = 249$MeV
    (right). To disentangle systematic and medium effects we compare
    the Bayesian $T>0$ spectra from all $N_\tau=12$ data points
    (colored solid) to the spectra obtained from the first $\tau_{\rm
      max}=12$ points of the $T\simeq0$ correlator (blue dashed). To
    verify the presence of a bound state the spectra from
    noninteracting correlators are plotted (solid gray).
  } \label{Fig:CompFreeIntP}
\end{figure*}
In Fig. \ref{Fig:CompFreeIntP} we present such a comparison for the
P-wave channel at the lowest temperature (left) $T=140$MeV
($\beta=6.664$) and the highest (right) $T=249$MeV
($\beta=7.280$). While the colored solid curves represent the result
of the finite $T$ reconstruction from $N_\tau=12$ data points, the
dark blue, dashed curve represents the reconstruction from the first
$\tau_{\rm max}=12$ data points of the $T\simeq0$ correlator. Just as
in the S-wave case, the reconstructed interacting spectra at
$T\simeq0$ and $T>0$ are very similar. Nevertheless at the lowest
temperature $T=140$MeV distinguishing between the ground state peak
and the numerical ringing in the free spectra poses no difficulty, as
their amplitudes differs by almost two orders of magnitude. At the
highest temperature $T=249$MeV we find that the reconstructions take
on similar values at frequencies above $\omega=13.5$GeV but a
significant difference (at least a factor of three) remains for the
lowest lying peak. We have checked that another choice of default
model used in the reconstruction of the free spectra does not change the
lowest lying peak significantly, as shown in Appendix
\ref{sec:systematics}. Based on the outcome of this systematic
comparison we can establish the survival of the $\chi_{b1}$ state up
to at least a temperature of $T=249$MeV.
\begin{figure*}
  \centering
  \includegraphics[scale=0.3,angle=-90]{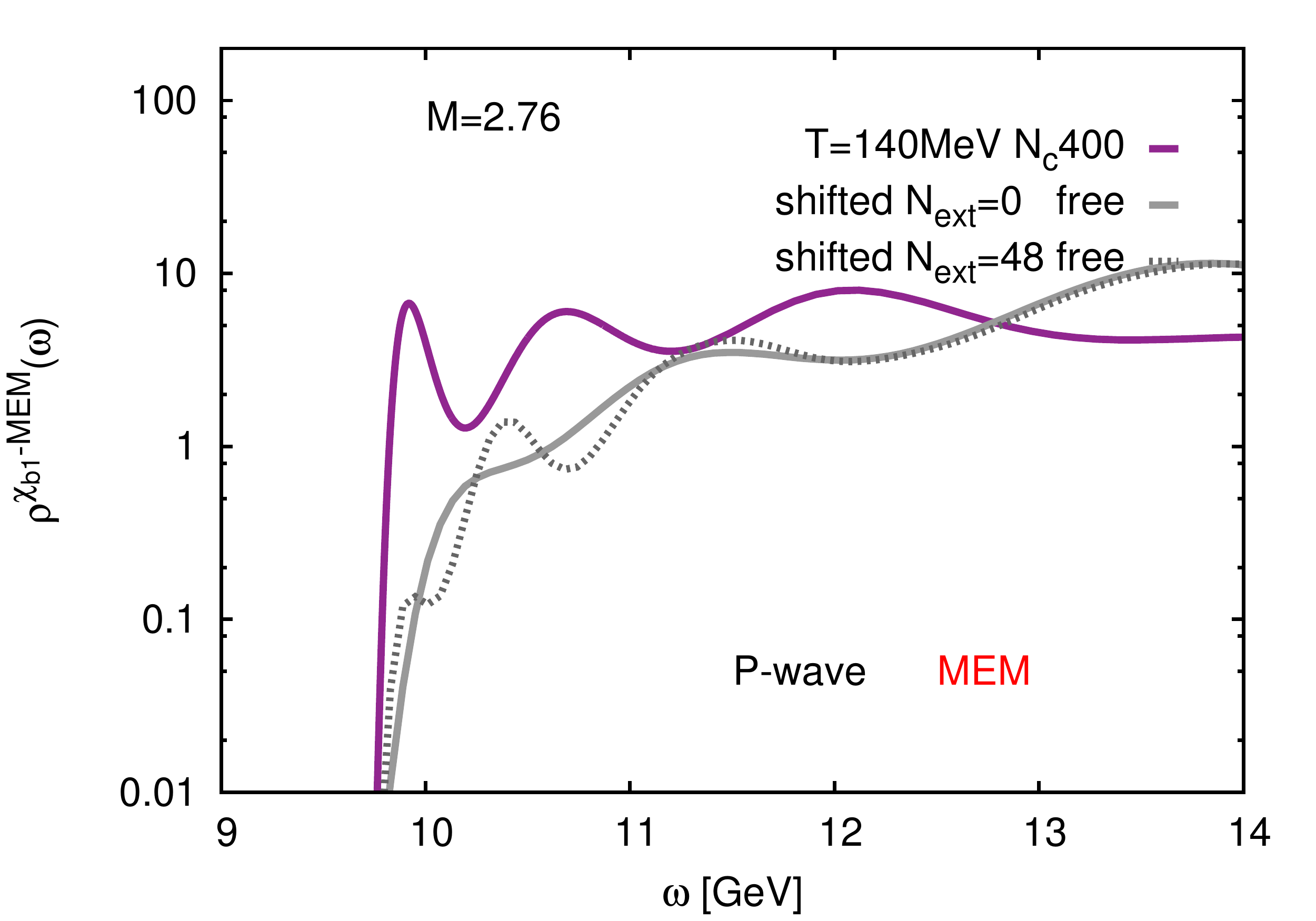}
  \includegraphics[scale=0.3,angle=-90]{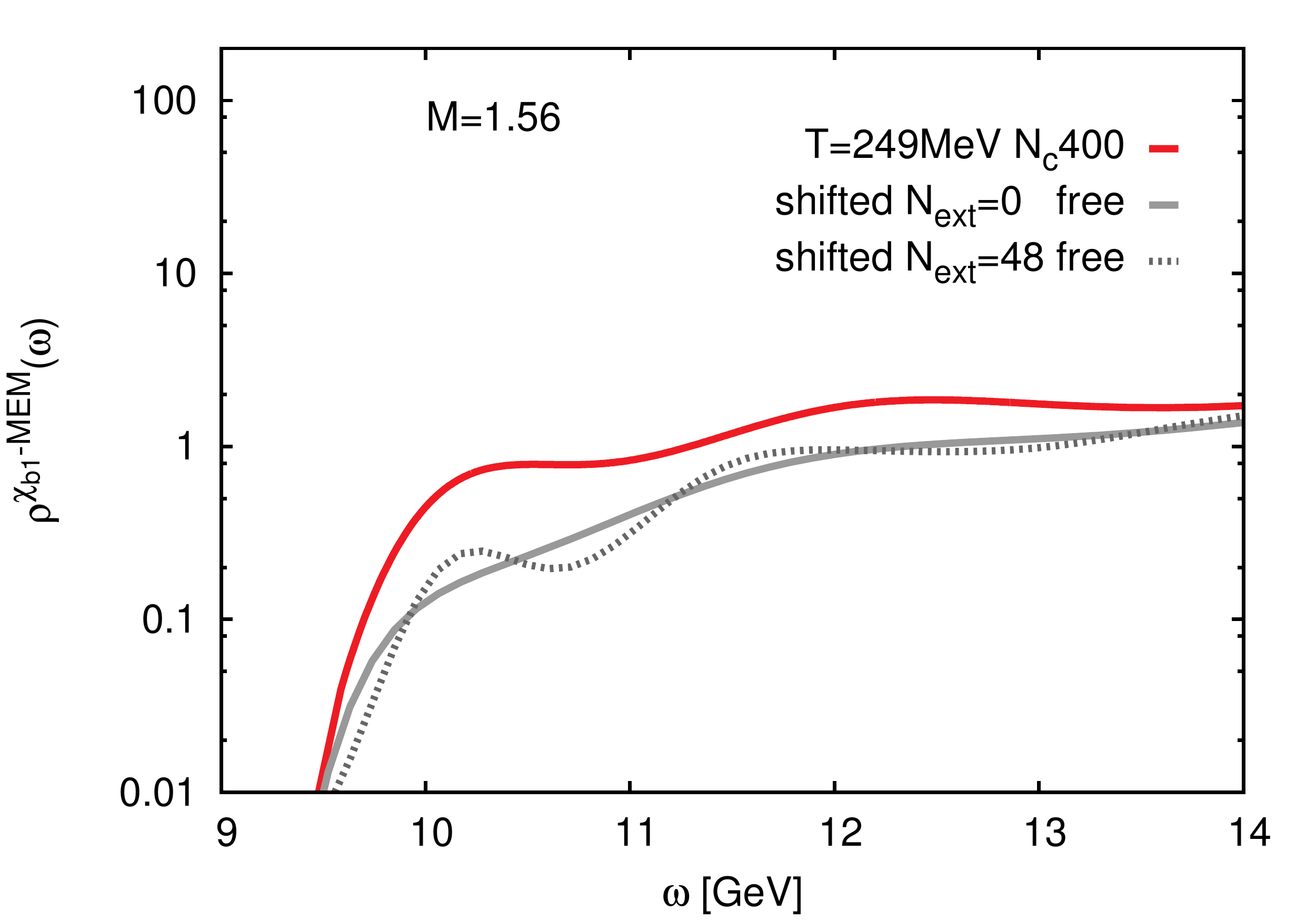}
  \caption{P-wave MEM spectral reconstructions at $T = 140$MeV (left)
    and $T = 249$MeV (right). Finite T spectrum (colored) as well as
    free spectrum with standard (solid gray) and extended $(N_{\rm
      ext}=48)$ search space (dashed gray).
  } \label{Fig:CompFreeIntPMEM}
\end{figure*}

We close this section with a comparison of the MEM reconstruction of
the interacting (colored gray) and free spectra (gray) for the P-wave
shown on the right-hand side of Fig. \ref{Fig:CompFreeIntPMEM}. As was
already found in Fig. \ref{Fig:ThermalSpectraOverviewPwave}, the lowest
lying peak is very shallow even at the lowest temperature $T=140$MeV
and completely absent at $T=249$MeV. The combination of a lower
signal-to-noise ratio together with the small number of available data
points does not allow the MEM to show oscillating behavior even if the
number of basis functions is increased to $N_{\rm ext}=48$. Hence,
although there seems to persist a difference at high $T$ between the
free and interacting spectra between $9.5$ and $11$GeV, the limited
resolution of the MEM for the interacting spectrum does not allow us
to relate it to a surviving $\chi_{b1}$ state.

\section{Conclusion} \label{sec:conclusion}
The in-medium modification of heavy quark bound states, elucidated by
lattice QCD methods, provides first principles insight into the
physics of the strong interactions under extreme conditions. Due to
the role of bottomonium as probe for the QGP in relativistic heavy-ion
collisions, understanding bottomonium behavior in the presence of a
thermal medium is of direct phenomenological relevance. In this study
we investigated the spectral properties of the $\Upsilon$ ($^3S_1$) as
well as the $\chi_{b1}$ ($^3P_1$) channel at fourteen different
temperatures around the deconfinement transition $140{\rm MeV} (=0.911
T_c) \le T \le 249{\rm MeV} (=1.61 T_c)$ in a medium with $N_f=2+1$
light quark flavors.

The underlying $48^3 \times 12$ isotropic lattices were provided by
the HotQCD collaboration and are based on the HISQ action. Since
temperature is changed by varying the lattice spacing, a narrowly
spaced temperature scan was achieved. However, a zero temperature
calibration of the absolute energy scale at each lattice spacing was
required and has been performed. A lattice regularization of the
effective field theory (NRQCD) was deployed to calculate the
bottomonium correlators. This allows us to utilize the full number of
data points in temporal direction, since the periodic boundary
conditions of relativistic field theory are absent. The relation
between correlator and spectrum in this case reduces to a convolution
over a simple exponential kernel.

We extracted spectral information both with the standard MEM and a
recently developed Bayesian approach. This novel method differs
significantly from the MEM, and features a prior functional which
enforces the positive definiteness of the spectrum, independence of
the end result from the choice of units for $\rho$ and favors smooth
spectra for the energy region where data do not imprint peaked
structures. As a general observation we find that the spectra
reconstructed using the new Bayesian approach are far superior to
those produced by the MEM as seen in
Figs. \ref{Fig:ThermalSpectraOverviewSwave} and
\ref{Fig:ThermalSpectraOverviewPwave}. Not only are we able to 
resolve much narrower peak widths but the functional form of the bound
state peaks is reproduced in Lorentzian form, as expected for a
particle of finite lifetime. We see by systematically investigating
the effect of removing data points in
Figs. \ref{Fig:T0DatPRemovalEffectS}-\ref{Fig:6664DatapointDependeceS} and
Figs. \ref{Fig:T0DatPRemovalEffectP}-\ref{Fig:6664DatapointDependeceP}
that our current data set does 
not allow us to quantitatively disentangle the effects of the medium
from the degradation of the reconstruction quality due to a smaller
extent in $\tau$. Nevertheless, due to the high resolution of the new
Bayesian approach we are able to give stringent upper bounds on the
in-medium modification of both $\Upsilon$ and $\chi_{b1}$ states.

From a systematic comparison of the reconstructed spectra from finite
temperature correlators to those reconstructed on noninteracting
lattices we furthermore conclude that the ground state $\Upsilon$
survives up to at least $T=249 {\rm MeV} (=1.61 T_c)$, the highest
temperature investigated. A similar comparison carried out for the
P-wave channel shows that even though the deviation between
interacting ground state peak and numerical ringing in the free
spectra is smaller, even at $249 {\rm MeV}$ at least a factor of three
difference between the interacting spectra and the free spectra
remains. This suggests to us the survival of the ($^3P_1$) bottomonium
ground state up to this temperature (well into the QGP phase).

The conclusion on the S-wave ground state survival agrees
qualitatively with those obtained by the MEM. However the restricted
search space of the MEM does not allow us to find a well-defined peak
for the ground state in the $\chi_{b1}$ channel at the highest three
temperatures. Since the new Bayesian approach is not affected by this
well-known deficiency of the standard MEM (i.e. changing $\omega_{\rm
  min}$ to larger negative values leads to a systematic broadening and
eventual disappearance of any reconstructed peaked structures), we are
confident that the observed presence of the P-wave peak up to $249 {\rm
  MeV}$ is not an artifact due to the spectral reconstruction method.

We have checked that our findings do not suffer from a possibly
inadequate choice of lattice NRQCD discretization by repeating the
analysis for the values of the parameter $n=2,3$ and $4$ described in
Appendix \ref{sec:systematics2}. While the high frequency behavior of
both the free and the fully interacting spectra do change as expected,
we confirm that the ground state peak and its features remain
virtually unchanged.

We also investigated additional systematic uncertainties of the
spectral reconstruction itself in Appendix \ref{sec:systematics} and
find that they are in general smaller than the errors introduced due
to the finite number of data points. These dependencies appear
stronger than what has been found in previous MEM studies. This
however is related to the fact that the new Bayesian approach is able
to resolve much narrower structures and thus does not hide the default
model dependency in an artificially broad reconstructed width.

The results obtained here from the combination of NRQCD correlators
and the new Bayesian method are promising. It appears that
deficiencies of the MEM can be overcome at least in principle, while
the small number of data points still precludes us from a quantitative
determination of possible in-medium mass shifts and a width
broadening. Carrying out dynamical lattice QCD simulations with a
larger number of temporal lattice sites, be it in an isotropic or
anisotropic setting is thus called for.

Incremental progress on the reconstruction of spectra can be expected within the
ongoing programs for gauge configurations generation, i.e. $N_\tau=16$ or $N_\tau=24$.
For a quantitative determination of the in-medium modification, especially the width broadening,
it will be necessary to start a dedicated generation program using anisotropic lattices with $N_\tau>64$.
Once the temporal extend becomes as large as $N_\tau>64$ the default model dependence will
also reduce significantly, as the high frequency regime, encoded in the correlator at small $\tau$
is more highly resolved. Due to the fact that the reconstruction success depends on the
physical temporal extend, what exact number of lattice points are needed will ultimately be
connected to the melting temperature of the individual state.

We hope that the availability of the new Bayesian approach for the
determination of heavy quarkonium in-medium spectral features will
benefit the understanding of bottomonium suppression in heavy-ion
collision and look forward to future studies with lattices of larger
temporal extent and higher statistics.

\section*{ACKNOWLEDGEMENTS}
A.R. thanks Y. Burnier for many fruitful discussions. S.K. is supported
by the National Research Foundation of Korea funded by the
Korean government (MEST) Grant No.\ 2010-002219 and in part by Grant No.
NRF-2008-000458. P.P. is supported by the U.S.Department of Energy under
Award No.DE-AC02-98CH10886. A.R. was partly supported by the Swiss
National Science Foundation (SNSF) under Grant No. 200021-140234.

\appendix

\section{Tests of the Bayesian Spectral Reconstruction}
\label{sec:systematics}

Any reconstruction of $N_\omega\gg N_\tau$ parameters from a noisy set
of $N_\tau$ data points through an inversion of
Eq. \eqref{Eq:SpecConv} remains an ill-defined problem. Besides the
obvious fact that the results depend on the properties of the measured
data themselves (i.e. the number of available data points and their
signal-to-noise ratio), we have to control how the inclusion of prior 
information affects the final outcome. Prior information enters
implicitly e.g. through the choice of the underlying frequency
interval, as it spans only the region we deem relevant for the
spectral function we wish to reconstruct. Explicit prior information
on the other hand enters through the choice of prior functional $S$
and the default model $m(\omega)$ it contains. In the following, all
of these factors are considered and are independently varied to
estimate the systematic uncertainties of the reconstructed spectral
function. The main outcomes of these tests are the follwing:
\begin{itemize}
\item The reconstruction of the ground state peak suffers most from
  discarding points close to $\tau=\beta$. We however do not observe
  any abrupt changes when including or discarding the last data point
  at $\tau=\beta-a_\tau$ in contrast to other studies.
 \item The ground state peak of the $\chi_{b1}$ spectra in general
   suffers more strongly from the systematic uncertainties, as the
   signal-to-noise ratio of the corresponding correlator is
   consistently smaller than in the $\Upsilon$
   case. ($M_{\chi_{b1}}>M_{\Upsilon}$)
 \item The size of the statistical error on the correlator data limits
   how well we can reconstruct the width of the ground state
   peak. However, with the inclusion of more than $N_{\rm conf}=360$
   configurations for the averaging, the dependence on $N_{\rm conf}$
   appears to be small.
 \item The dependence on the choice of $\omega^{\rm num}_{\rm
   max}=\omega_{\rm max} a$ ($a$ is the lattice spacing) is small as
   long as the interval is chosen large enough $\omega^{\rm num}_{\rm
     max} >20$ to accommodate all relevant peak structures. The
   dependence on $\omega_{\rm min} a$ is a bit more pronounced,
   especially if not enough frequencies in the negative range are
   included (this dependence is significantly smaller than that in the
   case of the MEM, where no stable plateau is found).
 \item Using different default models gives roughly twice of the
   jackknife error bar to the first peak position in $\Upsilon$
   spectral function and ten times uncertainty to the first peak
   width. In the $\chi_{b1}$ case, due to a worse signal-to-noise
   ratio, already the default model effect on the peak position may be a
   factor of eight larger than the statistical errors.
\end{itemize}

Finally we also take a look at the systematics of the MEM,
specifically the fact that reconstruction success depends
strongly on the choice of frequency interval. Indeed we find that
simply moving $\omega_{\rm min}$ to larger negative values melts any
otherwise visible peaked structures.

\subsection*{Dependence on the $\tau$ range}

\begin{figure}
 \includegraphics[scale=0.3,
   angle=-90]{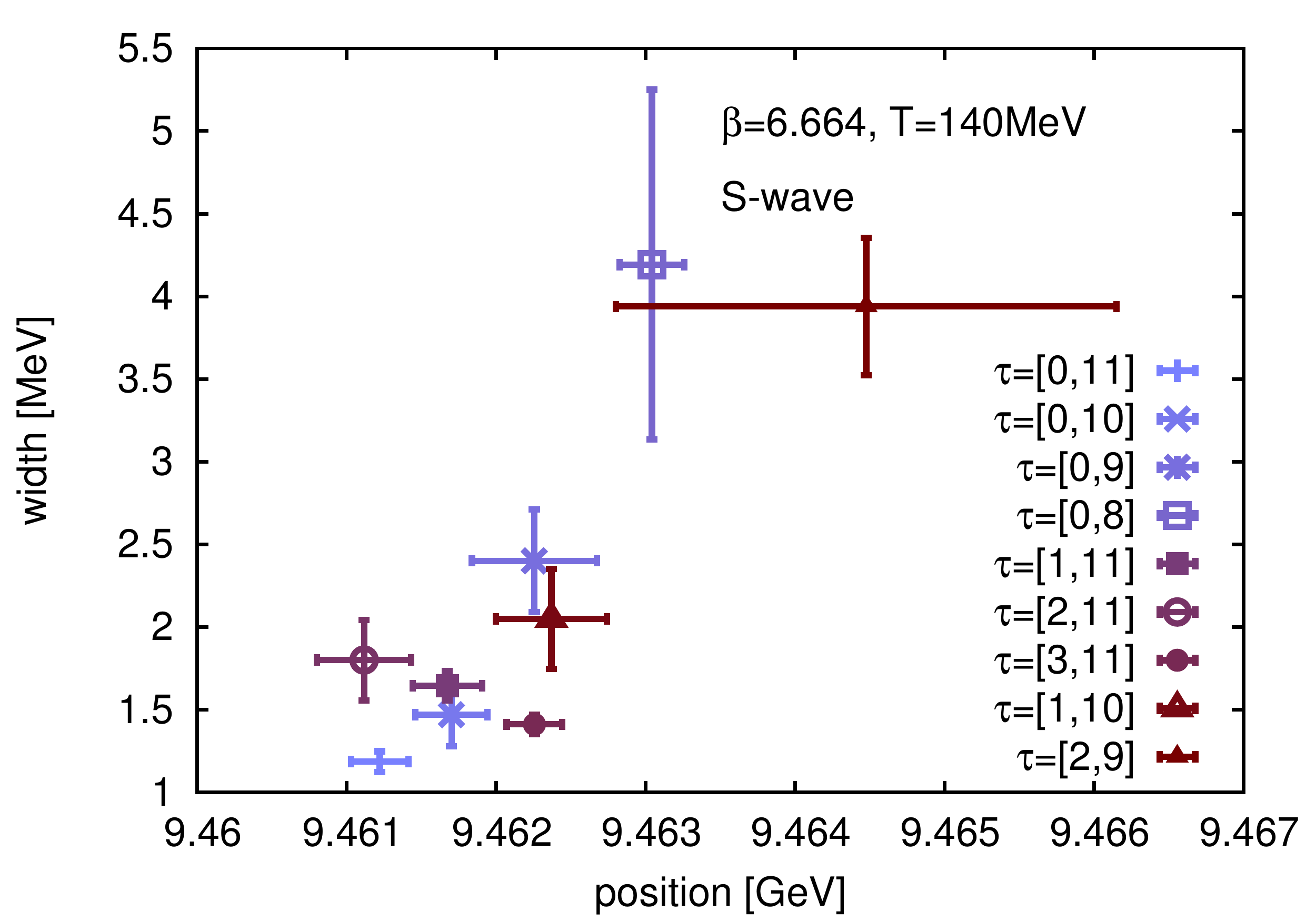}
 \includegraphics[scale=0.3,
   angle=-90]{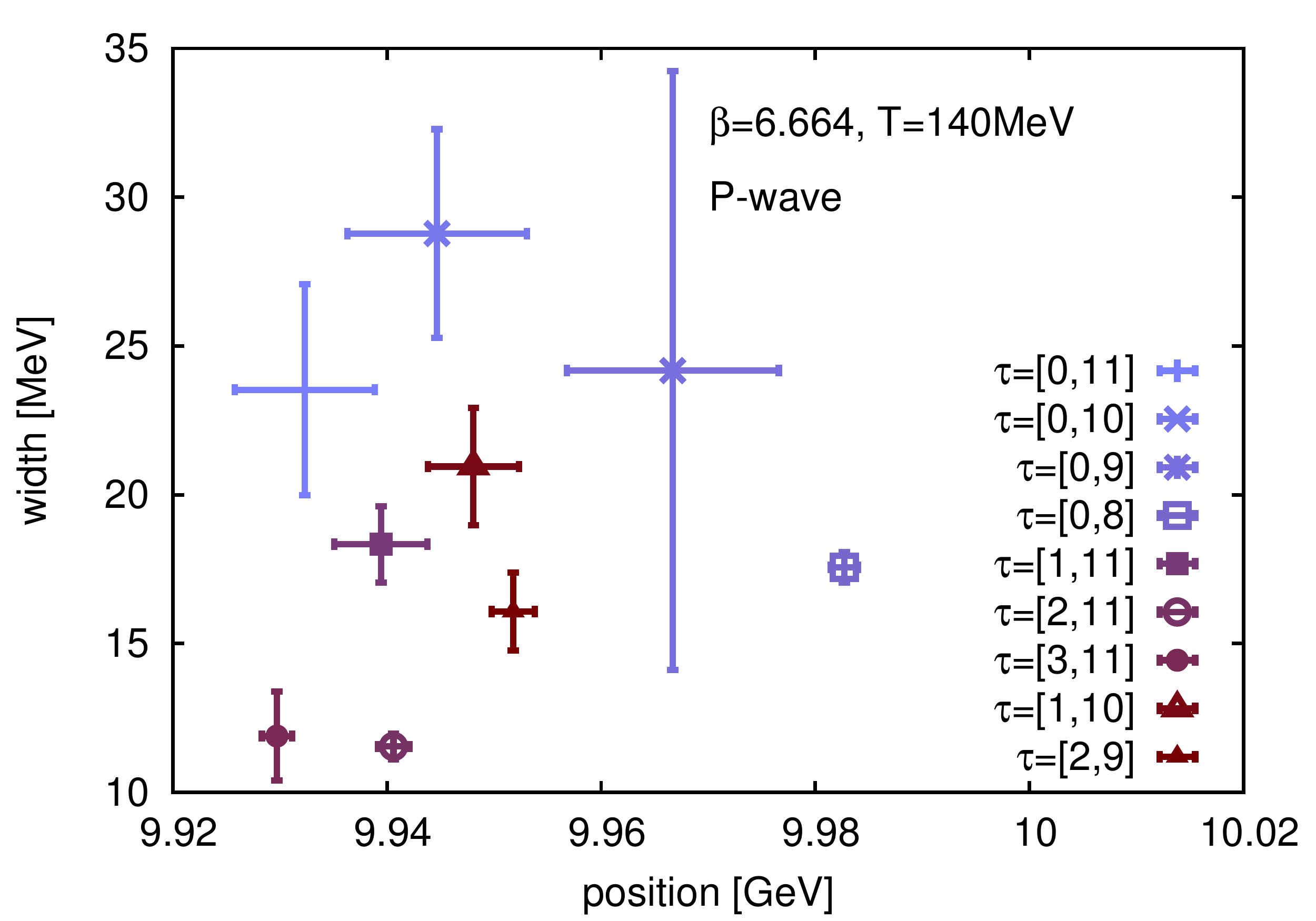}
 \caption{At $140$MeV: Dependence of the reconstructed lowest peak
   position and peak width on choosing different subsets of
   data points along the $\tau$ axis. In general the peak position is
   less susceptible than the peak width with the $\Upsilon$ (top) showing
   consistently less dependence than the $\chi_{b1}$
   (bottom).}\label{Fig:6664DatapointDependece}
\end{figure}
\begin{figure}
 \includegraphics[scale=0.3,
   angle=-90]{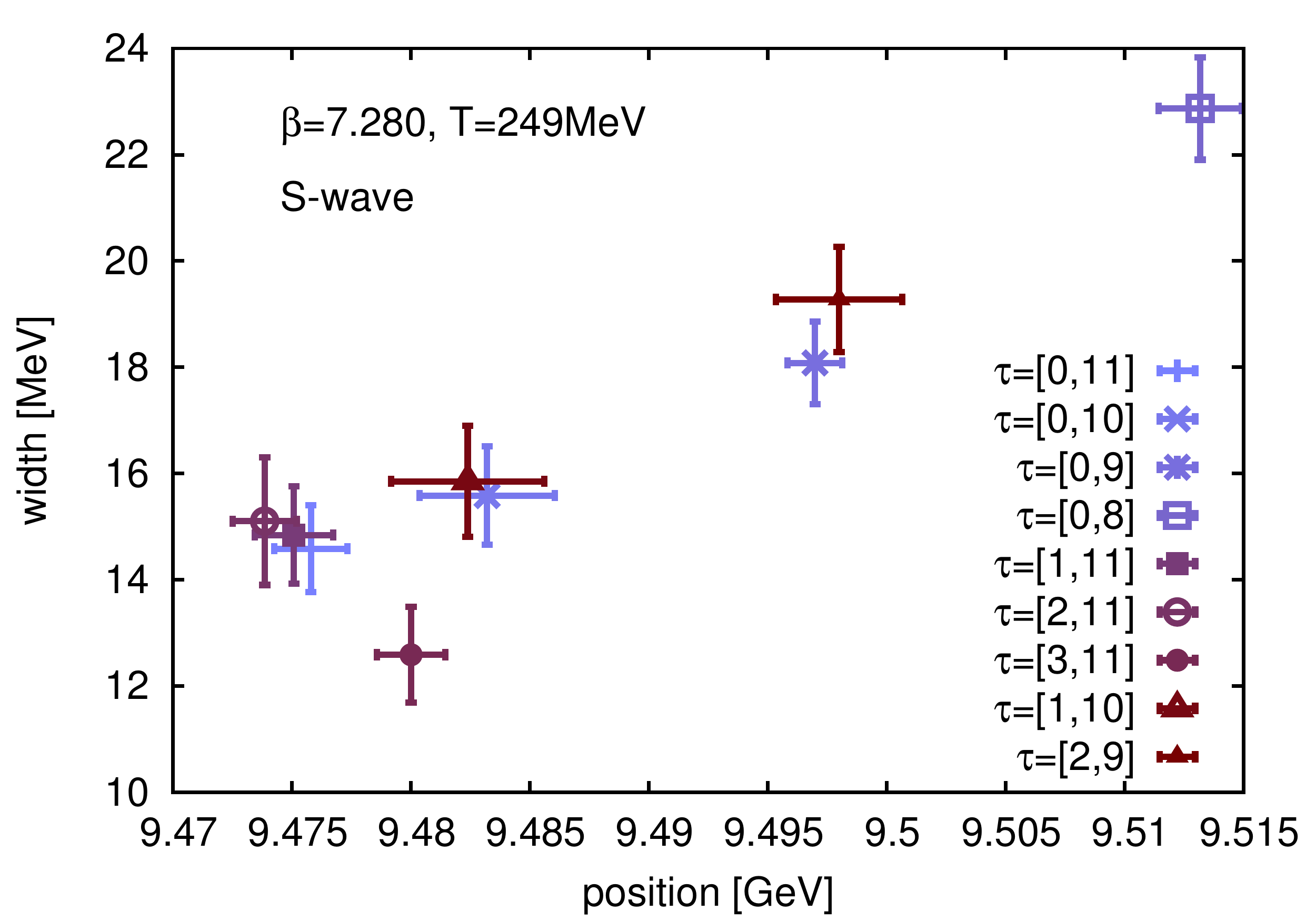}
 \includegraphics[scale=0.3,
   angle=-90]{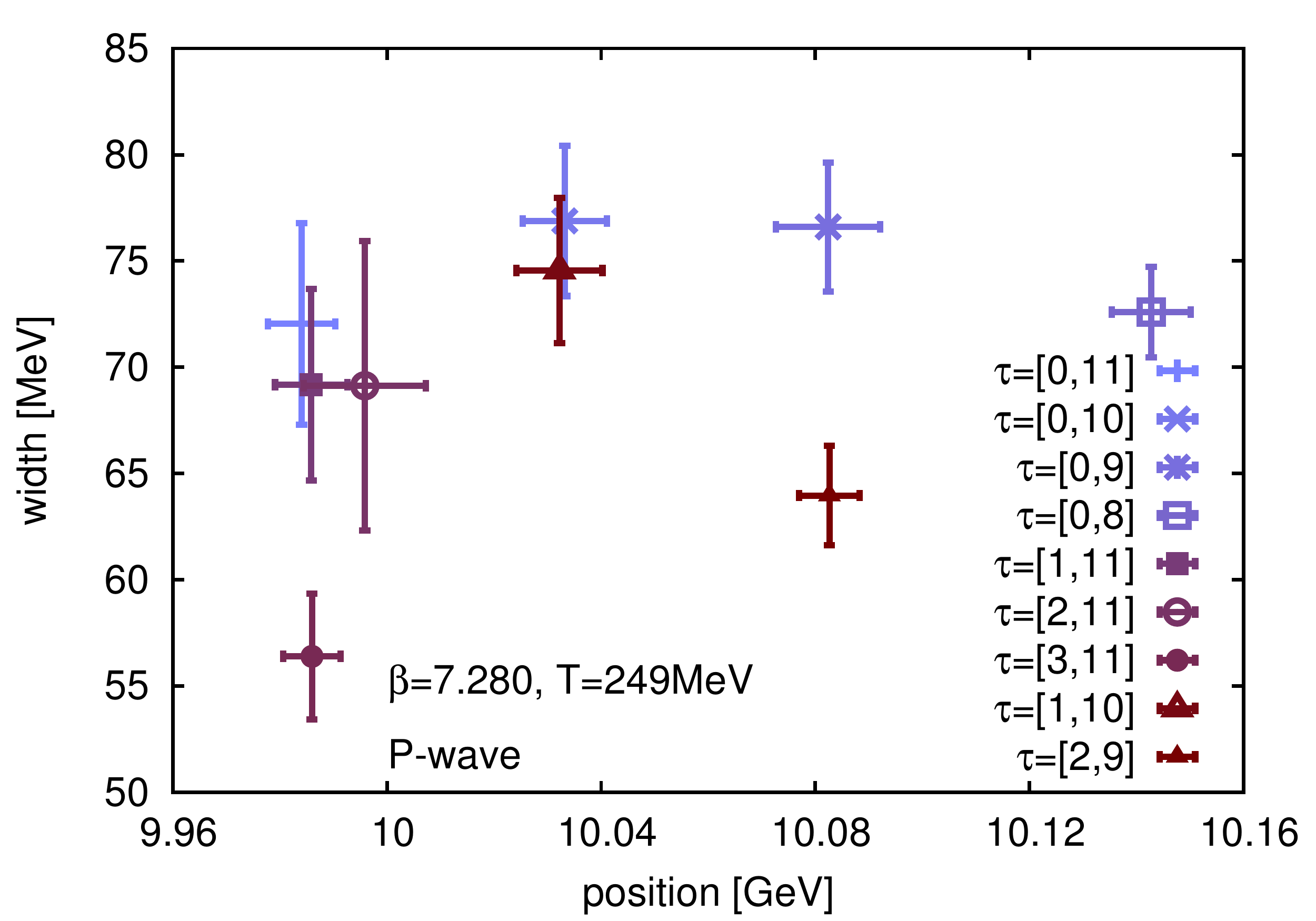}
 \caption{At $249$MeV: Dependence of the reconstructed lowest peak
   position and peak width on choosing different subsets of
   data points along the $\tau$ axis. In general the peak position is
   less susceptible than the peak width with the $\Upsilon$ (top) showing
   consistently less dependence than the $\chi_{b1}$
   (bottom).}\label{Fig:7280DatapointDependece}
\end{figure}

Using a relatively small number of data points, $N_\tau=12$, we expect
the reconstructed spectral function to suffer significantly from
removing even more of them. We find as elaborated on in the main text
that since low frequency structures dominate late $\tau$ times,
removing data points close to $\tau = N_\tau$ indeed leads to a
significant increase in reconstructed peak width beyond the
statistical error bars as well as a systematic shift of the extracted
peak position to larger values as can be seen in
Figs. \ref{Fig:6664DatapointDependece}-\ref{Fig:7280DatapointDependece}. On
the other hand it is also 
clear that we do not observe any abrupt changes in the reconstruction
if the last data point is included, as has been reported by previous
studies.

\subsection*{Dependence on the signal-to-noise ratio}

\begin{figure}[t]
 \includegraphics[scale=0.3,
   angle=-90]{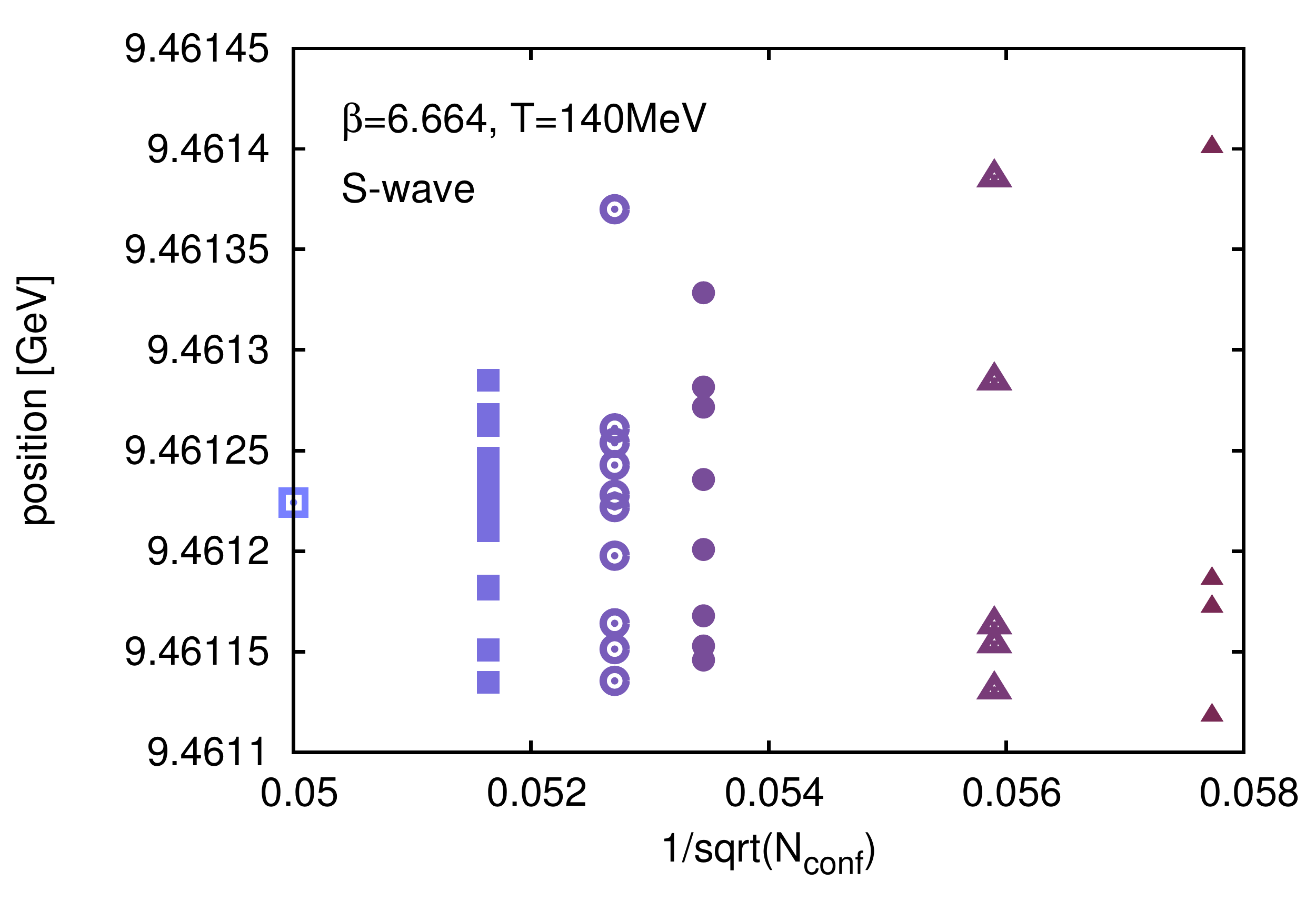}
 \includegraphics[scale=0.3,
   angle=-90]{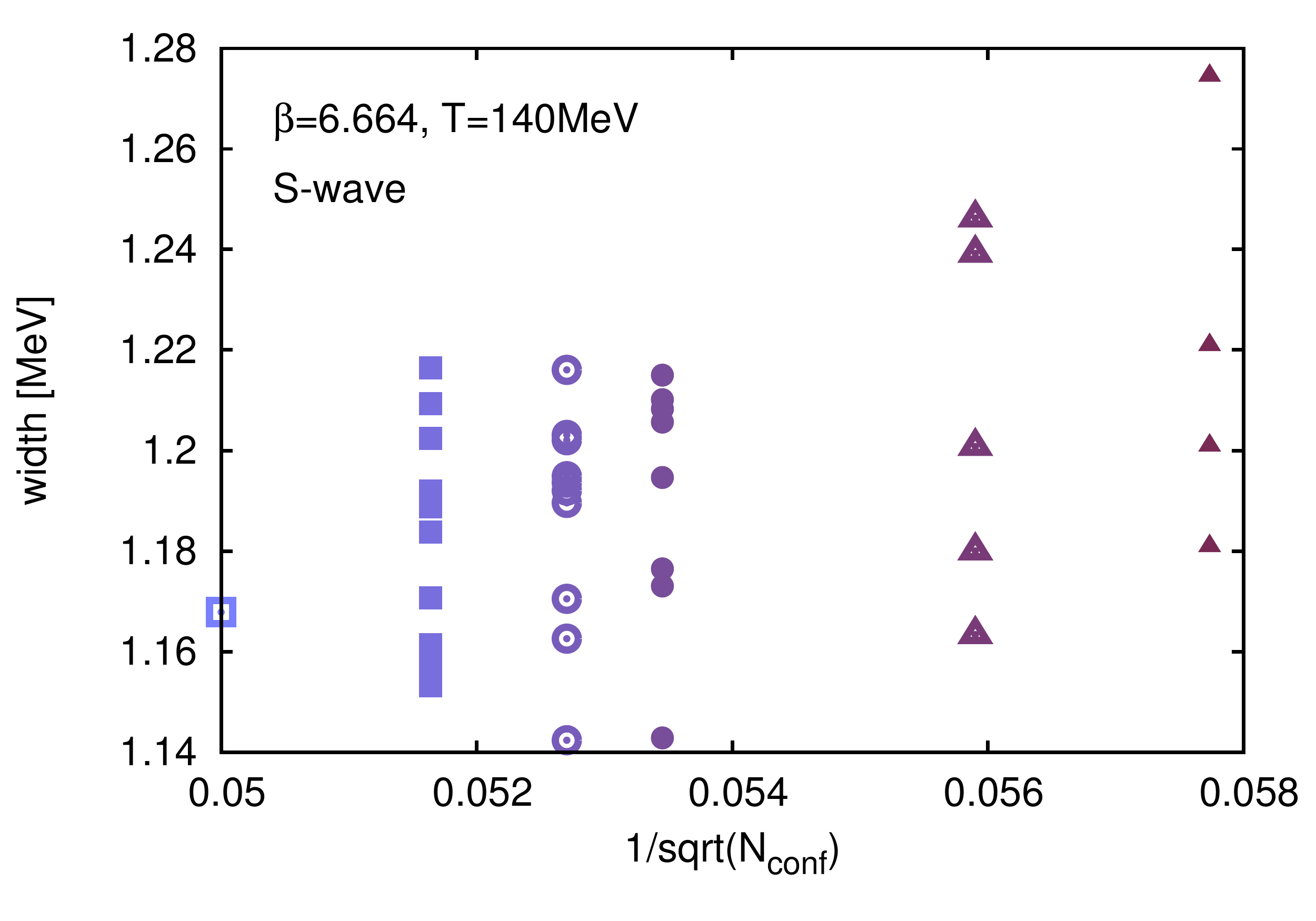}
 \caption{At $140$MeV: The raw $\Upsilon$ reconstructed peak positions
   (top) and peak widths (bottom) for different choices of the number
   of jackknife bins and thus of the number of measurements
   contributing to the correlator average. Each plotted point
   corresponds to one spectral reconstruction for an individual
   jackknife bin. Their spread is a direct measure of the statistical
   uncertainty and reduces slightly with increasing number of jackknife
   bins. No systematic trend appears in reconstructed peak position,
   while the peak width seems to move to a smaller value until a
   plateau is reached at $N_{\rm
     conf}=350$.}\label{Fig:6664BinDepSWave}
\end{figure}

\begin{figure*}[t]
 \includegraphics[scale=0.3,
   angle=-90]{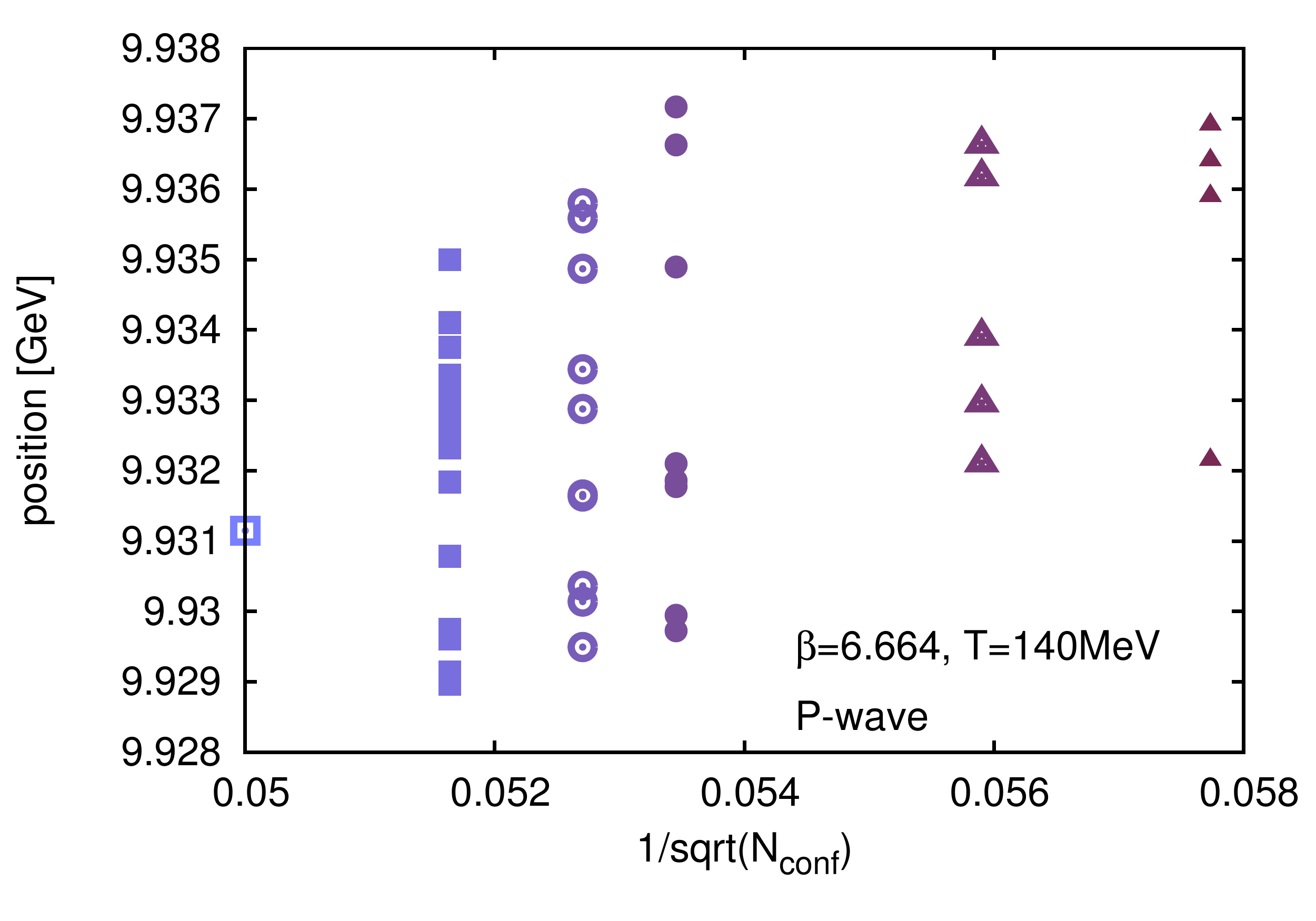}
 \includegraphics[scale=0.3,
   angle=-90]{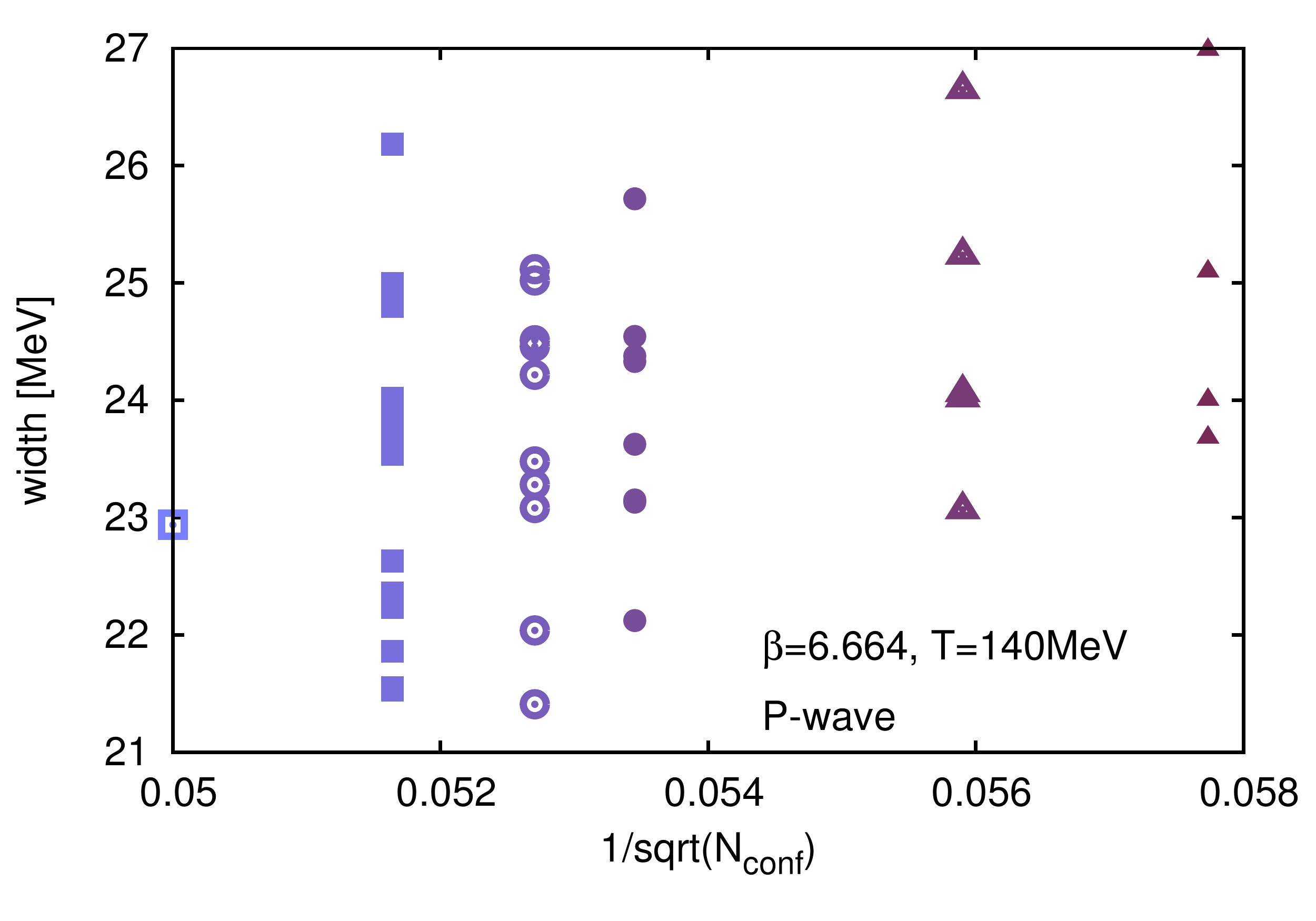}
 \caption{At $140$MeV: The raw $\chi_{b1}$ reconstructed peak positions
   (left) and peak widths (right) for different choices of the number
   of jackknife bins and thus of the number of measurements
   contributing to the correlator average. A slight trend to smaller
   values of the reconstructed peak position is visible, which however
   lies within the statistical uncertainty. No systematic trend
   appears for the peak width.}\label{Fig:6664BinDepPWave}
\end{figure*}

\begin{figure*}[t]
 \includegraphics[scale=0.3,
   angle=-90]{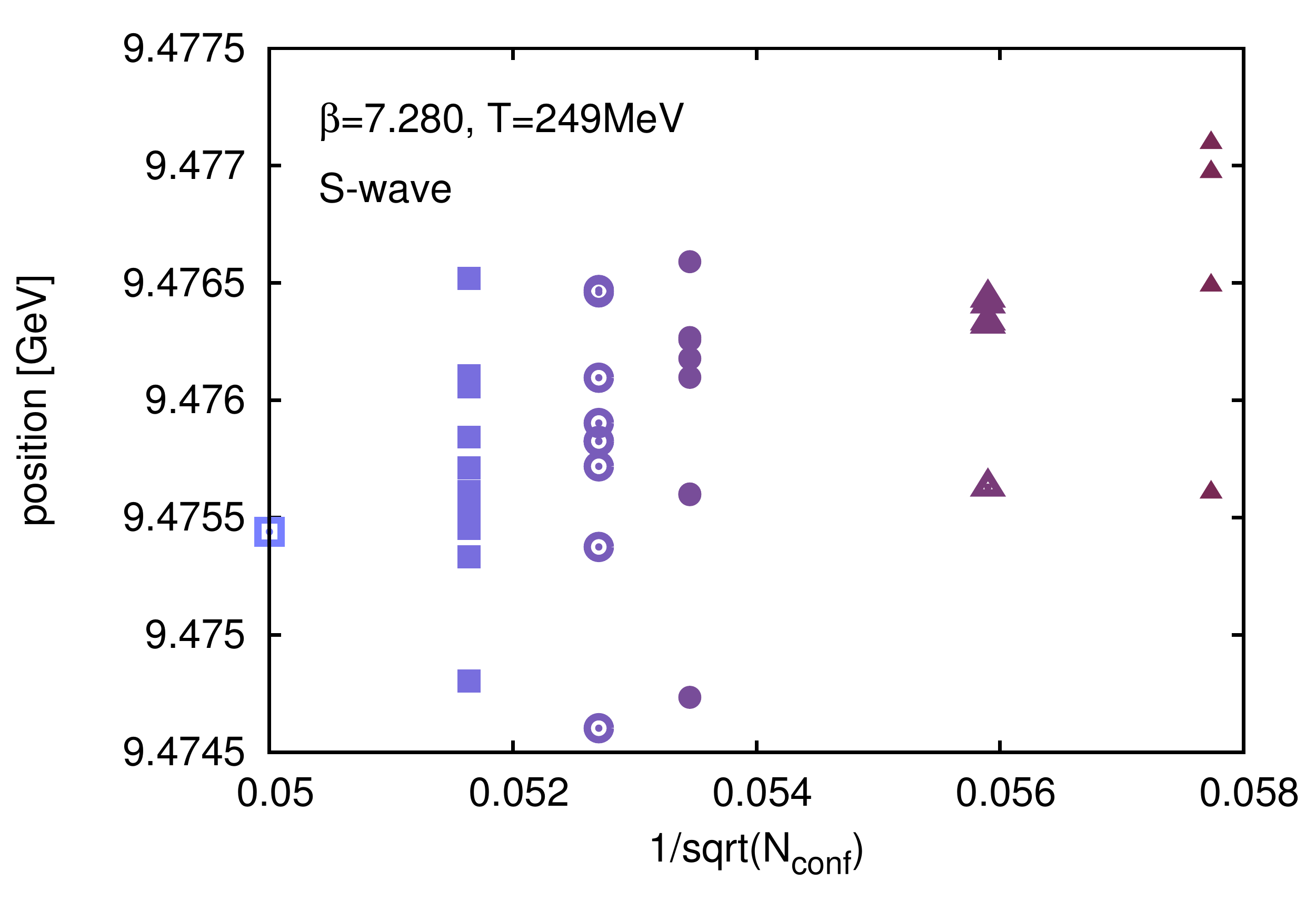}
 \includegraphics[scale=0.3,
   angle=-90]{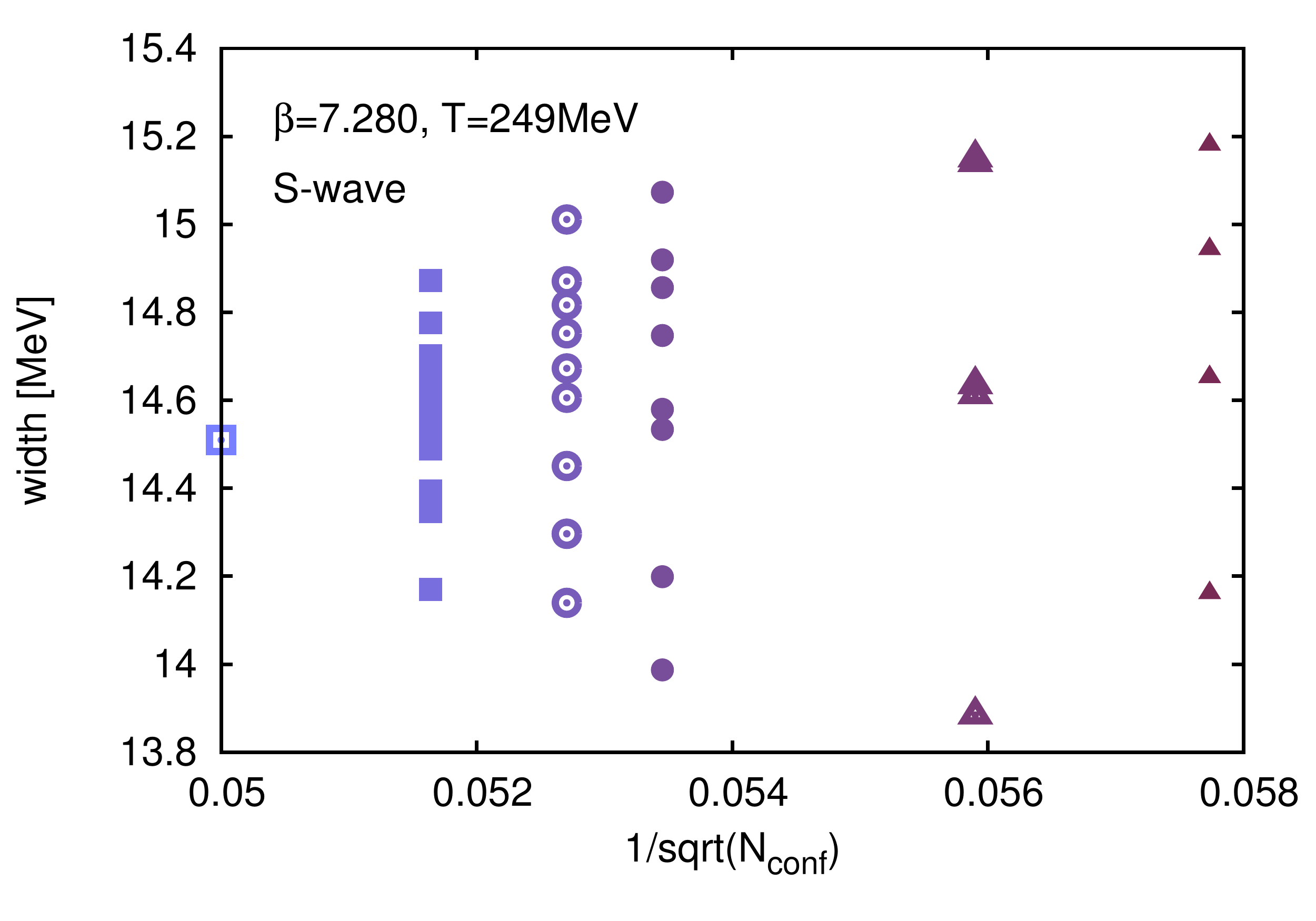}
 \caption{At $249$MeV: The raw $\Upsilon$ reconstructed peak positions
   (left) and peak widths (right) for different choices of the number
   of jackknife bins and thus of the number of measurements
   contributing to the correlator average. No systematic trend appears
   in reconstructed peak position and width. In the reconstructed
   width the reduction of statistical uncertainty with increasing
   number of used measurements is clearly
   visible.}\label{Fig:7280BinDepSWave}
\end{figure*}

\begin{figure*}[h]
 \includegraphics[scale=0.3,
   angle=-90]{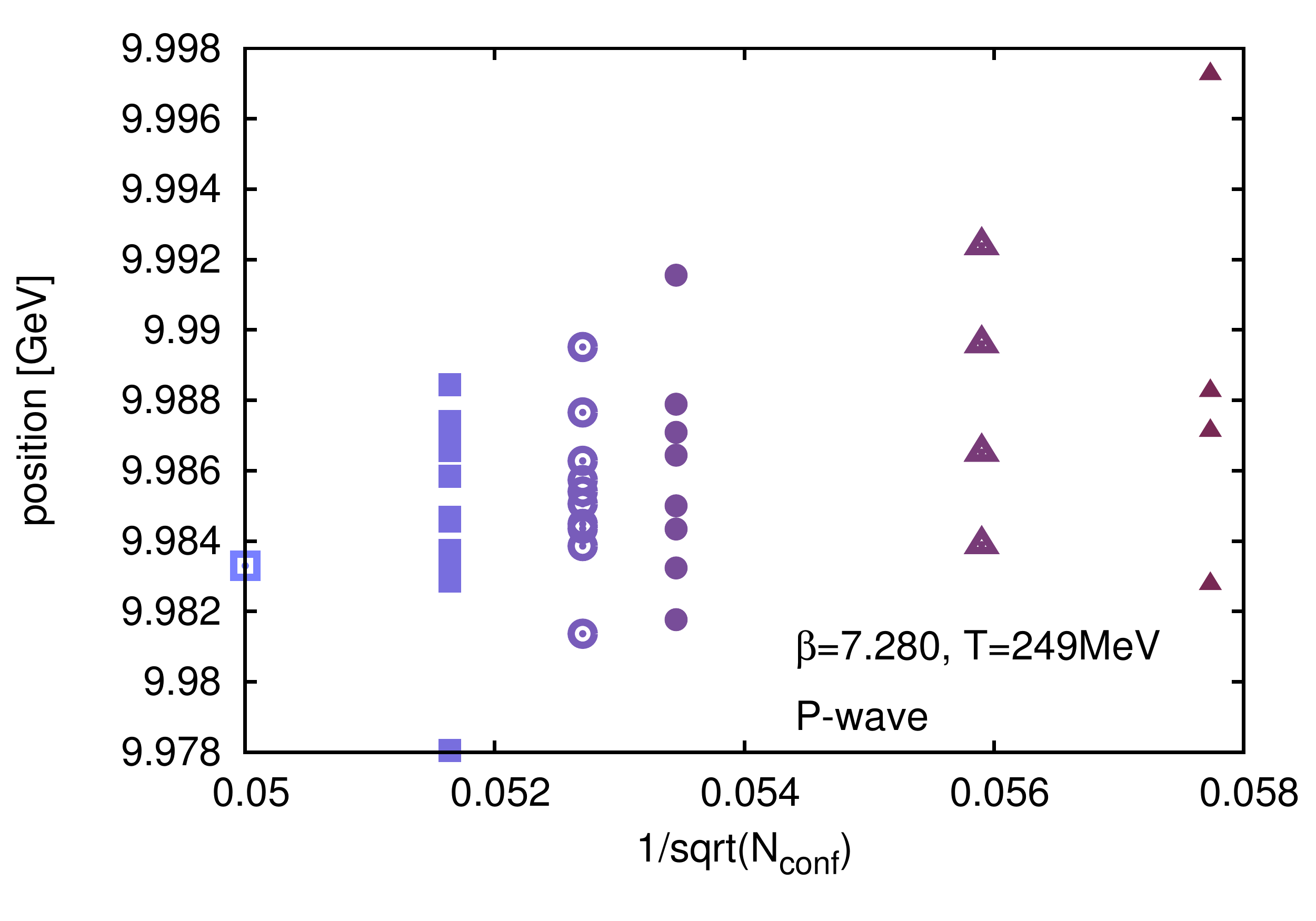}
 \includegraphics[scale=0.3,
   angle=-90]{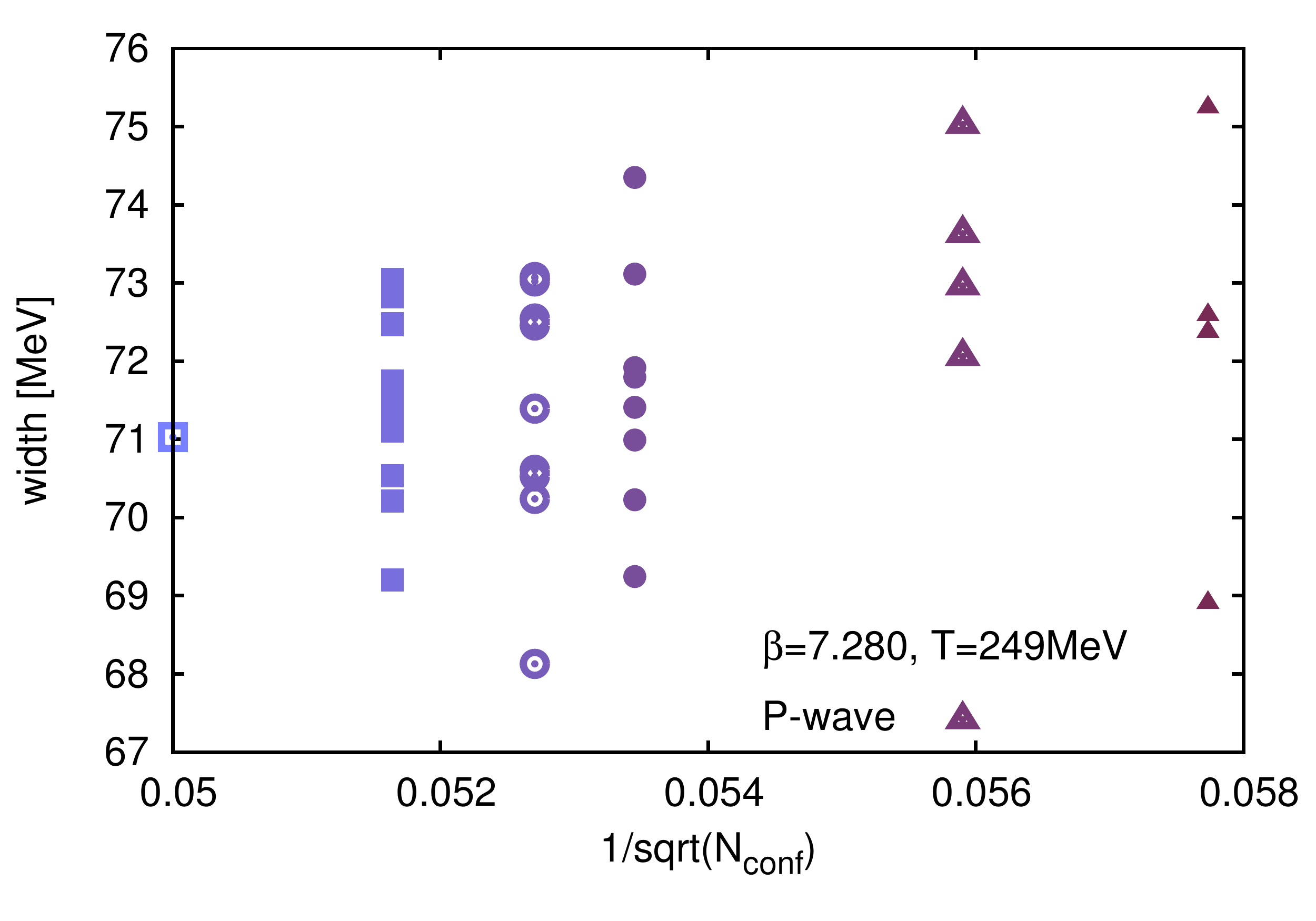}
 \caption{At $140$MeV: The raw $\chi_{b1}$ reconstructed peak positions
   (left) and peak widths (right) for different choices of the number
   of jackknife bins and thus of the number of measurements
   contributing to the correlator average. A slight trend to smaller
   values of the reconstructed peak position is visible, which however
   lies within the statistical uncertainty. No systematic trend
   appears for the peak width.}\label{Fig:7280BinDepPWave}
\end{figure*}

In contrast to the Fourier transform, where the signal-to-noise ratio
strongly affects the highest possible frequency one can resolve, the
inverse Laplace transform reacts to degrading signal with increasing
peak widths. Thus a systematic trend towards smaller reconstructed
widths is expected if more and more configurations contribute to the
averaged underlying correlator. Such kind of behavior can be found for
the smallest numbers of used configurations but eventually it seems
that a stable plateau is reached at least for the $\Upsilon$ channel
(we choose a number of jackknife bins $N_J$ and reconstruct the
spectra for each of these with a corresponding number of $N_{\rm
  conf}-N_{\rm conf}/N_J$ configurations. The peak positions and widths
for each bin are plotted against the number of configurations in
Figs. \ref{Fig:6664BinDepSWave}-\ref{Fig:7280BinDepPWave}). The
reconstruction of the peak position in the $\Upsilon$ channel is
robust, while in the $\chi_{b1}$ channel one can see a slight
dependence on the number of used configurations with a trend to lower
values.

\FloatBarrier
\subsection*{Dependence on the default model}

\begin{figure}
 \includegraphics[scale=0.3,
   angle=-90]{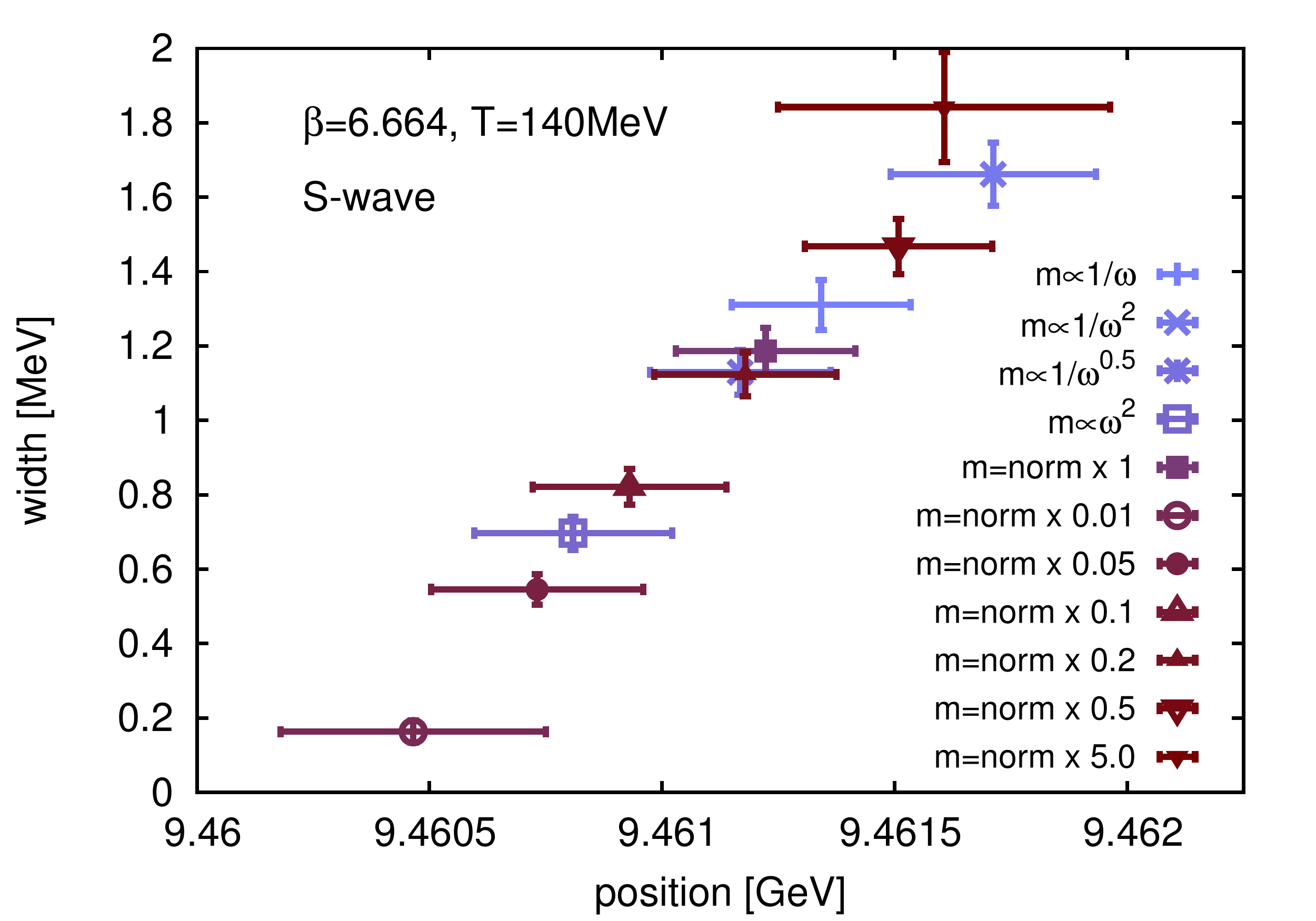}
 \includegraphics[scale=0.3,
   angle=-90]{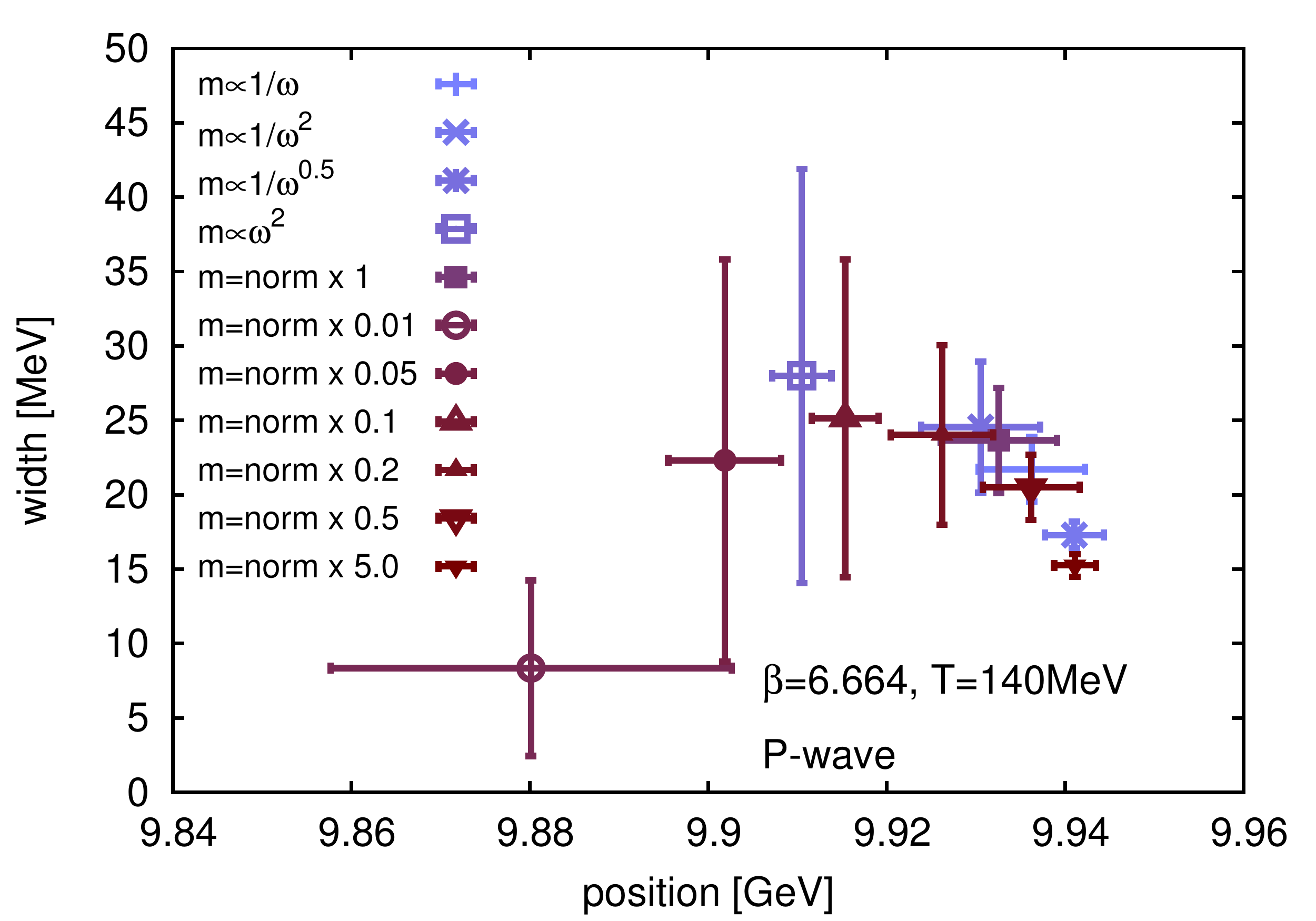}
 \caption{At $140$MeV: Dependence of the reconstructed peak position
   and width on different choices of the default model for the $\Upsilon$
   (top) and $\chi_{b1}$ (bottom) channel. Changing the functional form has
   a similar effect than moderately changing the overall
   normalization.}\label{Fig:6664PriorDep}
\end{figure}

\begin{figure}
 \includegraphics[scale=0.3,
   angle=-90]{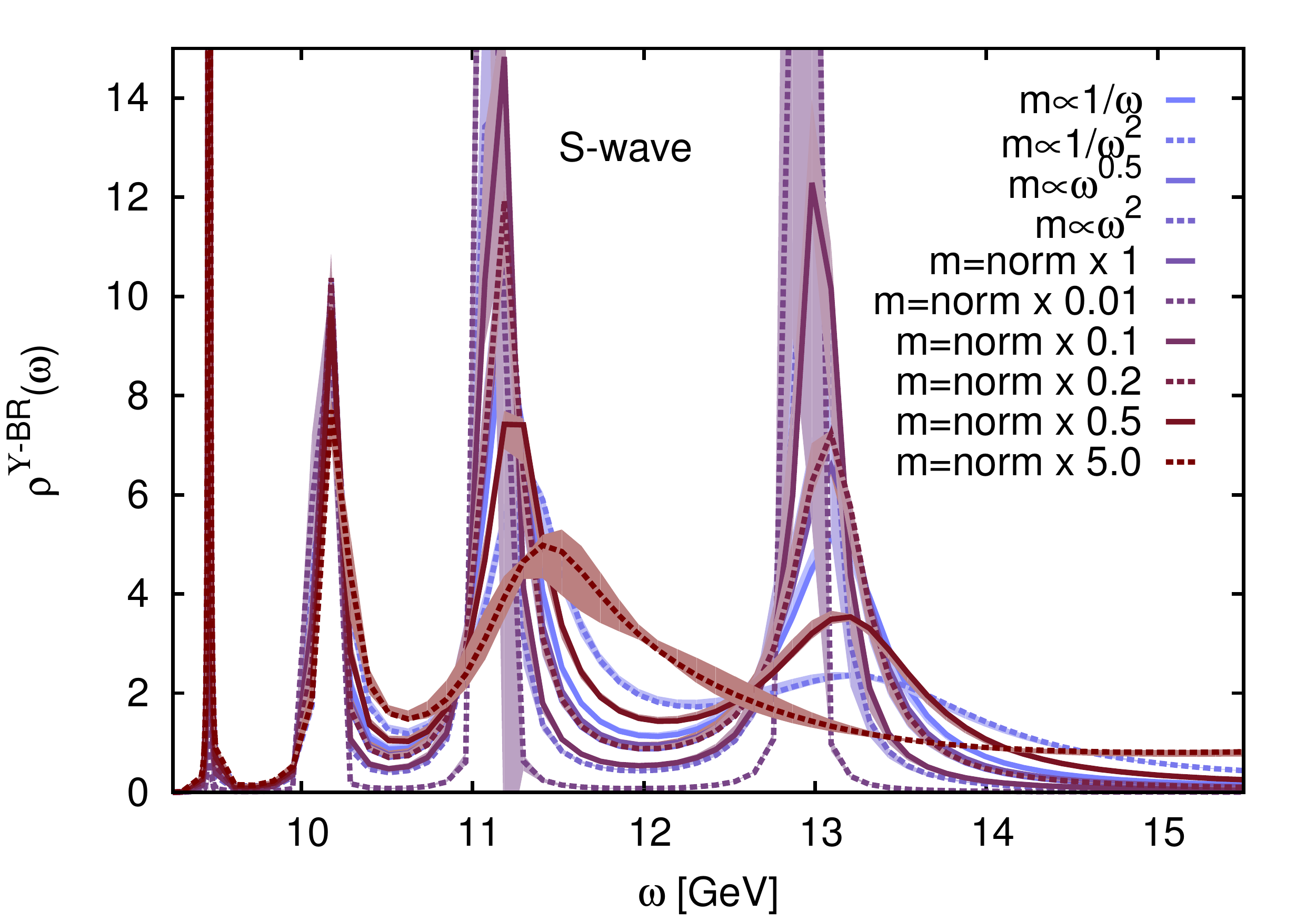}
 \includegraphics[scale=0.3,
   angle=-90]{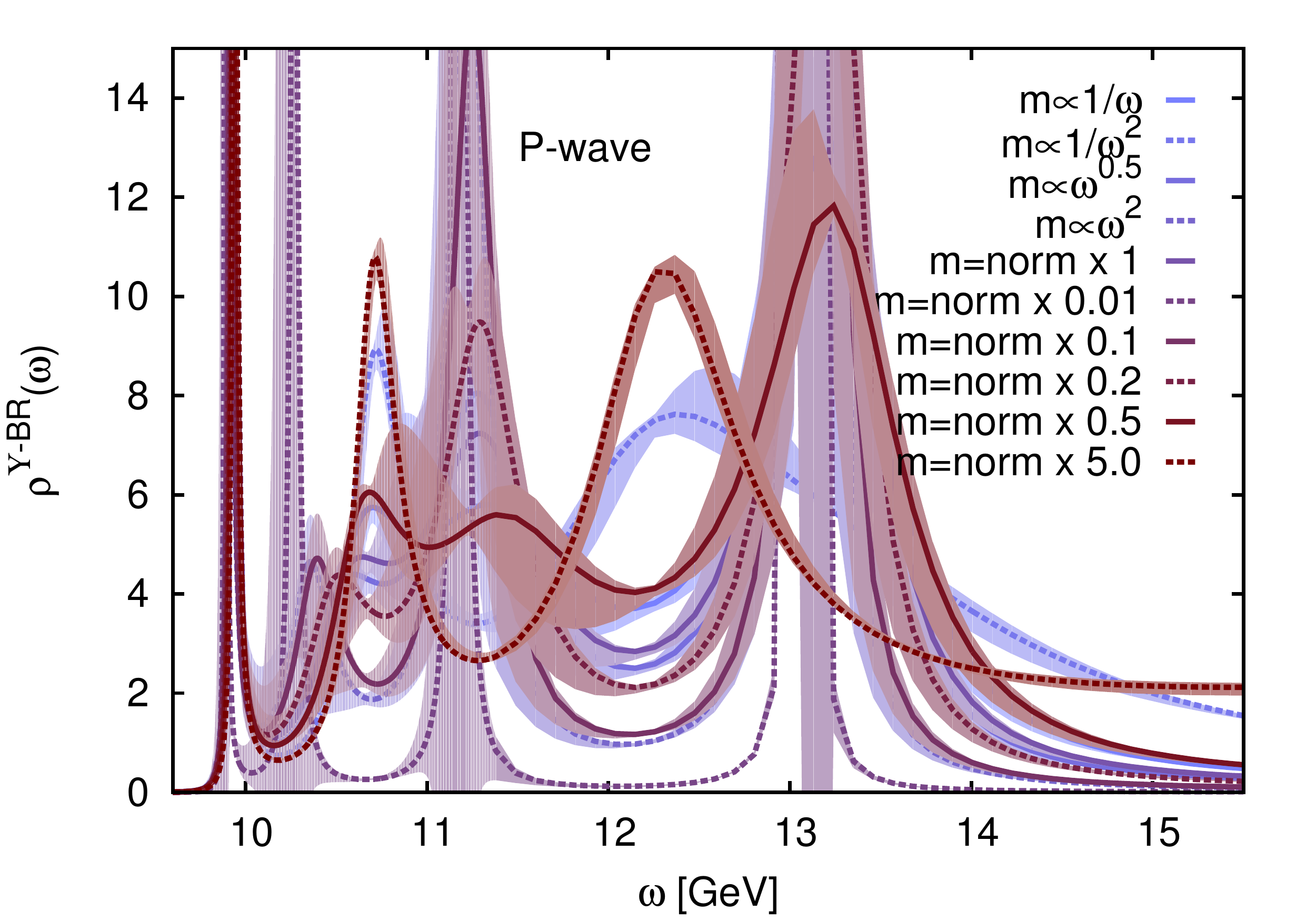}
 \caption{At $140$MeV: The actual spectral reconstructions for each of
   the choices of the prior model from which the values in
   Fig. \ref{Fig:6664PriorDep} have been determined. The S-wave is
   shown on the left, the P-wave channel on the
   right.}\label{Fig:6664PriorDepSpec}
\end{figure}

\begin{figure}
 \includegraphics[scale=0.3,
   angle=-90]{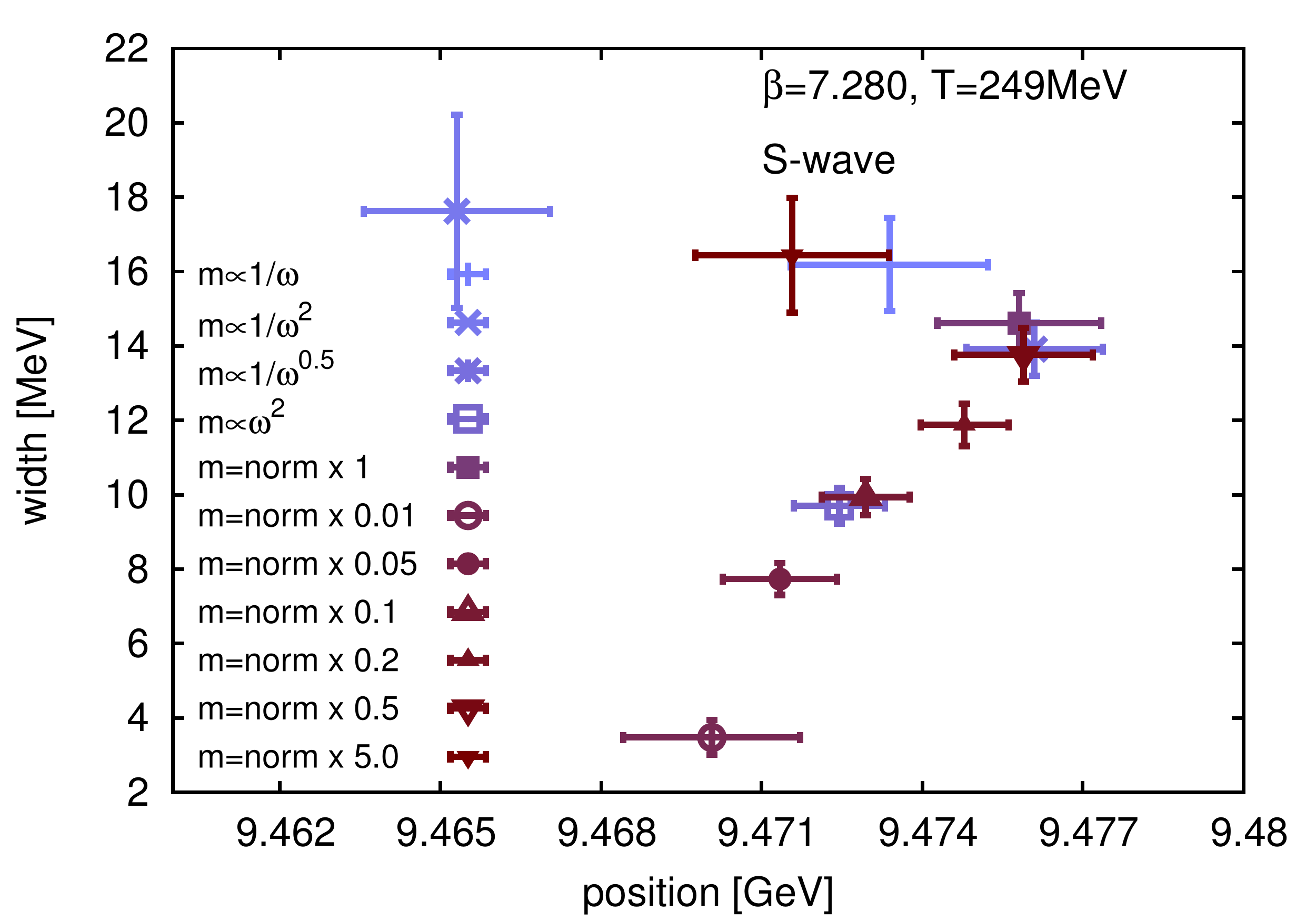}
 \includegraphics[scale=0.3,
   angle=-90]{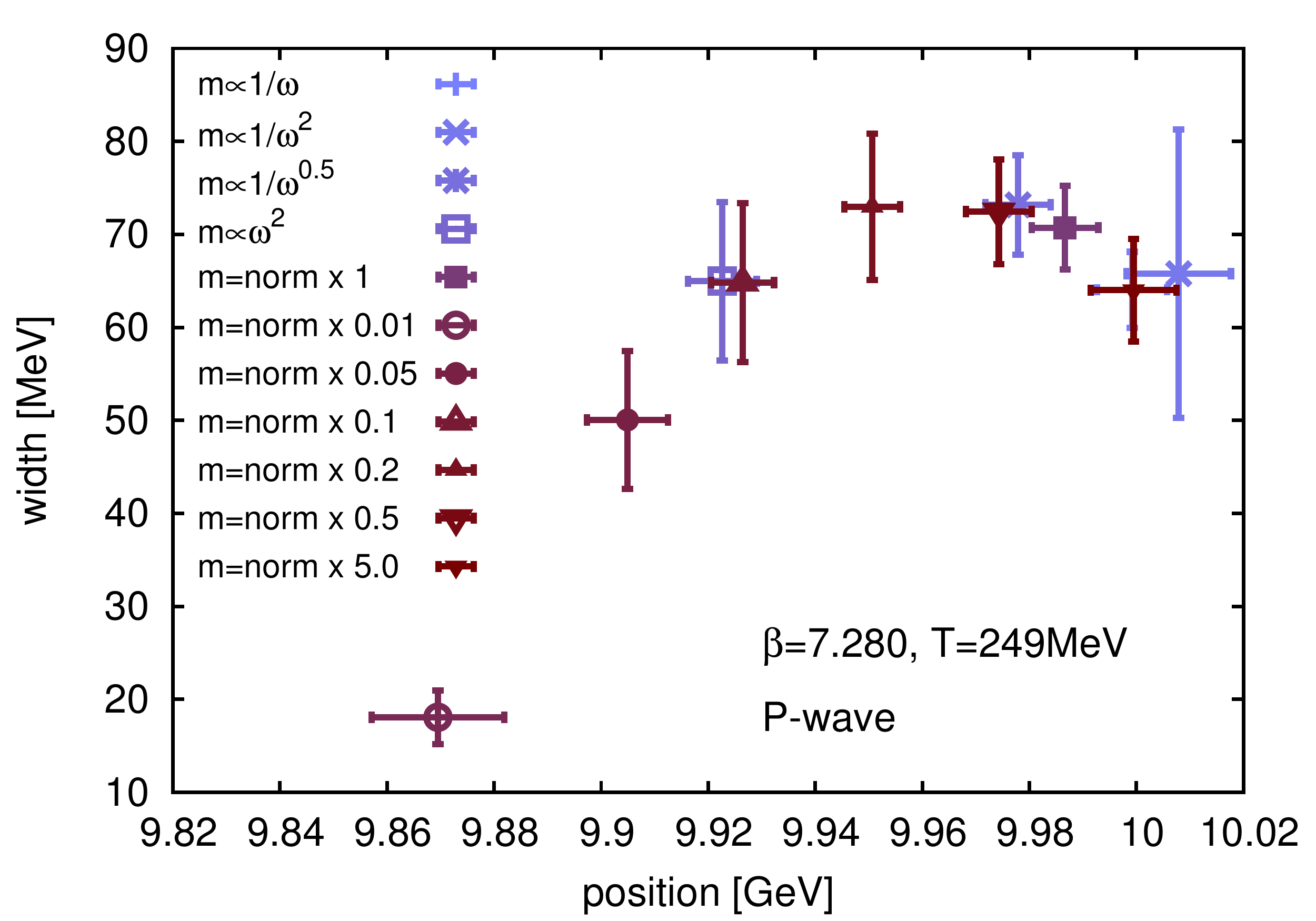}
 \caption{At $249$MeV: Dependence of the reconstructed peak position
   and width on different choices of the default model for the $\Upsilon$
   (top) and $\chi_{b1}$ (bottom) channel. Changing the functional form has
   a similar effect to moderately changing the overall
   normalization.}\label{Fig:7280PriorDep}
\end{figure}

The explicit dependence of the finite temperature spectral
reconstructions on prior information can be assessed by changing the
default model residing in the prior functional $S$. We change either
the functional form $m(\omega)\propto\omega^k$ with
$k=\{-1,-2,0.5,2\}$, while leaving the normalization intact or change
the overall magnitude of the previously normalized constant prior by
one of the factors indicated in
Figs. \ref{Fig:6664PriorDep}-\ref{Fig:7280PriorDep}. Note that in the
derivation of the prior 
functional $S$ the constant prior was singled out as being the most
neutral default model.

We find that there is a significant dependence of the reconstructed
peak position and width on the choice of default model, by which we
mean that the induced changes go beyond the statistical error bars
estimated from the jackknife. For the $\Upsilon$ channel peak position
this systematic uncertainty amounts to up to twice the statistical
error-bars, while for the width it can be a factor of ten. In the
P-wave channel we have a worse situation, since the lower
signal-to-noise ratio leads to the systematics of the peak width being
around eight times that of the jackknife errors while the position
only shows a factor of four.

\begin{figure}
 \includegraphics[scale=0.3,
   angle=-90]{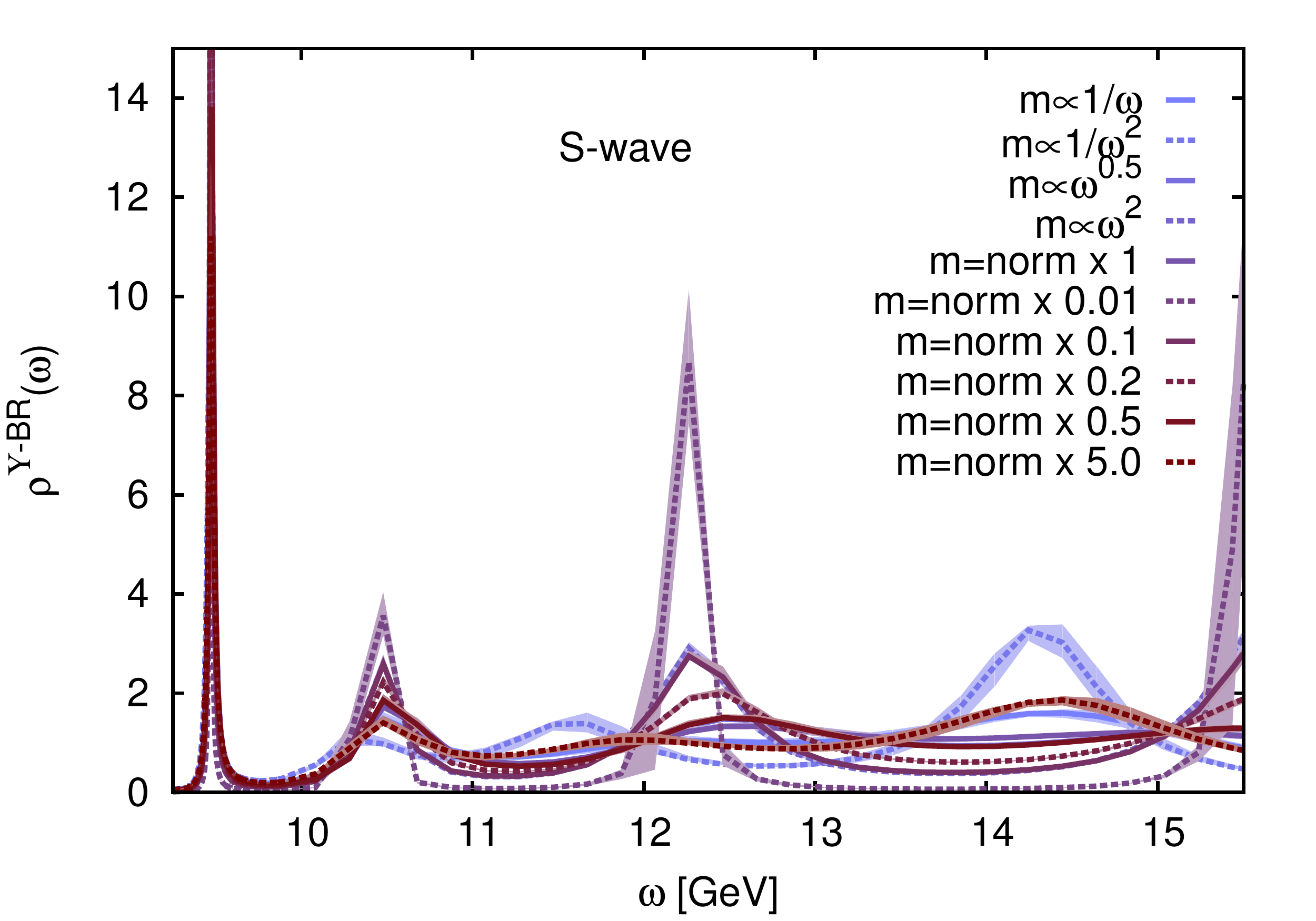}
 \includegraphics[scale=0.3,
   angle=-90]{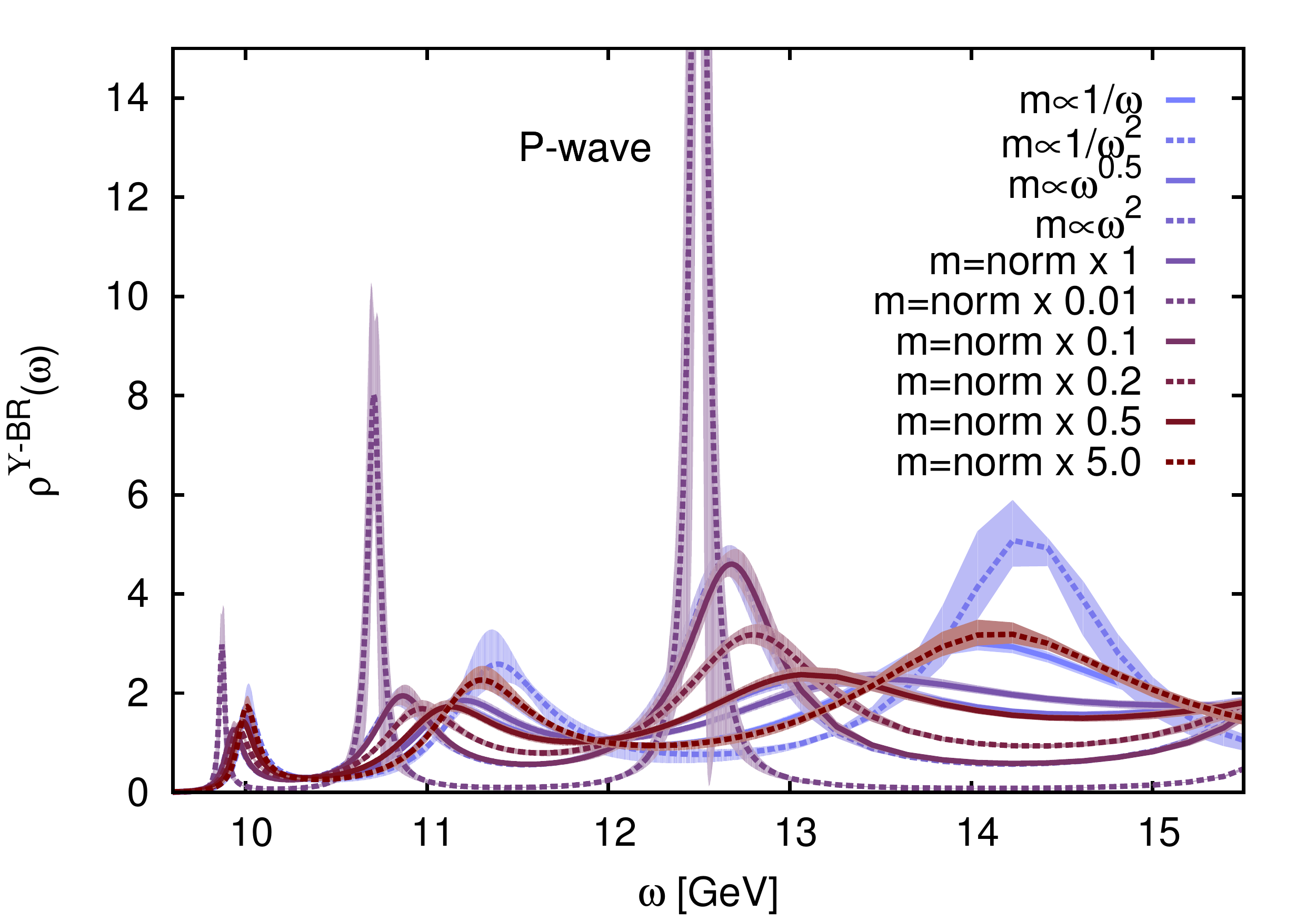}
 \caption{At $249$MeV: The actual spectral reconstructions for each of
   the choices of the prior model from which the values in
   Fig. \ref{Fig:7280PriorDep} have been determined. The S-wave is
   shown on the left, the P-wave channel on the
   right.}\label{Fig:7280PriorDepSpec}
\end{figure}

In the case of changing the normalization of the constant prior, the
P-wave shows a sudden change once the artificial prefactor reaches
$0.01$ (Figs. \ref{Fig:6664PriorDepSpec} and
\ref{Fig:7280PriorDepSpec}). In this case three rather sharp 
peaks appear, which are not found with any other choice of default
model. This outcome is the combined result of the weak constraint of
the result by the data and the corresponding stronger influence of the
prior functional. If the normalization of $m(\omega)$ is changed away
from the correct value more than two orders of magnitude the prior
probability functional $S$ will favor spectra with much less
integrated area as encoded in the data and the accuracy of the
reconstructed peak structure suffers. In the S-wave case the dominance
of the likelihood due to a better signal-to-noise ratio prevents the
incorrect normalization to distort the final outcome within the
parameter tested here. Note that it is however always possible to
choose numerical parameters for $m(\omega)$ such that they are highly
incompatible with the data and will lead to a distorted
reconstruction.

These systematic dependencies appear stronger than what was previously
found in the MEM. This is not surprising since the limited number of
available degrees of freedom in the MEM washes out features that we
are able to resolve with the new method. Consequently the default
model dependence is hidden from the MEM inside e.g. the larger widths,
which make the result appear robust against changes in $m(\omega)$ but
do not allow us to assign a high accuracy.

\begin{figure*}
 \includegraphics[scale=0.3,
   angle=-90]{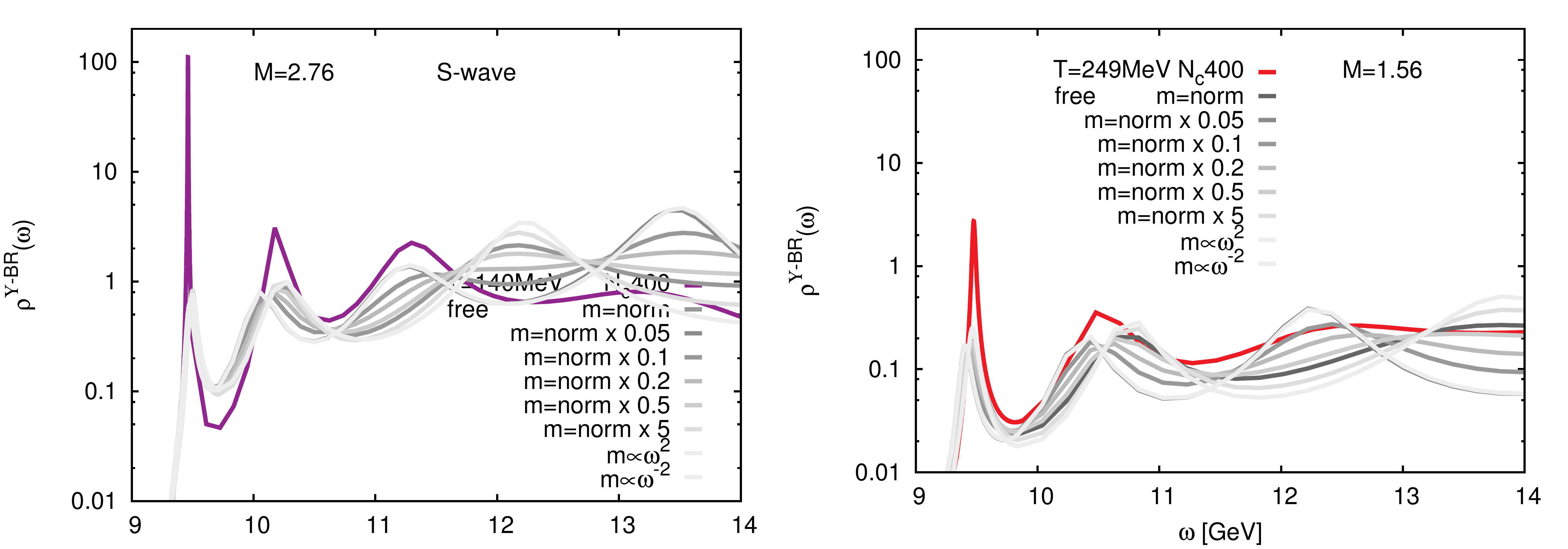}
 \caption{Free S-wave spectra (gray) for different choices of the
   default model $m(\omega)$, reconstructed from noninteracting NRQCD
   correlators using the effective mass values from the $T=140$MeV
   (left) and $T=249$MeV (right) lattices. In addition the interacting
   spectrum at the same temperature is shown (colored solid).
 }\label{Fig:S-waveFreePriorDepSpec}
\end{figure*}

\begin{figure*}
 \includegraphics[scale=0.3,
   angle=-90]{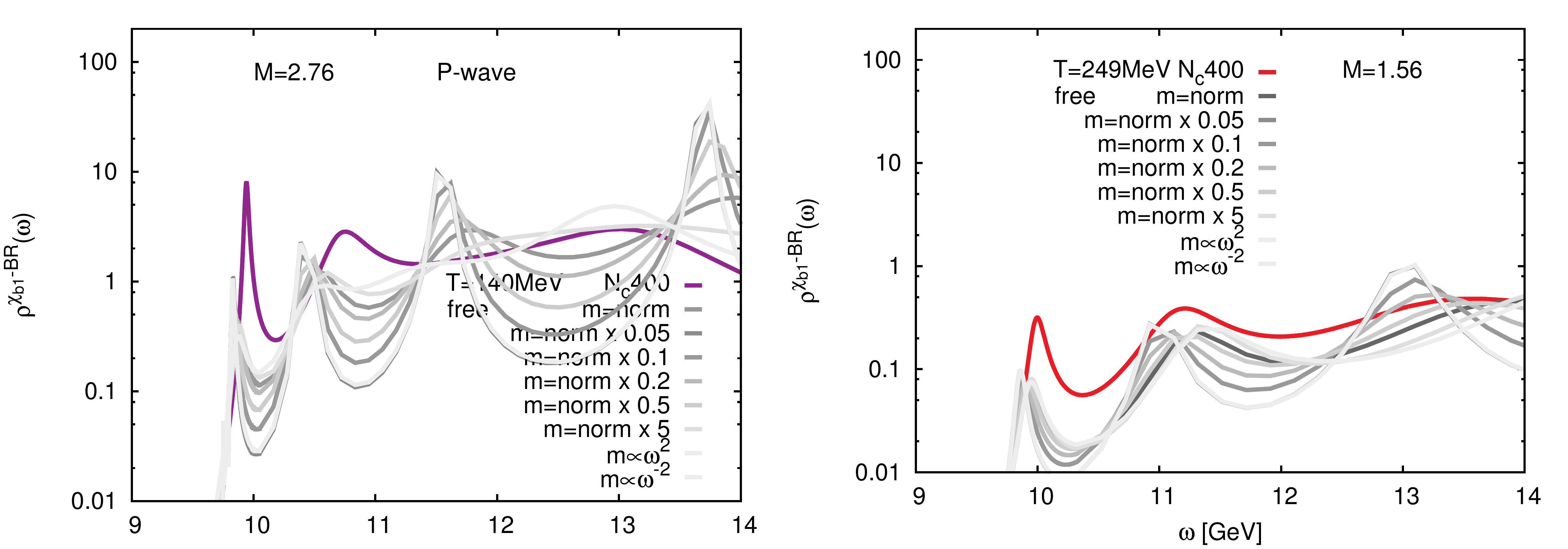}
  \caption{Free P-wave spectra (gray) for different choices of the
    default model $m(\omega)$, reconstructed from noninteracting
    NRQCD correlators using the effective mass values from the
    $T=140$MeV (left) and $T=249$MeV (right) lattices. In addition the
    interacting spectrum at the same temperature is shown (colored
    solid). }\label{Fig:P-waveFreePriorDepSpec}
\end{figure*}

For the determination of possible ground state survival or melting, we
compare the interacting spectral functions to those from free NRQCD
correlators. For this test to be meaningful we also need to understand
the default model dependence of the free spectra. In
Fig. \ref{Fig:S-waveFreePriorDepSpec} we show the reconstructions of
the free S-wave spectra based on a similar selection of default models
deployed in the finite temperature case above. We find that while the
higher lying wiggly features depend on the form of $m(\omega)$ the
lowest lying peak is quite stable. In the case of the free P-wave
spectra shown in Fig. \ref{Fig:P-waveFreePriorDepSpec} we see that the
lowest peak is slightly more susceptible to a change in the default
model, its shape is however robust enough for our conclusion about a
difference between interacting and free spectrum at $T=249$MeV to
hold.

The fact that the lowest lying artificial ringing structures are
stable against default model changes should not be
disconcerting. Similar to the Fourier transform, the presence and
strength of the Gibbs ringing depends solely on the number of
available data points. Adding noise might wash out some of these
features or make them more susceptible to the form of the default
model but with the size of the statistical errors used here, the
lowest lying wiggle indeed remains stable.

\FloatBarrier
\subsection*{Dependence on the choice of $\omega$ interval}

A form of implicit prior information lies in the choice of the
discretization interval $\omega\in[\omega_{\rm min},\omega_{\rm max}]$
along the frequency axis. If we choose it to be too small, not all relevant
frequencies encoded in the data are accessible and the reconstruction
fails. In the case of the standard implementation of the MEM, choosing
the lower limit of the interval to be too low adversely affects the
reconstruction result, as the limited number of basis functions only
contain an oscillatory part close to $\omega_{\rm min}$ before damping
away at higher frequency. This artificial limitation is completely
absent in the new Bayesian method.

\begin{figure*}
 \includegraphics[scale=0.3,
   angle=-90]{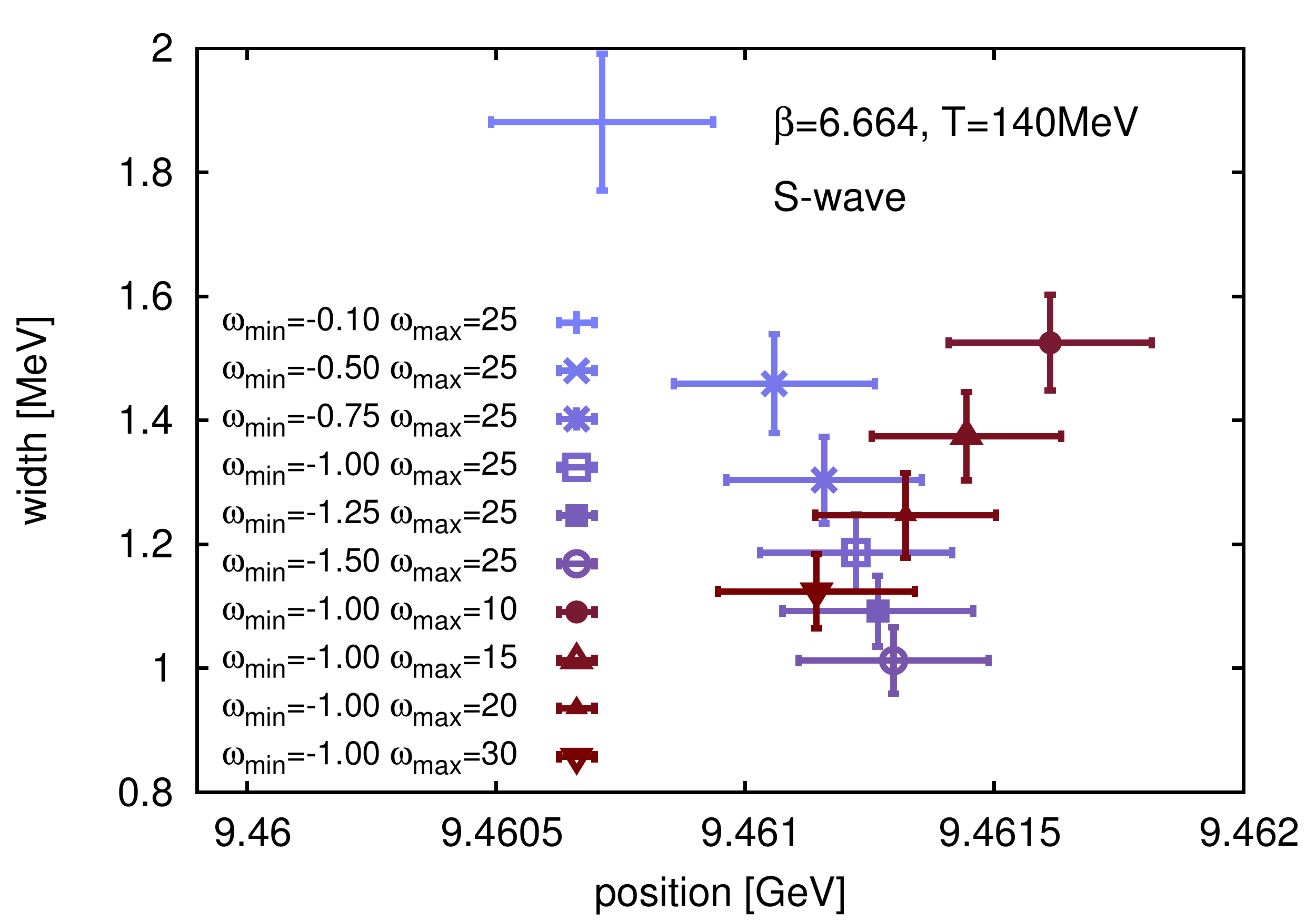}
 \includegraphics[scale=0.3,
   angle=-90]{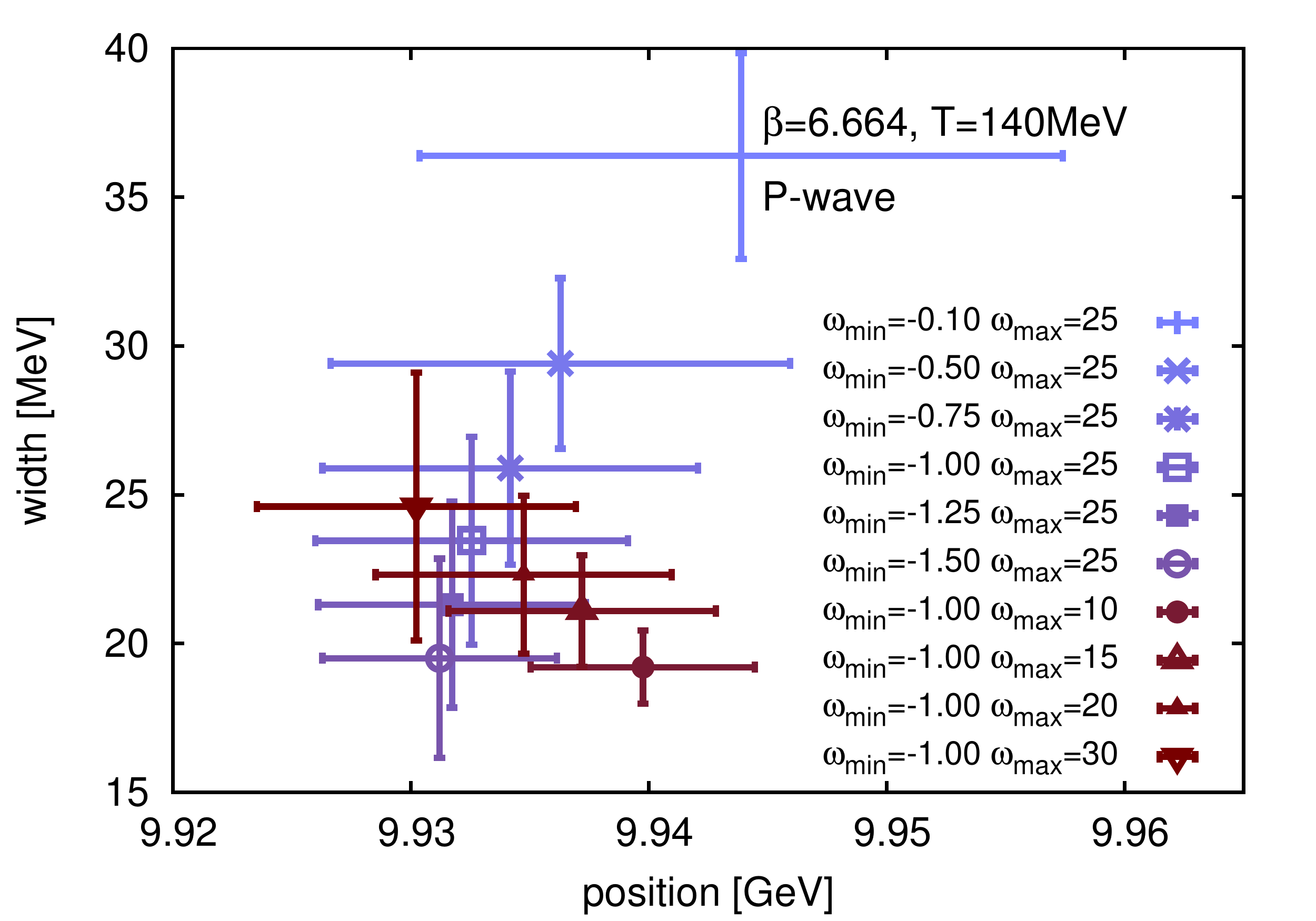}
 \caption{At $140$MeV: Dependence of the reconstructed peak position
   and peak width on different choices of the frequency interval for
   the $\Upsilon$ (left) and $\chi_{b1}$ (right) channel. While a too
   narrow choice of $[\omega_{\rm min},\omega_{\rm max}]$ leads to
   significant changes in the reconstructed values, peak positions
   stabilize to a plateau at large enough interval lengths. The peak width
   is slightly less stable, since it depends more strongly on the default
   model normalization, which changes with each different choice of
   $[\omega_{\rm min},\omega_{\rm max}]$.}\label{Fig:6664OmegaIntDep}
\end{figure*}

\begin{figure*}
 \includegraphics[scale=0.3,
   angle=-90]{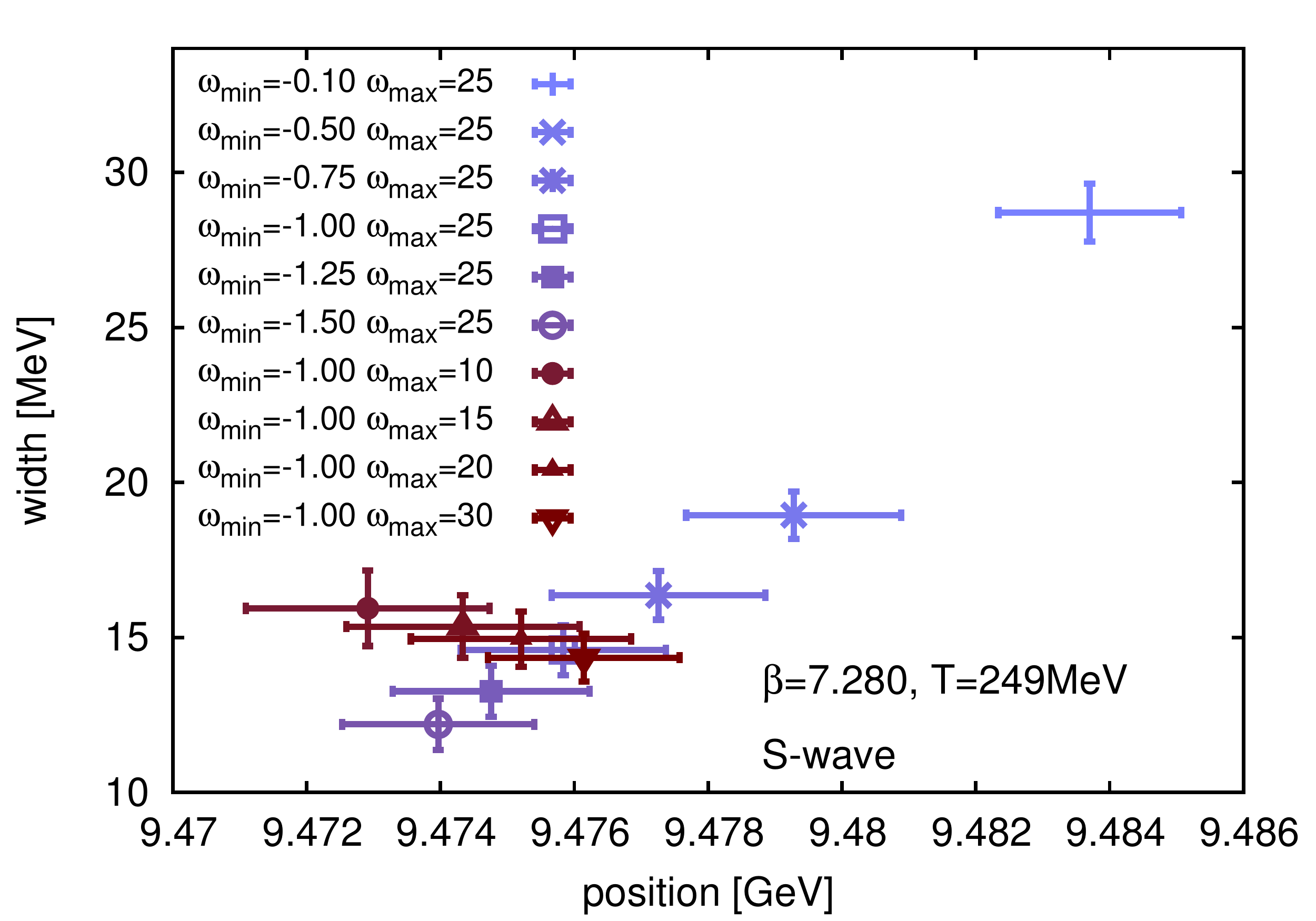}
 \includegraphics[scale=0.3,
   angle=-90]{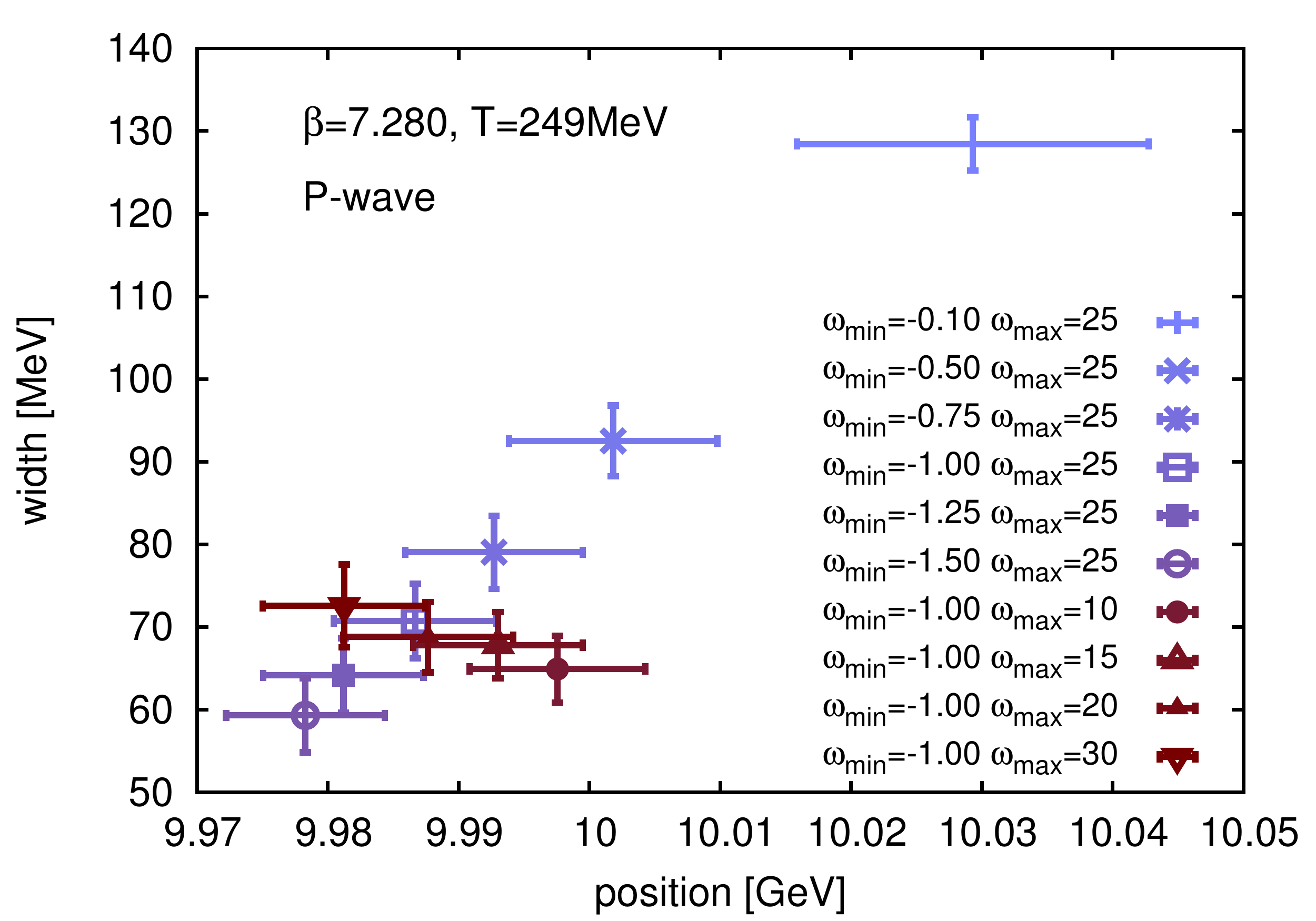}
 \caption{At $249$MeV: Dependence of the reconstructed peak position
   and peak width on different choices of the frequency interval for
   the $\Upsilon$ (left) and $\chi_{b1}$ (right) channel. While a too
   narrow choice of $[\omega_{\rm min},\omega_{\rm max}]$ leads to
   significant changes in the reconstructed values, peak positions
   stabilize to a plateau at large enough interval lengths. The peak width
   is slightly less stable, since it depends more strongly on the default
   model normalization, which changes with each different choice of
   $[\omega_{\rm min},\omega_{\rm max}]$.}\label{Fig:7280OmegaIntDep}
\end{figure*}

We find that a too high choice of $\omega_{\rm min}$ and a too low
choice of $\omega_{\rm max}$ affects the outcome of our reconstruction
significantly beyond the statistical error bars. However it is evident
from Figs. \ref{Fig:6664OmegaIntDep}-\ref{Fig:7280OmegaIntDep}
that by allowing the reconstruction to proceed on an increasingly
larger interval one finds that the peak position stabilizes while the
peak width is still weakly affected due to the change in prior
normalization occurring between different choices of the frequency
interval length.

\subsection*{Systematics of the MEM}
\label{sec:comparison}

In order to compare our study's results to previous work based on the
MEM, we have carried out the reconstruction of spectral information
based on the popular implementation by Bryan. In this approach the
search space from which the spectra are chosen is restricted to have
dimension equal to the number of data points. Furthermore the basis
functions of this search space are constructed from a singular value
decomposition of the convolution kernel.

Again we use $N_\omega=1200$ and a numerical value of $\tau_{\rm
  max}^{\rm Num}=20$ but here the frequency interval is taken to span
$[-0.15,25]$ in order for the oscillatory part of the restricted
search space basis functions to extend into the positive frequency
domain. Our minimizer is set to accept an extremum of the $Q=L-\alpha
S$ functional if the step size becomes less than $5\times
10^{-9}$. The standard alpha integration is carried out for each
jackknife bin to arrive at the spectral functions shown in
Figs. \ref{Fig:ThermalSpectraOverviewSwave}--\ref{Fig:ThermalSpectraOverviewPwave}. 

One finds that the reconstruction is not only less stable but also
shows rather washed out spectral features. The ground state peak is
not a Lorentzian but instead appears more Gaussian with its width
roughly an order of magnitude larger compared to the new Bayesian
reconstruction method. It is difficult to compare the obtained peak
positions and widths as their values depend crucially on the choice of
the frequency interval.

\begin{figure*}
 \includegraphics[scale=0.3,
   angle=-90]{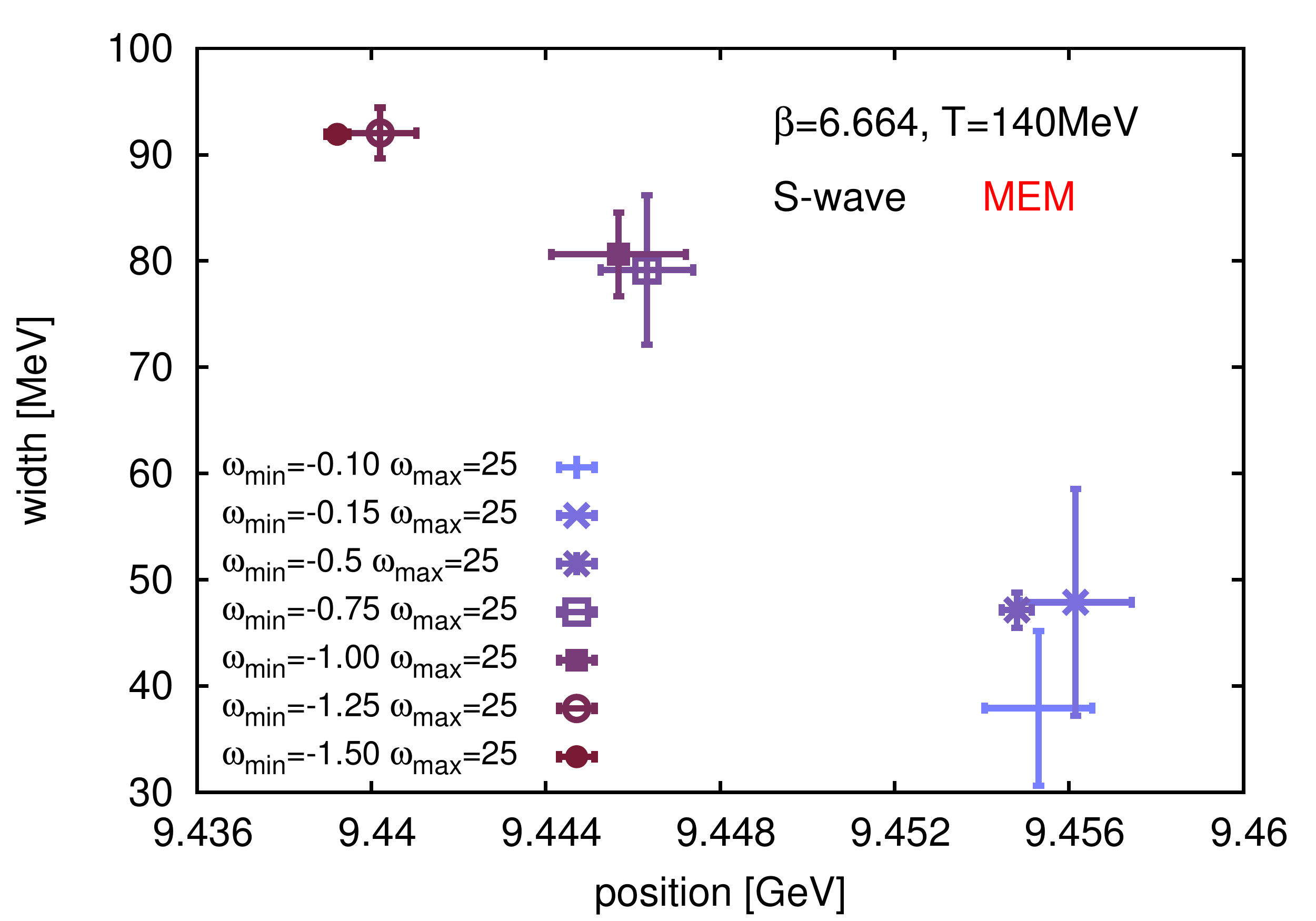}
 \includegraphics[scale=0.3,
   angle=-90]{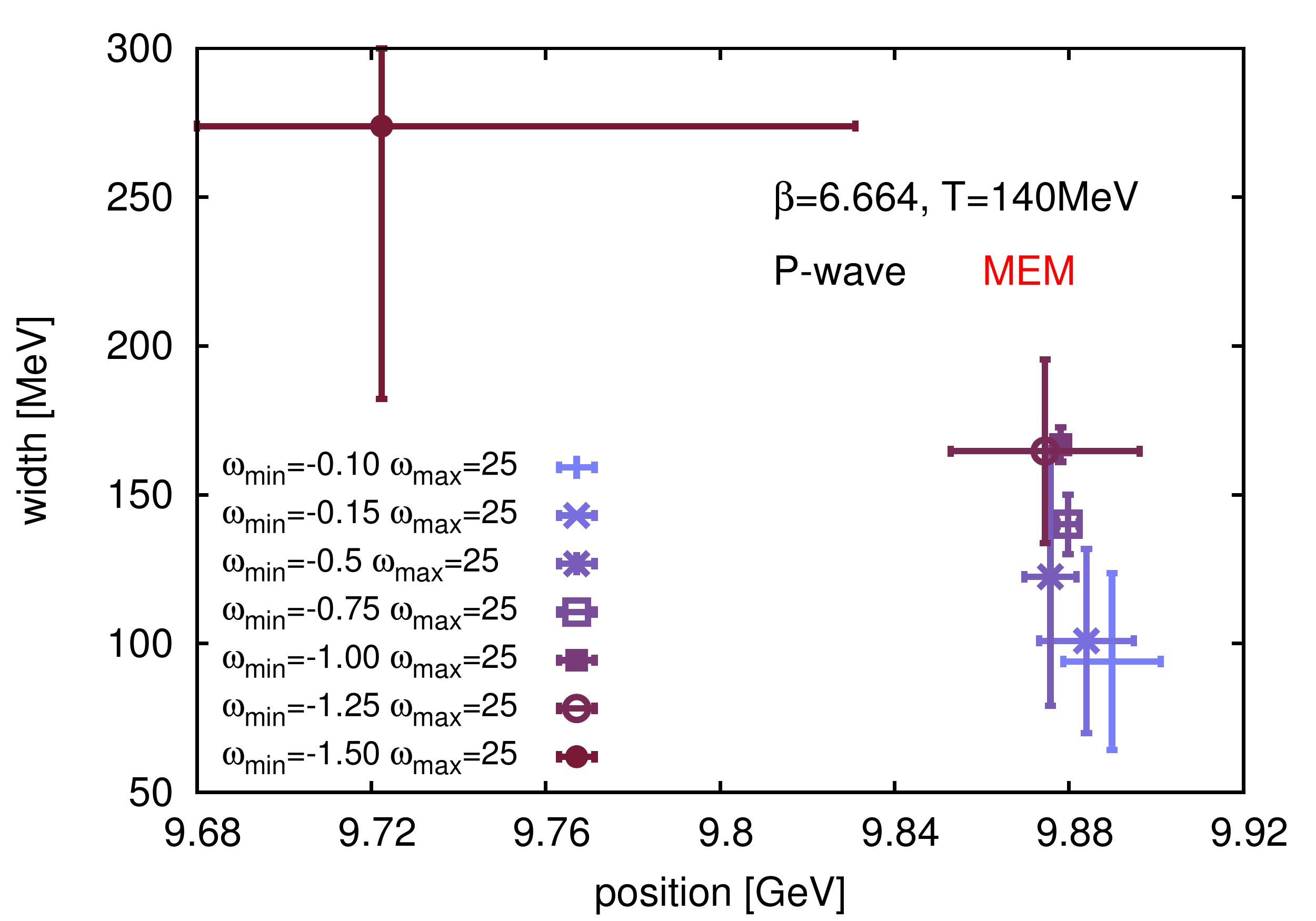}
 \caption{At $140$MeV: The dependence of the maximum entropy
   reconstructed peak width and position on different choices for the
   frequency interval in the $\Upsilon$ (left) and $\chi_{b1}$ (right)
   channel. We find as expected from the arguments laid out in the
   subsection on the choice of the $\omega$ interval that moving
   $\omega_{\rm min}$ to lower and lower values degrades the quality
   of the reconstruction. Both the peak positions move to smaller
   values while the width grows. Neither in the $\Upsilon$ nor in the
   $\chi_{b1}$ is a plateau reached within the inspected choices. Note
   that the error bars beyond $\omega_{\rm min}\leq-1$ cannot be
   trusted as the minimizer remains at values of $L>100$. This is a
   clear indication for a lack of oscillatory degrees of freedom in
   the search space basis functions.}\label{Fig:6664MEMIntDep}
\end{figure*}

\begin{figure*}
 \includegraphics[scale=0.3,
   angle=-90]{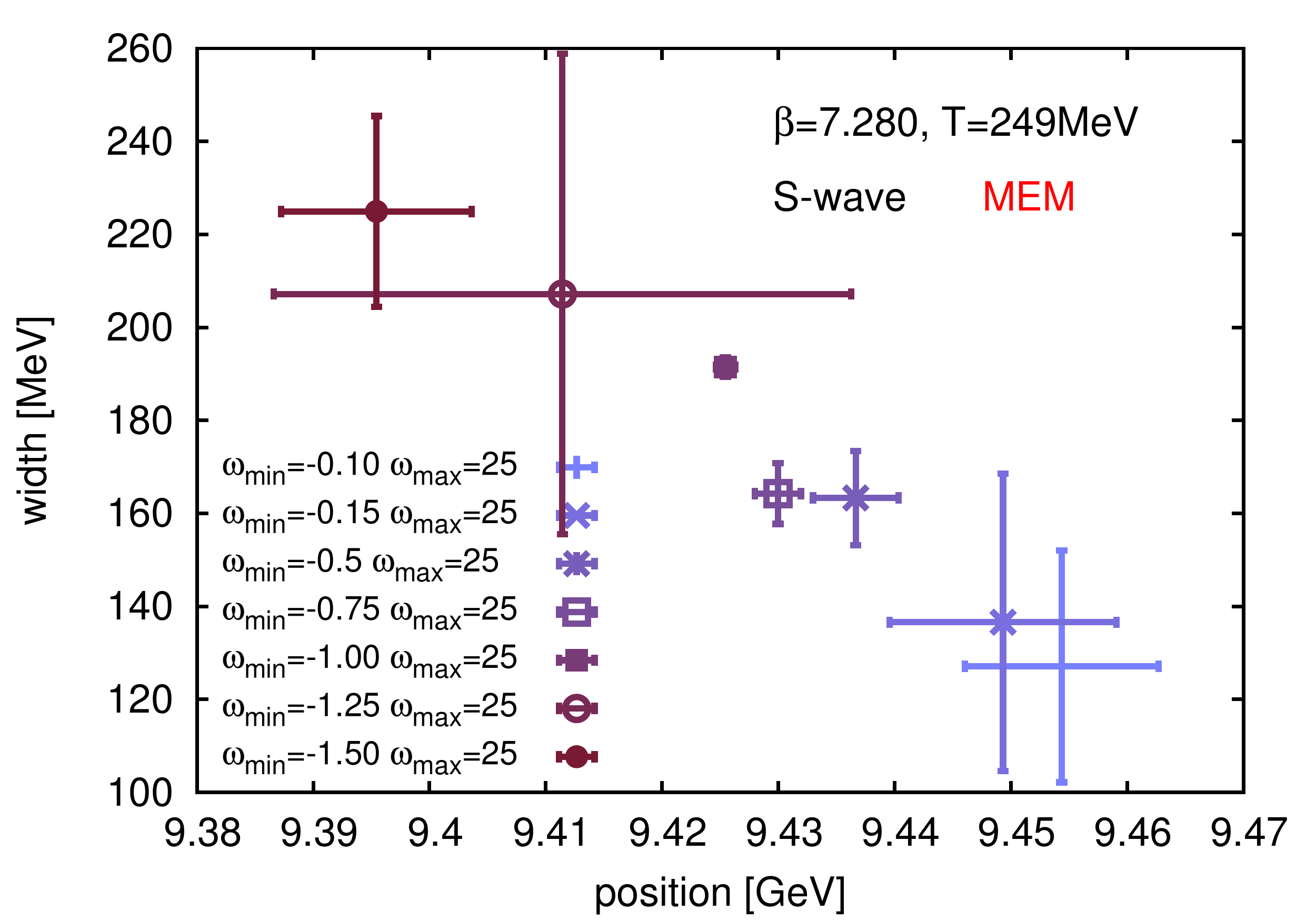}
 \includegraphics[scale=0.3,
   angle=-90]{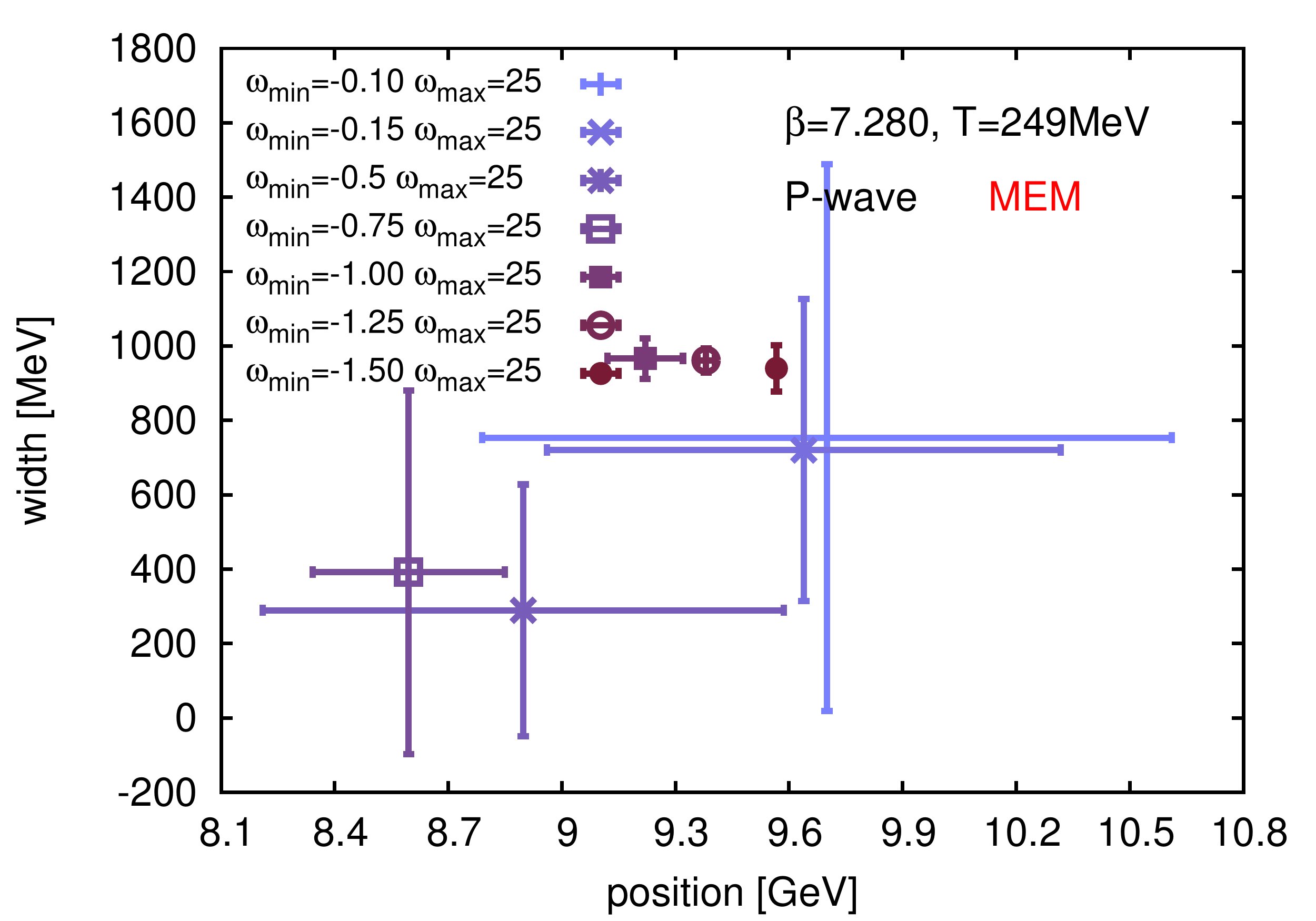}
 \caption{At $249$MeV: The dependence of the maximum entropy
   reconstructed peak width and position on different choices for the
   frequency interval in the $\Upsilon$ (left) and $\chi_{b1}$ (right)
   channel.}\label{Fig:7280MEMIntDep}
\end{figure*}

As shown in Figs. \ref{Fig:6664MEMIntDep}--\ref{Fig:7280MEMIntDep} the
reconstructed position can be shifted 
to significantly lower and the width to significantly higher values if
the frequency interval is extended further into the negative
regime. Essentially it is possible to melt bottomonium with the MEM
from a choice of frequency interval alone, which is in stark contrast
to the behavior of the new method which clearly shows a plateau of
the reconstructed values for large enough intervals.

\section{Tests of the NRQCD discretization}
\label{sec:systematics2}

The applicability of an effective field theoretical description relies
on the presence of a separation of scales. In the case of heavy quark
pairs in continuum NRQCD, it is characterized by $\Lambda_{\rm
  QCD}/M_b\ll1$, $T/M_b\ll1$ and $\mathbf{p}^2/2M_b\ll1$. On the
lattice the latter translates into a ratio between the discretized
lattice momenta
\begin{align}
 \hat{\mathbf{p}}^2=4 \sum_{i=1}^{3}{\rm sin}^2\Big(\frac{\pi n_i}{N_s}\Big), \quad n_i=-\frac{N_s}{2}+1,\ldots,\frac{N_s}{2}
\end{align}
of the first Brillouin zone and the effective mass parameter
$\hat{M}=2n \xi a_s M_b$, where $\xi$ denotes the physical lattice
anisotropy and $n$ characterizes the choice of Euclidean time
discretization in the NRQCD propagator equation of motion
\eqref{NRQCDEvolEq}.

\begin{figure*}[t]
 \includegraphics[scale=0.3,
   angle=-90]{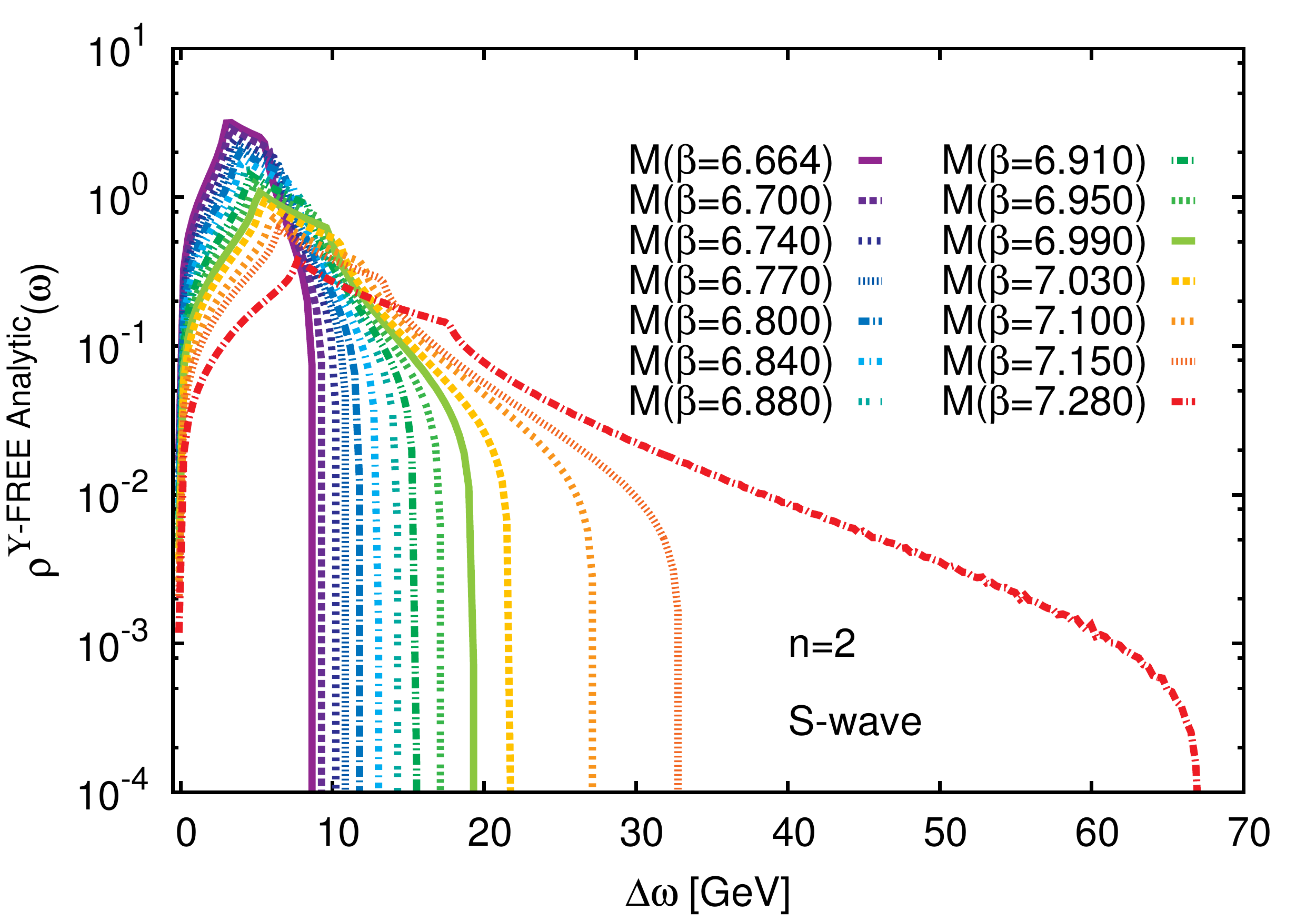}
 \includegraphics[scale=0.3,
   angle=-90]{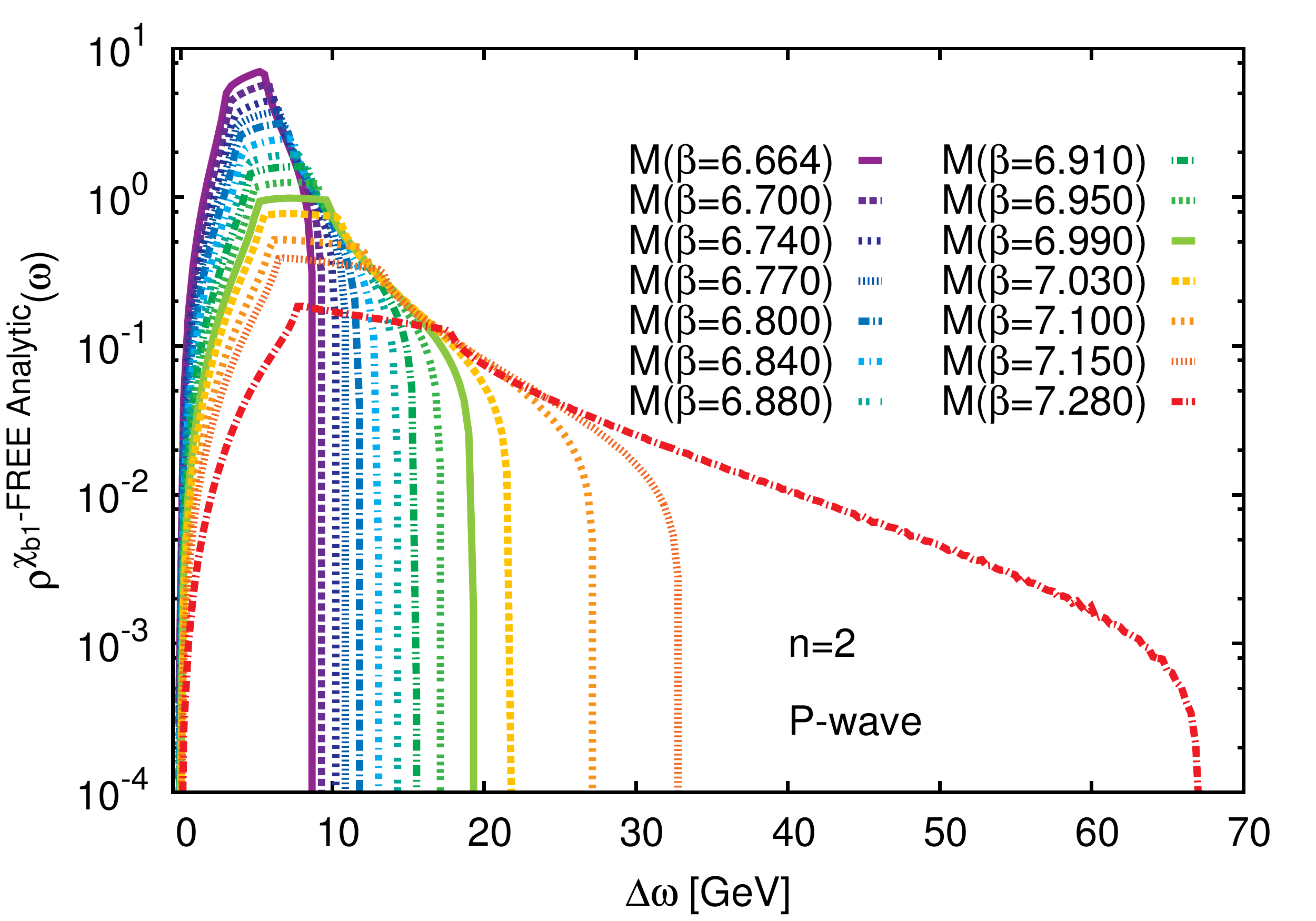}
 \caption{The analytic free S-wave (left) and P-wave (right) spectral
   functions according to Eq. \eqref{Eq:AnalytFreeSpec} for $n=2$ with
   the effective mass parameter chosen to agree with the values of the
   fully interacting lattices ($N_s=384$ used to obtain smooth
   curves). While lattice artifacts dominate the region above $8$GeV
   (i.e. above the first kink), the physics of the bound state resides
   at frequencies below $1$GeV.}\label{Fig:AnalyticFreeSpecN2}
\end{figure*}

\begin{figure*}[t]
 \includegraphics[scale=0.3,
   angle=-90]{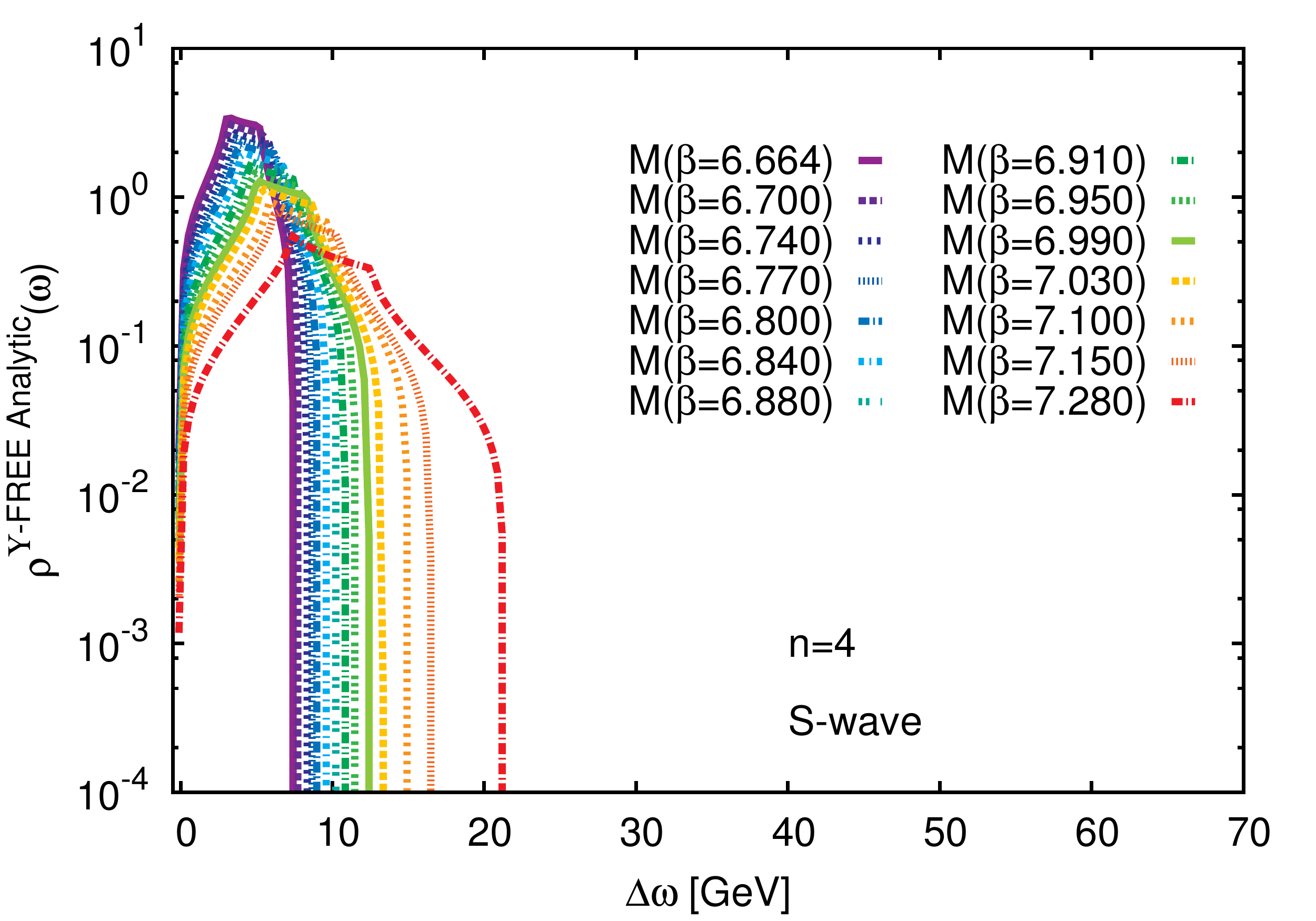}
 \includegraphics[scale=0.3,
   angle=-90]{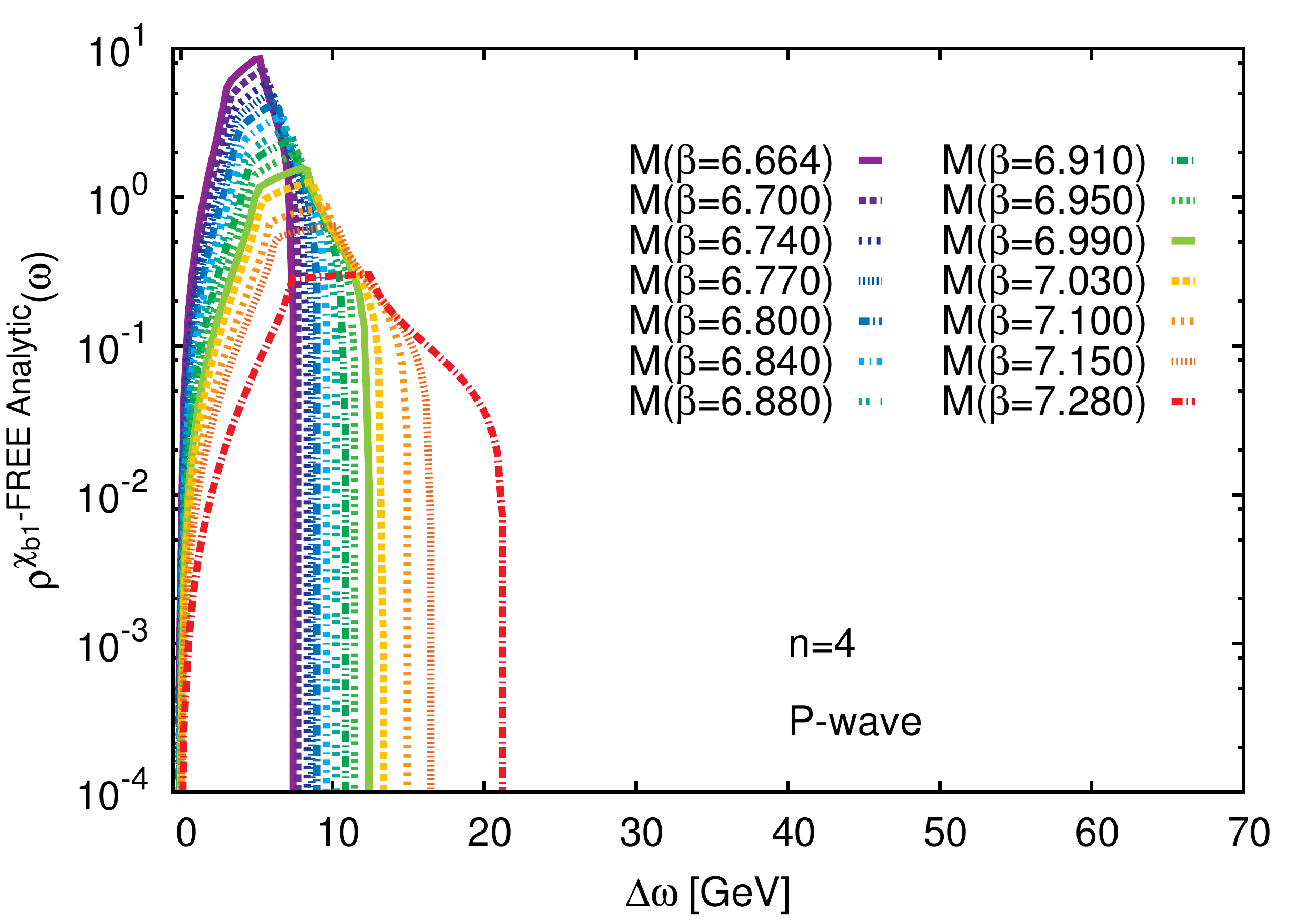}
 \caption{The analytic free S-wave (left) and P-wave (right) spectral
   functions according to Eq. \eqref{Eq:AnalytFreeSpec} for $n=4$ with
   the effective mass parameter chosen to agree with the values of the
   fully interacting lattices ($N_s=384$ used to obtain smooth
   curves). Note that in comparison to
   Fig. \ref{Fig:AnalyticFreeSpecN2} the high frequency range is much
   more limited, while the amplitude of the kink structure is
   increased. We also see that the region of relevant frequencies
   $w<1$GeV does not change
   appreciably.}\label{Fig:AnalyticFreeSpecN4}
\end{figure*}

Analyzing noninteracting lattice NRQCD Ref. \cite{Davies:1991py}
showed that for isotropic lattices $(\xi=1)$ the choice $n=1$ is
sufficient to obtain a stable high momentum behavior, as long as $a_s
M_b>3$, respectively $a_s M_b>1.5$ for $n=2$. Otherwise the expansion
in powers of the velocity (respectively inverse rest mass) breaks down
and the high frequency regime of the theory becomes ill defined. This
is a direct consequence of the EFT not possessing a naive continuum
limit. This constraint on $n$ related to the validity of the NRQCD
expansion is reflected also in the free dispersion relation
\begin{align}
a_{\tau}&E_{\hat{\mathbf{p}}} = 2n {\rm Log}\big[1-\frac{1}{2}\frac{\hat{\mathbf{p}}^2}{ 2 n\xi a_s M_b }\big] \nonumber \\
  +& {\rm Log}\big[1 + \frac{ (\hat{\mathbf{p}}^2)^2 }{ 16 n\xi (a_s M_b)^2 } +  \frac{ (\hat{\mathbf{p}}^2)^2 }{ 8 \xi (a_s M_b)^3 }  - \frac{ \hat{\mathbf{p}}^4 }{ 24 \xi (a_s M_b) } \big]\label{Eq:NRQCDDispRel}
 \end{align}
that follows from Eq. \eqref{NRQCDEvolEq}. Here the requirement of a
positive argument in the logarithm leads to e.g. $a_s M_b>1.5$ for
$n=2$. From Eq. \eqref{Eq:NRQCDDispRel} the free spectra are computed
via
\begin{align}
 \nonumber \rho_S(\omega)=\frac{4\pi N_c}{N_s^3}\sum_{\hat{\mathbf{p}}}\delta(\omega-2E_{\hat{\mathbf{p}}} ),\\
 \rho_P(\omega)=\frac{4\pi N_c}{N_s^3}\sum_{\hat{\mathbf{p}}} \hat{\mathbf{p}}^2\delta(\omega-2E_{\hat{\mathbf{p}}} ),\label{Eq:AnalytFreeSpec}
\end{align}
as laid out e.g. in the appendix of
\cite{Aarts:2011sm,Aarts:2014cda}. We note that a difference in the
overall normalization between the analytic and lattice regularized
free spectra exists, i.e. if we were to calculate from
Eq.\eqref{Eq:AnalytFreeSpec} the corresponding Euclidean correlator
its value at $\tau=0$ is larger than $D(0)$ obtained from
Eq.\eqref{NRQCDEvolEq} with unit links. One reason is that the full
dispersion relation that enters Eq.\eqref{Eq:AnalytFreeSpec} is only
applied to the lattice propagator from time step $\tau/a>1$.

\begin{figure*}[t]
 \includegraphics[scale=0.3,
   angle=-90]{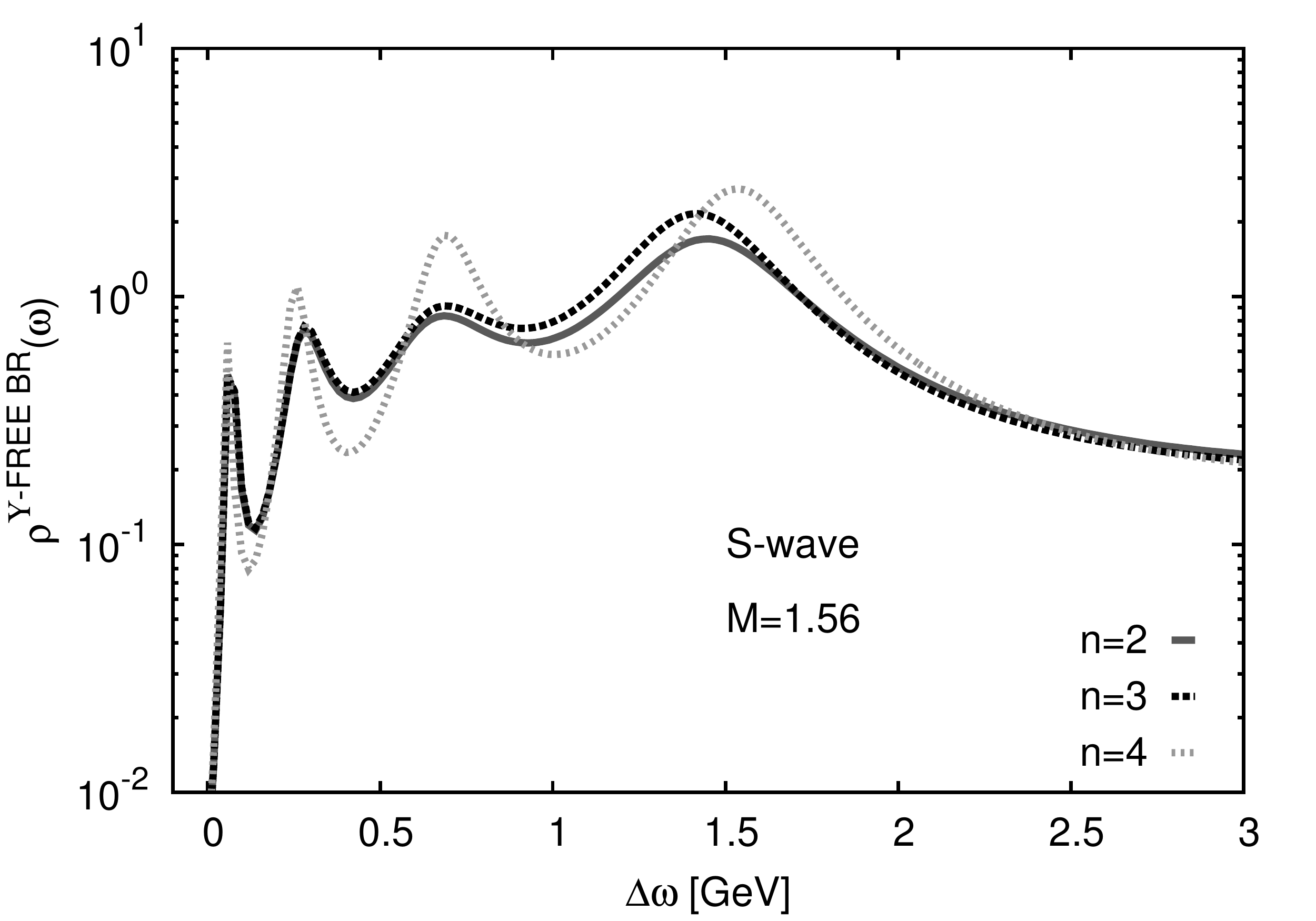}
 \includegraphics[scale=0.3,
   angle=-90]{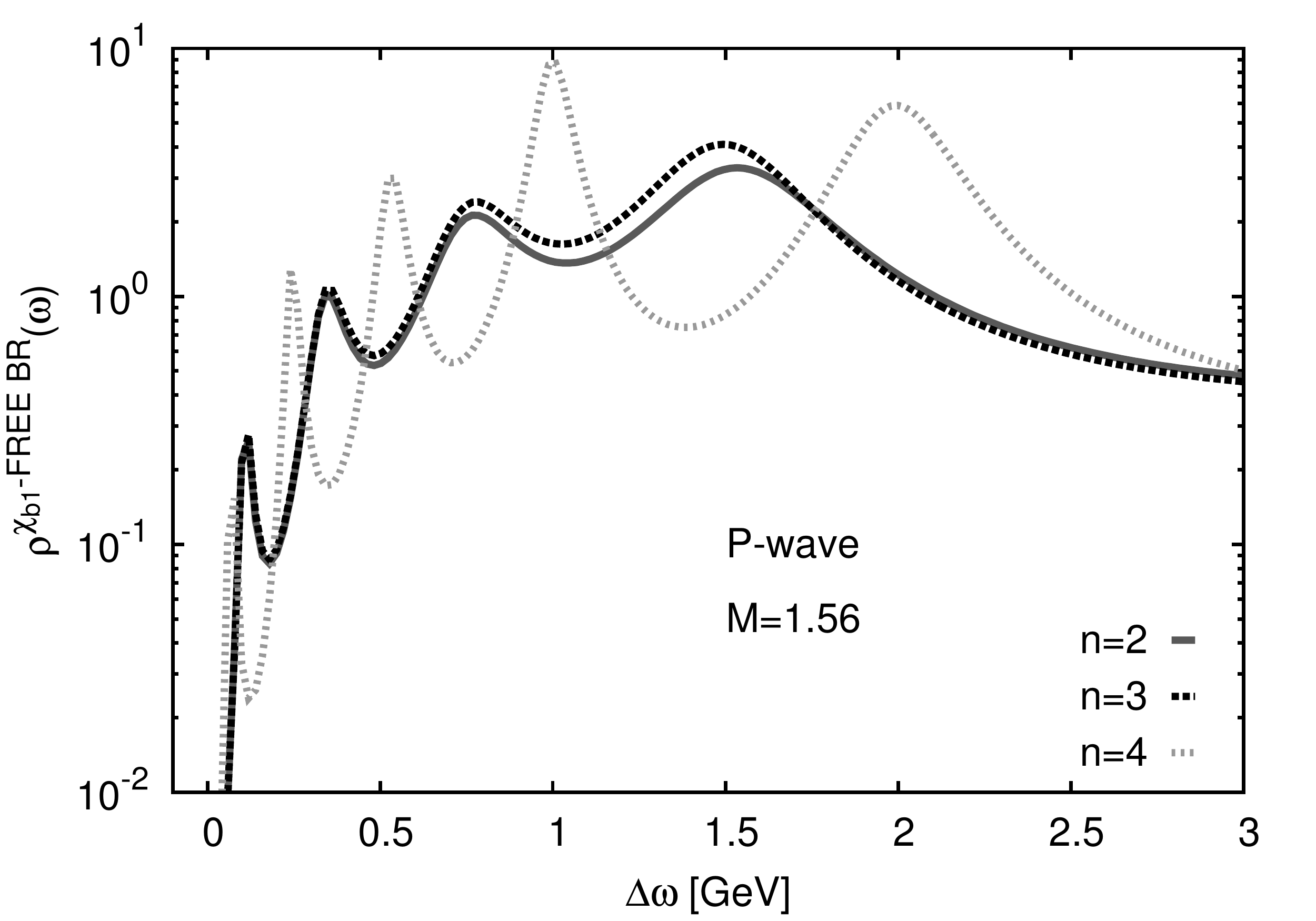}
 \caption{The free spectra for S-wave (left) and P-wave (right)
   extracted from the free NRQCD correlators, based on the effective
   mass parameters $a_s(\beta=7.280)M_b=1.56$ and three different
   values of the parameter $n=2,3,4$. Consistent with expectations,
   the spectra are increasing in amplitude at higher frequencies
   $\omega>1$GeV. In the S-wave channel the number of reconstructed
   wiggles remains stable, while there seems to be a slight increase
   in the amplitude of the ground state peak at $n=4$. In the P-wave
   channel for $n=4$, we obtain an additional wiggle but do not find
   indications that the strength of the low frequency structures
   changes.}\label{Fig:FreeNRQCDSpecLep}
\end{figure*}
\begin{figure*}[t]
 \includegraphics[scale=0.3,
   angle=-90]{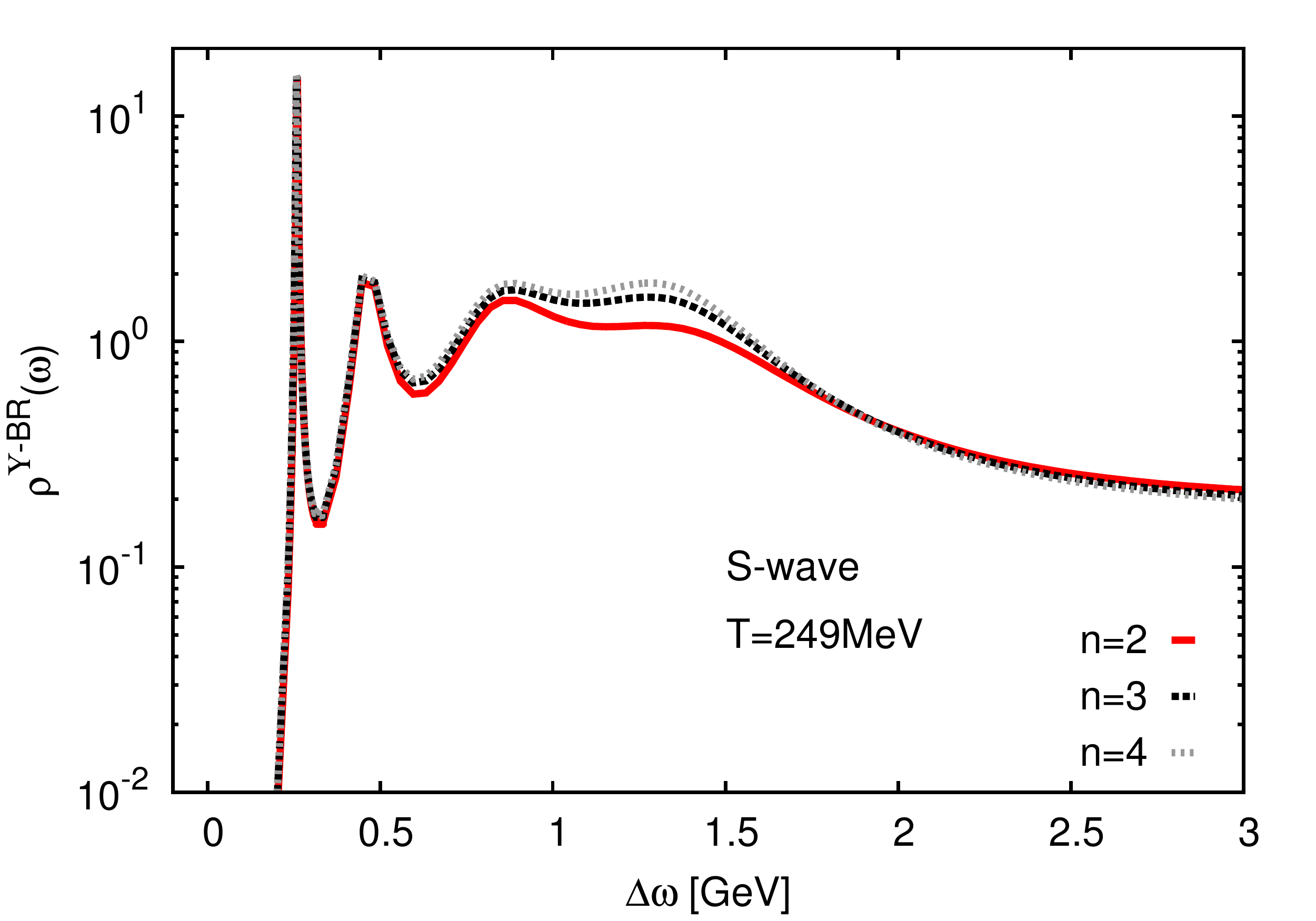}
 \includegraphics[scale=0.3,
   angle=-90]{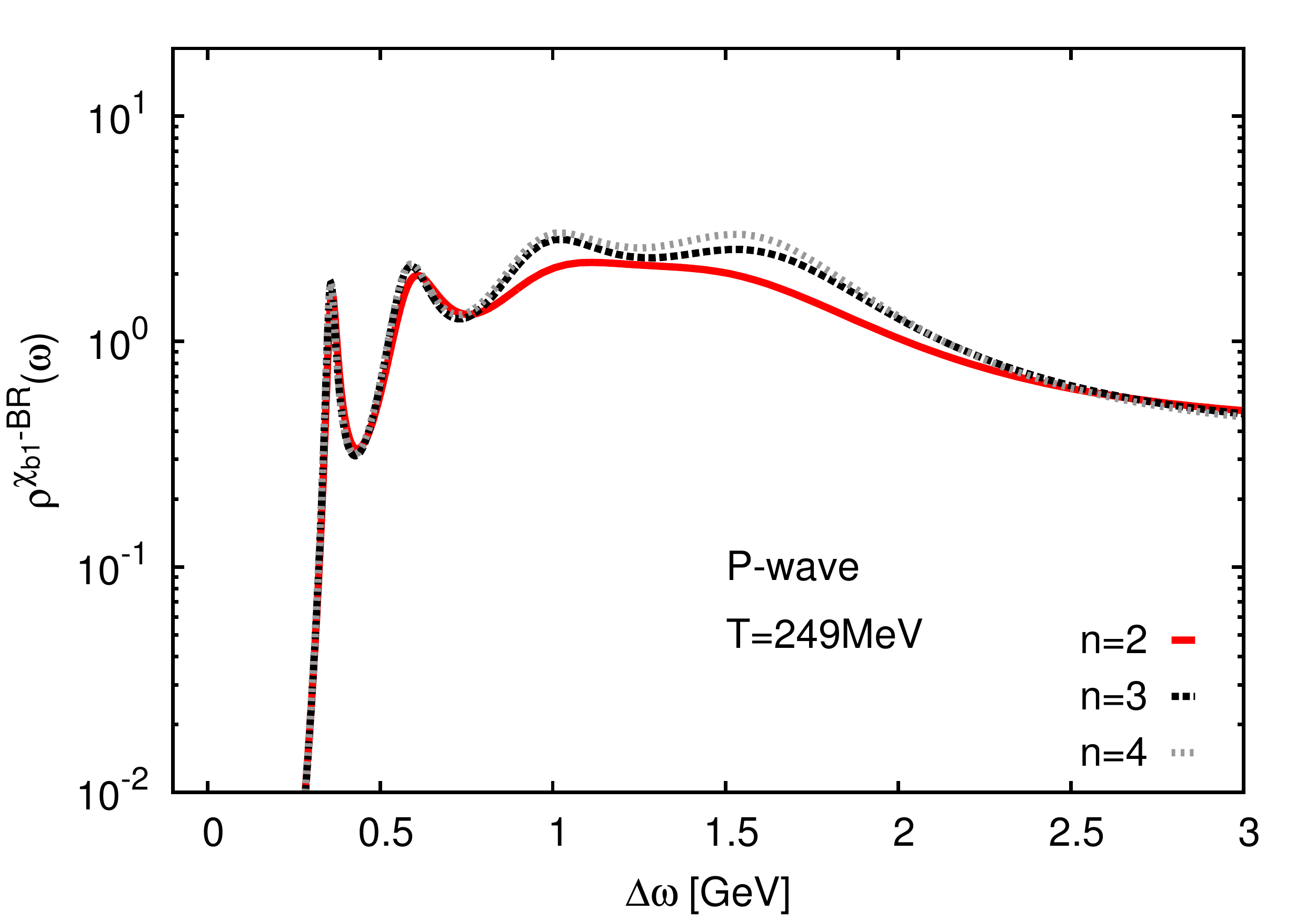}
 \caption{The interacting spectra for S-wave (left) and P-wave (right)
   extracted from the NRQCD correlators measured on the $T>0$ ,
   $\beta=7.280$ lattices at three different values of the parameter
   $n=2,3,4$. Again consistent with expectations the spectra are
   increasing in amplitude at higher frequencies $\omega>1$GeV. In
   both channels the ground state peak is virtually unaffected by the
   changes in the high frequency behavior of
   NRQCD.}\label{Fig:IntNRQCDSpecLep}
\end{figure*}

Since on the highest temperature lattices deployed in our study the
product $a_s(\beta=7.280) M_b=1.559$ lies only slightly above the
limiting value derived from noninteracting lattice NRQCD, we need to
ascertain, whether the results from our choice of $n=2$ are good
enough to capture the bound state physics we are interested in. This
question is related to the fact that all reconstructed spectra shown
in this manuscript do contain finite contributions beyond the inverse
lattice spacing, where in a relativistic description, one would expect
a relatively sharp cutoff. For NRQCD this is indeed not the case and
EFT induced artifacts can populate frequencies up to much larger (but
finite) values, as shown in the analytically determined free NRQCD
lattice spectral functions given in
Fig. \ref{Fig:AnalyticFreeSpecN2}. Note that in case of a breakdown of
the NRQCD approximation, the spectrum would be essentially unbounded
in $\omega$.

To check whether changes in the high frequency regime of the theory
affect the reconstruction of the ground state peak investigated in
this study, we either need to generate configurations with different
physical anisotropy or change the NRQCD temporal discretization
parameter to larger values than our standard setting $n=2$. We choose
the latter option and present the tests, comparing $n=2,3,4$ in the
following.

To understand the change induced by increasing $n$ to four, we inspect
the corresponding analytic free spectra plotted in
Fig. \ref{Fig:AnalyticFreeSpecN4}. The most obvious effect is a
significant reduction of the maximum frequency, up to which the
lattice EFT artifacts populate the spectrum. We find it to move
towards the origin by more than a factor of three. Interestingly at
the same time the amplitude of the kinked structure is slightly larger
than at $n=2$.  The most important fact however is that the region of
low frequencies $\omega<1$GeV in which the bound state physics is
located does not change appreciably with changing the NRQCD effective
temporal step size.

With this intuition at hand, we proceed to measure the lattice NRQCD
bottomonium correlation functions both on non- and fully interacting
lattices, using the different settings $n=2,3,4$ and subsequently
perform spectral reconstructions. We restrict ourselves to
$\beta=7.280$, since here the mass parameter $a_s M_b=1.56$ is closest to
the free NRQCD limit and the noninteracting spectra show the most
pronounced changes. Of interest is in particular, whether the change
in discretization affects the conclusion drawn about the presence of a
well-defined ground state peak in the P-wave channel at $T=249$MeV.

The free spectral functions are shown in
Fig. \ref{Fig:FreeNRQCDSpecLep}. We find that, as hinted at by the
analytically calculated free spectra, the amplitude of $\rho(\omega)$
goes up as we increase $n$ at frequencies above $1$GeV. In the S-wave
channel the number of reconstructed wiggles remains the same between
all choices of $n$ and the lowest lying peak only shows a minute
increase in strength for $n=4$, much smaller than the factor ten
separating it from the interacting spectrum as seen in
Fig. \ref{Fig:CompFreeIntS}. In the P-wave channel for $n=4$ the
reconstruction shows an additional wiggle, however there are no
indications that the overall strength at small frequencies
increases. In all, these findings indicate that the free spectral
functions obtained numerically with $n=2$ at the relevant (unshifted)
frequencies $\Delta\omega<1$GeV are robust against changes in the high
frequency behavior of NRQCD.

As final step we need to perform the same comparison with the spectra
based on NRQCD correlators from the fully interacting lattices,
evaluated with $n=2,3,4$. The results are shown in
Fig. \ref{Fig:IntNRQCDSpecLep}. Just as we saw in the case of the free
spectral functions, stepping up the value of $n$ increases the
amplitude of the spectrum at frequencies above $1$GeV. As is clearly
visible, both in the S-wave (left) and P-wave (right) channel, the
lowest lying peak structure remains virtually unchanged as we vary
$n$. This is a strong indicator that the spectra obtained at $n=2$
are robust against changes in the high frequency behavior of NRQCD and
thus are a reliable representation of the underlying QCD bound state
physics. Together with the observed robustness of the free spectral
function reconstruction, these findings reassure us that the
observation of a well-defined peak structure in the P-wave channel at
$T=249$MeV is not simply an artifact, neither of the Bayesian
reconstruction nor of the discretization prescription of the deployed
EFT.

\end{document}